\documentclass[12pt]{iopart}
\usepackage{iopams}
\usepackage{graphicx}
\usepackage[usenames,dvipsnames]{xcolor}
\usepackage{url,cite}

\def\lsim{\mathop{\hbox{${\lower3.8pt\hbox{$<$}}\atop{\raise0.2pt\hbox{$\sim$}}
$}}} \def\gsim{\mathop{\hbox{${\lower3.8pt\hbox{$>$}}\atop{\raise0.2pt\hbox{$
\sim$}}$}}}  
 3
 
\newcommand\eprint[1]{\href{http://arXiv.org/abs/#1}{#1}}

\providecommand\muK{\mu\mathrm{K}}
\providecommand\mv{{\mathrm v}}
\definecolor{dgreen}{rgb}{.1,.6,.1}
\definecolor{MyDarkRed}{rgb}{0.71,0.14,0.07}
\usepackage{hyperref}
\hypersetup{
    colorlinks=true,
    citecolor=blue,
    linkcolor=magenta,
    filecolor=cyan,      
    urlcolor=cyan,
}
\begin{document}
\title[Model--independent analyses of CMB non--Gaussianity]{Model--independent analyses of non--Gaussianity in Planck CMB maps using Minkowski Functionals}
\author{Thomas Buchert$^1$, Martin J. France$^1$, and Frank Steiner$^{1,2}$}
\smallskip
\address{$^1$Univ Lyon, Ens de Lyon, Univ Lyon1, CNRS, \\$\ \, $Centre de Recherche Astrophysique de Lyon UMR5574, F--69007, Lyon, France\\
$^2$Ulm University, Institute of Theoretical Physics, D--89069 Ulm, Germany
\\
Emails: buchert@ens--lyon.fr, martin.france@ens--lyon.fr, frank.steiner@uni--ulm.de}
\begin{abstract}
Despite the wealth of $Planck$ results, there are difficulties in disentangling the primordial non--Gaussianity of the Cosmic Microwave Background (CMB) from the secondary and the foreground non--Gaussianity (NG). For each of these forms of NG the lack of complete data introduces model--dependencies. 
Aiming at detecting the NGs of the CMB temperature anisotropy $\delta T$, while paying particular attention to a model--independent quantification of NGs, our analysis is based upon statistical and morphological univariate descriptors, respectively: 
the probability density function $P(\delta T)$, related to $\mv_{0}$, the first Minkowski Functional (MF), and the two other MFs, $\mv_{1}$ and $\mv_{2}$. From their analytical Gaussian predictions we build the discrepancy functions $\Delta_{k}$ (\textit{k=P,0,1,2}) which are applied to an ensemble of $10^{5}$ CMB realization maps of the $\Lambda$CDM model and to the $Planck$ CMB maps. In our analysis we use general Hermite expansions of the $\Delta_{k}$ up to the $12^{th}$ order, where the coefficients are explicitly given in terms of cumulants. Assuming hierarchical ordering of the cumulants, we obtain the perturbative expansions generalizing the $2^{nd}$ order expansions of Matsubara to arbitrary order in the standard deviation $\sigma_0$ for $P(\delta T)$ and $\mv_0$, where the perturbative expansion coefficients are explicitly given in terms of complete Bell polynomials. The comparison of the Hermite expansions and the perturbative expansions is performed for the $\Lambda$CDM map sample and the $Planck$ data.
We confirm the weak level of non--Gaussianity ($1$--$2$)$\sigma$ of the foreground corrected masked $Planck$ $2015$ maps.
\end{abstract}
\vspace{-16pt}
%
\vspace{-7pt}
\section{General context}
\label{context}
\vspace{-5pt}
In his March 26, 2010 talk (Observational Constraints on Primordial Non--Gaussianity) 
during the `Non--Gaussian Universe' workshop at the Yukawa institute (YITP), Eiichiro Komatsu \cite{komatsu1} concluded about 
the statistical analysis of the Cosmic Microwave Background (CMB):  
``So far, no detection of primordial non--Gaussianity of any kind by any method''\footnote{\url{http://wwwmpa.mpa-garching.mpg.de/~komatsu/talks.html} .}. This 
conclusion, strengthened by the analysis of the last available CMB data at the time ($WMAP$ 7yr)---and the reader may judge,
after reading this paper and in the future, whether we can say more after $Planck$---sounded 
like a challenge to the cosmological community. An ongoing challenge, because the high--precision $Planck$ CMB 
data are certainly not exhaustively analysed today. A challenge also as there are as many analytical 
definitions of `non--Gaussianity' as there are different statistical 
descriptors calling for the application of unified statistical analysis tools. 
But mostly a challenge, because in the frame of the $\Lambda$CDM model and various 
inflationary models, a non--detectable up to a non--negligible primordial non--Gaussianity may be 
expected \cite{martin1,martin2,planck1}, as measured by
bi-- and tri--spectra \cite{renauxpetel}. 
In various other models of primordial physics 
we may expect different kinds of primordial non--Gaussianity for the CMB as, e.g., in string gas cosmology 
\cite{brandenb1}, or even large non--Gaussianity such as in some ekpyrotic phase models \cite{ijjas}. 
Furthermore, it is a challenge given that a sufficiently high tensor--to--scalar ratio $r$ should allow for a slight 
detection of primordial gravitational waves with the non--Gaussianity of the CMB polarization (B--mode) 
and temperature maps correlation function ($\langle BTT \rangle$ bispectrum)\footnote{However, small angular scales 
(l$>$200) temperature anisotropies have to be analysed to reach the convergence power spectrum of the B--modes; unfortunately, 
at such scales the Sunyaev--Zel'dovich effect and reionization scattering pollute the temperature lensing reconstruction.} \cite{meerburg,seljak} and see (inter alia) the chapter ``$7$--Lensing and the CMB'' in the book by Durrer \cite{durrer}.

\smallskip

The observable part of the CMB covers, typically, 
depending on the frequency, up to 70$\%$ of the celestial sphere, and one has to assume what the 
properties of the remaining invisible CMB are. 
For this problem there is no model--independent analysis of the CMB---a model is required to 
infer and build the CMB regions hidden beyond the galaxy mask and beyond each field source out of 
the main foreground mask. It is interesting to see that some of the CMB anomalies reviewed in \cite{schwarz}, such as the cold 
spot or the hemispherical asymmetry, are very likely independent of the masked CMB reconstruction model. 
But, most of the various CMB anomalies, even of this magnitude, are consistent with a certain level of Gaussianity 
(p--value $> 99\%$). Some of these anomalies are detected with the 2--point correlation function; some anomalies could be remedied by a cutoff between 60 and 160$^{\circ}$ as in \cite{aurich1}, statistically favouring multi--connected universe models 
with finite volume \cite{aurich2,aurich3}.   
Besides assumptions on the topology of the Universe, an assumption is needed for the geometry of the support manifold, 
commonly thought of as being an ideal sphere. This is related to the estimate of the relative motion vector of the observer to the CMB. 
Will peculiar--velocity analyses of larger and larger catalogues converge to this 
motion, or is there a global dipole of a non--idealized space form? What is the correct definition of ``peculiar--velocities"?
There can be significant differential expansion of space that is not allowed for in a Newtonian model of 
structure formation \cite{bolejkoetal}.
Another challenge to the CMB non--Gaussianity is the way to link the specific CMB 
intensity, $I(\nu)$, to the CMB temperature anisotropy, $\delta T$
\cite{notari}, using the Taylor expansion at first order only; higher orders may eventually impair
the evaluation of the primordial non--Gaussianity. We shall explicitly address near--Gaussian expansions in the present paper.

\smallskip

Given the variety of potential contaminations in the cosmic microwave background map, a good strategy
to analyse it and discriminate primordial non--Gaussianity from secondary effects is not only to multiply 
the statistical methods, but to head for model--independent estimators. For that, integral geometry 
provides us with a general mathematical framework where a small set of descriptors allows for a complete 
morphological analysis over random fields such as the CMB temperature maps. The descriptive power of 
integral geometry relies on the polynomial of convex bodies in three dimensions, introduced by J. Steiner (1840) 
\cite{jsteiner}, its generalization to the mixed volume associated to a convex body by Minkowski 
\cite{minkowski}, and the Blaschke problem and diagram. Then, the Bonnesen enhanced isoperimetric 
inequality (1921), the Aleksandrov (Fenchel) inequalities (1937), the Hadwiger works and theorem 
(1955, 1957) \cite{hadwiger}, the studies by Santal\'o (1976) \cite{santalo}, the statistical predictions 
for random fields by Adler (1981) \cite{adler} and the work by Tomita (1986) \cite{tomita1,tomita2} 
bring the key mathematical foundations to the Minkowski Functionals (henceforth `MFs'). Interesting theoretical and 
applied developments are found in the mathematical reviews by Groemer \cite{groemer}, Schneider 
\cite{schneider} and Mecke \cite{mecke1}.   

\smallskip

The theory of $2D$ Gaussian random fields on the CMB 2--sphere was developed by Bond and Efstathiou (1987) \cite{bond}. 
They applied it to the number density of hot and cold spots and to the ellipticity of peaks. 
This has later been generalized for extrema counts and ellipticity contour lines by Aurich et al. \cite{aurich4} 
and also by Pogosyan et al. \cite{pichon1,pichon2,pichon3}.

\smallskip

Minkowski Functionals comprise the by now well--known set of scalar functionals, being rotation and translation 
invariant, Minkowski additive\footnote{Minkowski additivity assigns a functional of the union of bodies 
to the functionals of the individual bodies minus the functionals of their intersection.}, and 
conditionally continuous. This set of MFs describes the morphology of any convex body\footnote{Even more 
generally, the morphology of non--convex bodies is made possible using the property of Minkowski additivity, 
extending the analysis to the convex ring.} in a complete and unique way in the sense of Hadwiger's theorem. 
The explicit introduction of the MFs into cosmology (describing the morphology of galaxy distributions) was 
made in statistics of large--scale structure using the Boolean grain description by Mecke et al. (1994)
\cite{mecke2} with follow--up studies of galaxy catalogues \cite{kerscher1,kerscher2,wiegand}; Kerscher wrote a review on the MFs including applications to cosmology \cite{kerscherreview}. 
In 1997, Schmalzing and Buchert introduced the MFs for the excursion set approach, suitable for any density 
or temperature contour maps \cite{schmalzingbuchert}. In 1998, Schmalzing and G\'orski are the first to apply 
the set of $3+1$ MFs of the CMB 2--sphere (curvature=$+1$) to COBE DMR data excursion sets on a quadriteralized 
spherical cube tesselation ($6~[2^{2(N-1)}]$ pixels) \cite{schmalzinggorski}, implementing also the Gaussian 
premises predicted in 1990 by Tomita \cite{tomita3}. (See also \cite{schmalzingPHD}.)
Also, galaxy catalogues have been analysed with the excursion set approach \cite{schmalzingshapef,hikage}, and a generalization to 
vector--valued MFs has been proposed \cite{beisbart} and applied to the morphological evolution of galaxy clusters \cite{beisbartcluster}.

\smallskip

At present, many fundamental tools are available to take up the challenge in the broad sense of several different but unified statistical descriptors
to ask:  is the cosmic microwave background Gaussian?     
Looking at CMB non--Gaussianity within the MF approach was first undertaken by Winitzki and Kosowsky \cite{winitzki}, and by Novikov et al \cite{novikovetal}.
Following work by Takada {\it et al.} \cite{takada} on the detectability of the weak lensing with the two--point correlation function, a further study predicts that the weak lensing effect could be detected directly 
with the MFs $\mv_{1}$ and $\mv_{2}$ \cite{schmalzing:lensing}. Not only this capability of the MFs was confirmed in later work, but also the lensing--induced morphology changes in modified gravity theories
could be detected. Furthermore, the lensing and the Sunyaev--Zel'dovich non--Gaussianity can be separately detected with MFs, all of this for CMB temperature maps
at high resolution (up to $\ell = 3000$) \cite{munshi}. Specific, regional morphological features like the {\it Cold Spot} could be detected by local analyses with MFs \cite{zhao}.
 
Since then the majority of works rely on model assumptions, either by testing a given 
inflationary model as in \cite{novikovetal}, or explicitly replacing the analytic Gaussian premises for the MFs 
by the MFs of a grid of $\Lambda$CDM model maps: the so--evaluated deviations from non--Gaussianity yield what we below call the `difference of the normalized MFs' (abridged by ${\rm Df}$)
(see \cite{komatsu2,eriksen,ducout,modest,planck2,planck3}, to mention only a few works in this context). These authors argue 
that the noise, the mask and the pixel effects are better taken into account by using a model ensemble rather than using analytical predictions for evaluating the data map non--Gaussianity. 

\smallskip

Ade et al. \cite{planck3} propose a rather exhaustive (claimed model--independent) investigation
of the CMB isotropy and statistics and neither reach a clear rejection nor a confirmation of the standard FLRW cosmological model.
It is then a natural next step to investigate perturbative models at a FLRW background (here the works by Matsubara \cite{matsu1,matsu2,hikage2} stand out as a sustained such attempt). In this paper we follow another route. We focus on and specify model--independent methods to quantify non--Gaussianity,
and we compare with what is obtained when applying the standard model--dependent perturbative ansatz. General perturbative expansions 
are based on series of terms, some of which are solely specified in terms of the chosen model \cite{juszkiewicz,blinnikov,bernardeau}, leaving a certain degree of arbitrariness in the application of perturbation theory. Beyond perturbation theory taken in this broad sense, we can fundamentally explore the non--Gaussianity only when using non--perturbative expansions. We shall introduce these latter paying careful attention to some methodological details that may have a strong impact on the extremely weak level of CMB non--Gaussianity.

\smallskip
\noindent
This article is structured as follows: we investigate in section~\ref{sec:v0} specific descriptors of the map ensemble, such as the probability density function, section~\ref{PDF}, and derived descriptors of non--Gaussianity such as the discrepancy functions and Hermite expansions, section~\ref{discrepancyandhermite}, putting the first Minkowski Functional into perspective in section~\ref{subsec:v0}.
For the Hermite expansion coefficients we give closed expressions in terms of the cumulants using the complete Bell polynomials.
In section~\ref{accuracy} we demonstrate in general terms the accuracy of Hermite expansions of the discrepancy functions that apply to any non--Gaussianities of arbitrary magnitude, and we show that the Hermite expansions approach the discrepancy functions with a mean square error as small as needed. 
We then demonstrate that the Hermite expansions can be truncated to yield the perturbative expansions with
hierarchical ordering of the cumulants, usually considered in the literature, in section~\ref{HO}. We discuss our descriptors for the remaining morphological Minkowski Functionals in section~\ref{sec:v1v2}, and
proceed by comparing the discrepancy function approach with other approaches in the literature in 
section~\ref{discrepancy}.
We apply our descriptors to the $Planck$ $2015$ data with and without masks in section~\ref{planck15},
and we discuss the results by addressing the issue of the origin of non--Gaussianities in 
section~\ref{discussion NGs}. Some conceptual problems related to the range of analysis are raised in 
section~\ref{discussion sigma0}, while a short situational analysis of the $f_{\rm NL}$ results is given in
section~\ref{discussion fNL}.
In section~\ref{conclusion} we give a short conclusion.
Basic definitions and general remarks on uncertainties, the construction of model maps, and the problem of discretization are addressed in \ref{appA}. 
Some important mathematical facts about cumulants, the Hermite and Edgeworth expansions, and the derivation of closed expressions for the Hermite expansion coefficients are given in \ref{appB} and \ref{appC}.

\section{Probability Density Function and Minkowski Functionals---Descriptors for CMB non--Gaussianity and Application to Planck~2015 Data}
\label{sec:v0}

The CMB anisotropy $\delta T$ is described by a scalar random field on the surface of last scattering. The CMB support manifold is idealized by the constant curvature $2-$sphere ${\cal S}^{2}$, but 
different supports of variable curvature may be envisaged for the CMB. Furthermore, the thickness of the surface of last scattering is not taken into account. The manifold ${\cal S}^{2}$ is very convenient and universally adopted and we use it in the present work too. Upon this chosen manifold, any statistical descriptor such as the Minkowski Functionals is otherwise model--independent.    

\subsection{The probability density function of the CMB anisotropy}
\label{PDF}

An important statistical descriptor of the random variable $\delta T$ is its 
{\em probability density function} (PDF) or {\em frequency function} $P(\tau) \ge 0$, where we assume that $\tau$ can take on any real value, and 
$P(\tau)$ is continuous. (The more general case will be discussed below.) $P(\tau) {\mathrm d}\tau$ gives the probability of finding $\delta T$ in between $\tau$ and $\tau + {\mathrm d}\tau$. Hence, $P(\tau)$ is normalized to unity. 

\begin{figure}[!htb]
\includegraphics[height=0.6\textheight,width=1.00\textwidth]{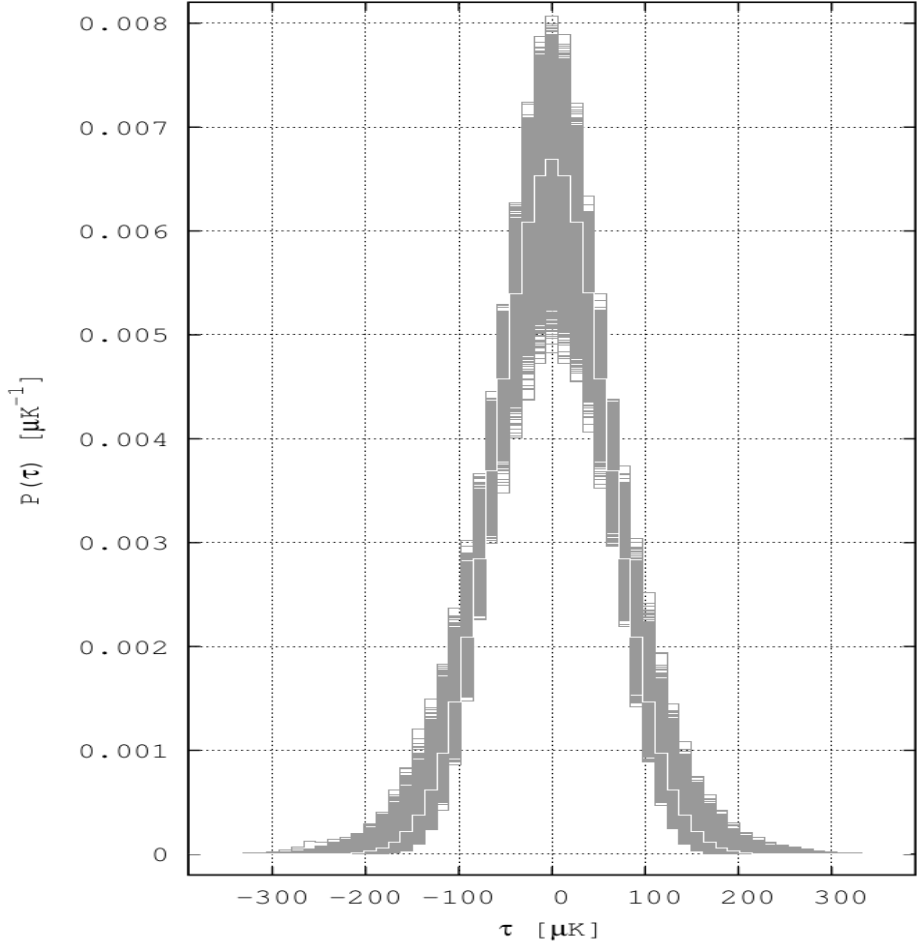}
\caption{$\Lambda$CDM map sample, $N_{\rm side}$=128, $\ell_{\rm range}$=[2,256] without mask, 2$^{\circ}$ fwhm, for a 13$\mu$K temperature bin width over the largest temperature range of the sample ($\pm$396.5$\mu K$). We plot the envelope of P$(\tau)$ histograms for the ensemble of $10^5$ simulation maps. Each PDF histogram is normalized to $1$ (the cosmic variance being not shown here as it is not normalized to $1$). The small asymmetries of this envelope reflect a non--vanishing skewness. The average of all the individual PDFs is the white histogram from which the centre $\tau_{\ell}$ of each white horizontal segment serves as comparison point with the Gaussian hypothesis (of the averages), used to calculate the discrepancy function. (Model and computational details are given in \ref{appA}).}
\label{PFR_Fig1}
\end{figure}

\noindent
Figure \ref{PFR_Fig1} shows the $10^{5}$ histograms envelope of the individual PDFs in the $\Lambda$CDM sample over a total temperature range $\pm$396.5$\mu K$, divided into $61$ bins of $13\mu K$. Detailed informations regarding the $\Lambda$CDM map sample generation, the numerical methodology, and the conventions adopted for the analysis all along this work are given in \ref{appA}.

\noindent
From $P(\tau)$ one obtains the {\em cumulative distribution function} $F(\tau)$ (CDF),
\begin{equation}
\label{F}
F(\tau) := {\rm prob}(\delta T < \tau) = \int_{-\infty}^{\tau}  P(\tau') {\mathrm d}\tau' \;\;,
\end{equation}
respectively, the {\em complementary cumulative distribution function} $F_C (\tau)$ (CCDF)\footnote{It will turn out that $F_C (\tau)$ is identical to the {\em Minkowski Functional} $\mv_0$ (see below).},
\begin{equation}
\label{FC}
F_C (\tau) := {\rm prob}(\delta T \ge \tau) = \int^{\infty}_{\tau}  P(\tau') {\mathrm d}\tau' = 1 - F(\tau)\;\;,
\end{equation}
satisfying $F(-\infty) = 0$, $F(\infty) =1$, respectively, $F_C ( - \infty) =1$, $F_C (\infty) = 0$.
Note that $F( \tau)$ is non--decreasing and ${\rm prob} (\tau_1 \le \tau < \tau_2 ) = F(\tau_2 ) - F(\tau_1 )$.
 
For a continuous function $f: {\mathbb R} \rightarrow {\mathbb R}$, $f (\delta T)$ is again a random field whose {\em expectation value}
is defined as:
\begin{equation}
\label{E}
\left\langle f (\delta T) \right\rangle : = \int_{-\infty}^{\infty} f(\tau) P(\tau) {\mathrm d}\tau \;\;,
\end{equation}
if the integral exists. An important role is played by the {\em moments} $\alpha_n$ of $\delta T$,
\begin{equation}
\label{alphan}
\alpha_n : = \left\langle (\delta T)^n \right\rangle : = \int_{-\infty}^{\infty} \tau^n P(\tau) {\mathrm d}\tau \quad;\quad n = 0,1,2, \cdots \;\;,
\end{equation}
the {\em central moments} $m_n$ (i.e., the moments about the {\em mean} $\mu : = \alpha_1$),
\begin{equation}
m_n : =  \left\langle (\delta T - \mu)^n \right\rangle : = \int_{-\infty}^{\infty} (\tau - \mu)^n P(\tau) {\mathrm d}\tau \;\;,
\end{equation}
and the {\em cumulants} $\varkappa_n : = \left\langle (\delta T)^n \right\rangle_C $. The {\em generating function of the moments} $\alpha_n$ is
given by:
\begin{equation}
\label{M}
M (x) : = \left\langle \e^{x \delta T} \right\rangle =  \int_{-\infty}^{\infty} \e^{x \tau} P(\tau) {\mathrm d}\tau = \sum_{n =0}^{\infty} \alpha_n \frac{x^n}{n!} \;\;,
\end{equation}
from which one obtains the {\em generating function of the cumulants} $\varkappa_n$, 
\begin{equation}
\label{C}
C (x) : = \ln M(x) = \sum_{n =1}^{\infty} \varkappa_n \frac{x^n}{n!} \;\;.
\end{equation}
The first few moments and cumulants are $\alpha_0 = m_0 = 1$, the {\em mean values} $\mu : = \alpha_1 = \varkappa_1$, respectively, 
$m_1 =0$, and the {\em variance} $\sigma_0^2$, 
\begin{equation}
\sigma_0^2 : = \left\langle (\delta T - \mu)^2 \right\rangle = m_2 = \alpha_2 - \mu^2 = \varkappa_2 \;\;,
\end{equation}
respectively, the {\em standard deviation} (uncertainty) $\sigma_0 = \sqrt{\alpha_2 - \mu^2}$.
There are the following {\em recurrence relations}:
\begin{equation}
m_n = \sum_{r=0}^{n} \left( \begin{array}{c} n \\ r \end{array} \right)\, (-1)^{n-r} \mu^{n-r} \alpha_r \quad;\quad n = 0,1,2, \cdots \;\;,
\end{equation}
and, for $n \ge 2$:
\begin{equation}
\label{kappan}
\varkappa_n = m_n - \sum_{r=1}^{n-1} \left( \begin{array}{c} n-1 \\ r-1 \end{array} \right)\, \varkappa_{r} m_{n-r} \;\;.
\end{equation}
Any odd non--vanishing central moment $m_{2n +1}$ of the CMB anisotropy $\delta T$ is a measure of the {\em skewness} of $P (\tau)$; the
simplest of these is 
\begin{equation}
m_3 = \varkappa_3 = \alpha_3 - 3\mu \alpha_2 + 2\mu^3 \;\;,
\end{equation}
respectively, the dimensionless {\em skewness coefficient} 
\begin{equation}
\label{skewcoeff}
\gamma_1 : = \frac{m_3}{\sigma_0^3}\;\;.
\end{equation}
This latter indicates a possible asymmetry of $P(\tau)$, i.e., whether the left tail ($\gamma_1 < 0$) or the right tail ($\gamma_1 > 0$) 
is more pronounced. Another important measure of the ``tailedness'' of $P(\tau)$ is the dimensionless {\em excess kurtosis} (sometimes simply called kurtosis or excess), 
\begin{equation}
\label{gamma2}
\gamma_2 : = \frac{m_4}{\sigma_0^4} - 3 \;\;.
\end{equation}
It should be stressed that the basic equations (\ref{F}), (\ref{FC}), and (\ref{E}) will in general only hold for the theoretical continuous limit
distributions of the CMB anisotropy in the sense of ensemble averages in the limit of an infinite ensemble (infinitely many realizations).
For a single realization, as it is the case for the data obtained by $WMAP$ or $Planck$, or in computer simulations using a large but finite number of realizations, the distributions will in general not be continuous but rather discrete and, thus, the PDF will not satisfy $P(\tau) = {\mathrm d}F / {\mathrm d}\tau = - {\mathrm d} F_C / {\mathrm d}\tau$, as is implied by (\ref{F}), respectively (\ref{FC}). Instead, the above Riemann integrals have to be replaced by Riemann--Stieltjes integrals such that, for example, the {\em expectation value} (\ref{E}) is given by:
\begin{equation}
\left\langle f (\delta T) \right\rangle : = \int_{-\infty}^{\infty} f(\tau) {\mathrm d}F(\tau) \;\;.
\end{equation}
The last relation even holds in cases where $F(\tau)$ is ill--behaved, for example, if $F(\tau)$ has at most enumerably many jumps at the discrete points $\tau_{\ell}$, and as long as $F(\tau)$ is a CDF. In this case one has (if the integral and the sum over $\ell$ converge):
\begin{equation}
\left\langle f (\delta T) \right\rangle : = \int_{-\infty}^{\infty} f(\tau) F' (\tau) {\mathrm d}\tau + \sum_{\ell} f(\tau_{\ell}) \lbrack F(\tau_{\ell} + 0) -F(\tau_{\ell} - 0 )\rbrack \;\;,
\end{equation}
where $F'(\tau)$ is the almost everywhere existing derivative of $F(\tau)$. 
An example is the case where $F(\tau)$ is given as a {\em histogram}, i.e., it is a step function and, thus, $F' (\tau) \equiv 0$ almost 
everywhere.

The non--Gaussianities, which are the subject of this paper, are defined as deviations from the analytically known Gaussian prediction
for the PDF and for the Minkowski Functionals that we shall introduce below. In the case of the PDF, the {\em Gaussian prediction} (G) is given by the normal distribution,
\begin{equation}
\label{G}
P^{\rm G} (\tau) = \frac{1}{\sqrt{2\pi} \sigma_0} \e^{-\frac{(\tau - \mu)^2}{2\sigma_0^2}} \;\;,
\end{equation}
which has mean $\mu$ and variance $\sigma^2_0$. From (\ref{G}) one derives with (\ref{M}) the moment generating function,
\begin{equation}
\label{MG}
M^{\rm G} (x) = \e^{\; \mu x + \frac{\sigma_0^2}{2}\, x^2 } \;\;,
\end{equation}
and with (\ref{C}) the generating function for the Gaussian cumulants,
\begin{equation}
\label{CG}
C^{\rm G} (x) = \mu x + \frac{\sigma_0^2}{2}\, x^2\;\;.
\end{equation}
Our numerical results for the CMB anisotropy are based on an ensemble of $10^5$ realizations (for details, see \ref{appA}).
It turns out that the mean value $\mu = \left\langle \delta T \right\rangle$ (calculated from pixels) is negligibly small, $\mu = \mathcal{O} (10^{-7} \mu K$) for sky maps that cover the full sky. Since typical values for the standard deviation $\sigma_0$ are $\sigma_0 \cong 59\mu K$, 
the ratio $\mu / \sigma_0$ is much smaller than $\tau / \sigma_0$ and, thus, we can put $\mu =0$ in (\ref{G}) when comparing with the full PDF. We then obtain from equations (\ref{M}) and (\ref{MG}) the well--known result that all odd moments of the Gaussian prediction $P^{\rm G} (\tau)$ vanish, $\alpha^{\rm G}_{2n+1}=0$, and that the even moments are given by:
\begin{equation}
\label{alphaG}
\alpha^{\rm G}_{2n} = \frac{2^n}{\sqrt{\pi}}\; \Gamma \left( n + \frac{1}{2}\right) \sigma_0^{2n} = \frac{(2n)!}{2^n n!} \sigma_0^{2n}\;\;,
\end{equation}
and increase with increasing $n$. (Note that $m_n^{\rm G} = \alpha_n^{\rm G}$.)
Equation (\ref{alphaG}) gives $\alpha_4^{\rm G} = 3\sigma_0^4$ and, thus, one obtains from (\ref{skewcoeff}) and (\ref{gamma2}) $\gamma_1^{\rm G} = \gamma_2^{\rm G} =0$, which shows that a non--vanishing value of the skewness coefficient $\gamma_1$ and/or of the excess kurtosis 
$\gamma_2$ are quantitative measures of non--Gaussianity. 
A comparison of equation (\ref{CG}) with equation (\ref{C}) shows that all Gaussian cumulants vanish apart from $\varkappa^{\rm G}_2 = \sigma_0^2$ and, therefore, any non--vanishing cumulant with $n \ge 3$ is a measure of non--Gaussianity.

\subsection{Discrepancy functions and Hermite expansions}
\label{discrepancyandhermite}

In the present paper we pursue a general model--independent approach to the PDF and to the Minkowski Functionals (MF) that depends,
in general, on all higher--order poly--spectra. As a measure of non--Gaussianity using the PDF, we consider the dimensionless {\em discrepancy function} $\Delta_P (\tau )$, defined by:
\begin{equation}
\label{Delta1}
\Delta_P (\tau ): = \frac{P(\tau) - P^{\rm G} (\tau)}{P^{\rm G} (0)} \;\;,
\end{equation}
with $P^{\rm G} (0) = (\sqrt{2\pi}\sigma_0)^{-1} = {\rm max} \lbrace P^{G}(\tau)\rbrace$. Here, $P(\tau)$ denotes the ensemble average over a large number of realizations ($10^5$ in our case), compatible with $\mu = \langle \delta T \rangle = 0$, and possessing the standard deviation $\sigma_0$.
The Gaussian prediction is defined by equation~(\ref{G}) for $\mu = 0$, and by identifying the standard deviation with the value $\sigma_0$ that is numerically obtained from $P(\tau)$, i.e., $\alpha_n^{\rm G} = \alpha_n$ for $n=0,1,2$, respectively $\varkappa_1 = 0$ and $\varkappa_2 = \sigma_0^2$. 

Although some models of inflation predict large non--Gaussianities for $P(\tau)$, there is clear evidence from 
$WMAP$ and $Planck$ data, \cite{komatsu2,eriksen,planck2,planck3}, that possible deviations from the Gaussian prediction $P^{\rm G} (\tau)$ are very small. Under very general conditions (for details see \ref{appC}), $\Delta_P (\tau)$ can be written as a product of a Gaussian and a function $h(\nu) \in L^2 (\mathbb{R}, w(\nu) {\mathrm d}\nu)$, where $w (\nu)$ denotes the weight function,
$w(\nu) = \exp (- \nu^2 /2)$, expressed in terms of the dimensionless scaled temperature variable $\nu : = \tau / \sigma_0 \in \mathbb{R}$.

The {\em Hermite polynomials} ${\rm He}_n (\nu)$ provide a complete orthogonal basis in the Hilbert space $L^2 (\mathbb{R}, w(\nu) {\mathrm d}\nu)$. Therefore, $h(\nu)$ possesses a convergent Hermite expansion (see \ref{appC}) and we are led to 
\begin{equation}
\label{Delta2}
\Delta_P (\tau ) = \e^{-\tau^2 / 2\sigma_0^2}\; \sum_{n=3}^{\infty} \frac{a_P (n)}{n!} {\rm He}_n \left(\frac{\tau}{\sigma_0}\right) \;\;,
\end{equation}
which describes the non--Gaussian ``modulations'' of the PDF. Possible non--Gaussianities are parametrized by the real dimensionless coefficients 
$a_P (n)$, where the non--vanishing of any of them is a clear signature of non--Gaussianity. We would like to point out that the 
${\rm He}_n (\nu)$ are the ``probabilist's Hermite polynomials'' that are different from the Hermite polynomials $H_n (\nu)$ commonly used in physics and which are defined with respect to the weight function $\exp (- \nu^2 )$. In the cosmology literature the ${\rm He}_n (\nu)$ are used, but unfortunately denoted as $H_n (\nu)$! The relation between the two is ${\rm He}_n (\nu) = 2^{-n/2} H_n (\nu / \sqrt{2}$), (see, e.g., \cite{abramowitz}).
Our condition on $h(\nu)$ is satisfied as long as $P(\tau)$ and, thus, also $\Delta_P (\tau)$ are piecewise continuous, and if $h(\nu) = o (\exp (\nu^2/4) / \sqrt{\vert\nu\vert})$ for $\vert\nu\vert \rightarrow \infty$ (see \ref{appC}). It is clear that the expansion  (\ref{Delta2}) is particularly useful if only a few terms have to be taken into account such that the series can be cut off at a low
value $n=N$, i.e., can be well--approximated by a polynomial of degree $N$.

In order to obtain a physical interpretation of the non--Gaussianity (NG) parameters $a_P (n)$, we insert $P(\tau) = P^{\rm G} (\tau) + 
[1/(\sqrt{2\pi}\sigma_0) ]\Delta_P (\tau)$ into the definition (\ref{M}) of the moment generating function $M(x)$, which in turn gives with 
(\ref{C}) the following {\em generating function of the cumulants $\varkappa_n$ of $P(\tau)$} (see \ref{appB}):
\begin{equation}
\label{CX}
C(x) = \frac{\sigma_0^2}{2} \ x^2 + \Delta C(x) \;\;,
\end{equation}
with 
\begin{equation}
\label{CX_}
\Delta C(x) := \ln \left[ 1+ \sum_{n=3}^{\infty} \frac{a_P (n) \sigma_0^n}{n!} \, x^n \right] \,=\, \sum_{n=3}^{\infty} \frac{\varkappa_n}{n!} \, x^n \;\;.  
\end{equation}
Then, the cumulants are $\varkappa_1 = 0$, $\varkappa_2 = \sigma_0^2$, and the higher cumulants, for $n \ge 3$, are uniquely determined by 
(\ref{CX_}). Here are the first coefficients $a_P (n)$, expressed in terms of the skewness coefficient $\gamma_1$, equation (\ref{skewcoeff}), the excess kurtosis $\gamma_2$, equation (\ref{gamma2}), and the dimensionless {\em normalized cumulants},
\begin{equation}
\label{CN_}
C_n : = \frac{\varkappa_n}{\sigma_0^n} = \frac{\langle (\delta T)^n \rangle_C}{\sigma_0^n} \;\;:
\end{equation}
\begin{eqnarray}
\label{AP1}
\fl
\qquad a_P (3) = \gamma_1 \quad;\quad a_P (4) = \gamma_2 \quad;\quad a_P (5) = C_5 \quad;\quad
a_P (6) = 10 \gamma_1^2 + C_6  \quad;\nonumber\\
\fl
\qquad a_P (7) = 35 \gamma_1 \gamma_2 + C_7 \quad;\quad
a_P (8) = 56 \gamma_1 C_5 + 35\gamma_2^2 + C_8 \quad;\nonumber\\
\fl
\qquad a_P (9) = 280 \gamma_1^3 + 126 \gamma_2 C_5 + 84 \gamma_1 C_6 + C_9  \quad;\nonumber\\
\fl
\qquad a_P (10) = 2100 \gamma_1^2 \gamma_2 + 120 \gamma_1 C_7 + 210 \gamma_2 C_6 + 126 C_5^2 + C_{10} \quad;\nonumber\\
\fl
\qquad a_P (11) = 5775 \gamma_1\gamma_2^2 + 4620 \gamma_1^2 C_5 + 165 \gamma_1 C_8 + 330 \gamma_2 C_7 + 462 C_5 C_6 + C_{11} 
\quad; \nonumber\\
\fl
\qquad a_P (12) = 15400 \gamma_1^4 + 5775  \gamma_2^3 + 27720 \gamma_1 \gamma_2 C_5 + \nonumber\\ \qquad 9240 \gamma^2_1 C_6 + 
220 \gamma_1 C_9 + 495\gamma_2  C_8 + 792 C_5 C_7 + 462 C_6^2 + C_{12} \;\;.
\end{eqnarray}
Note that there is the general closed expression in terms of the normalized cumulants ($n\ge 3$):
\begin{equation}
\label{A1Bell}
a_P (n) = B_n (0,0,\gamma_1 , \gamma_2 , C_5 , \cdots , C_n ) \;\;,
\end{equation}
where $B_n (x_1 , x_2 , \cdots , x_n )$ denotes the $n^{\rm th}$ complete {\em Bell polynomial} (see \ref{appB}).
An equivalent closed expression in terms of the moments is given in equation (\ref{APC}).

Table \ref{tablemnxn} shows the moments $\alpha_{n}$ (main term, equation (\ref{mainterm}) in \ref{appA:discretization}), 
the cumulants $\varkappa_{n}$ and the dimensionless normalized cumulants $C_{n}$ of $P(\tau)$. 
We here give the three first orders of each for the sample without mask, 
$\alpha_{0,1,2}=1,~ 0 \muK,~ 3558.31519 \muK^2$; $\varkappa_{0,1,2}=1,~ 0 \muK,~ 3558.31519 \muK^2$ and $C_{0,1,2}=1, 0, 1 $, and we obtain
$\sigma_{0}=59.65166\muK$, $\gamma_{1}=-5.1872\times10^{-4}$ and $\gamma_{2}=5.825\times10^{-5}$.

\bigskip

\begin{table}[!htb] 
\centering

\resizebox{\columnwidth}{1.9cm}{
  \begin{tabular}{|c|c|c|c|c|c|c|c|c|c|c|}
    \hline
    \multicolumn{11}{|c|}{(Units of $\mu K^{n}$) ~ ~ $\Lambda$CDM sample~ Full individual map range, no mask, 2$^{\circ}$fwhm, bin 13$\mu$K ~ {\em c.f.} \ref{appA}}\\
    \hline
    $n$ & $3$ & $4$ & $5$ & $6$ & $7$ & $8$ & $9$ & $10$ & $11$ & $12$\\
    \hline
    $\alpha_{n}$ & $-110.103727$ & $37985558.5$ & $-4040069.36$ & $6.8\times10^{11}$ & $-1.5\times10^{11}$ & $1.7\times10^{16}$ & $-6.6\times10^{15}$ & $5.4\times10^{20}$ & $-3.1\times10^{20}$ & $2.1\times10^{25}$\\
    \hline 
    $\varkappa_{n}$ & $-110.103727$ & $737.480513$ & $-122231.715$ & $-108751936.$ & $785344215.$ & $8.2\times10^{11}$ & $1.1\times10^{14}$ & $-1.1\times10^{16}$ & $2.4\times10^{18}$ & $8.4\times10^{20}$\\
    \hline 
    $C_{n}$ & $-5.187\times10^{-4}$ & $5.82\times10^{-5}$ & $-1.618\times10^{-4}$ & $-2.4138\times10^{-3}$ & $2.922\times10^{-4}$ & $5.1255\times10^{-3}$ & $1.14291\times10^{-2}$ & $-1.96036\times10^{-2}$ & $7.14927\times10^{-2}$ & $0.4147370$\\
    \hline 
    \hline
    \hline 
    \multicolumn{11}{|c|}{(Units of $\mu K^{n}$) ~ ~ $\Lambda$CDM sample ~ Equal temperature range ~ (ETR~$\pm$ 201$\mu K$), U73 mask, 2$^{\circ}$fwhm, bin 6$\mu$K}\\
    \hline
    $n$ & $0$ & $1$ & $2$ & $3$ & $4$ & $5$ & $6$ & $7$ & $8$ & $9$\\
    \hline
    $\alpha_{n}$ & $1.0001469$ & $-1.4912\times10^{-3}$ & $3515.73711$ & $-113.475560$ & $36119872.5$ & $-2963938.41$ & $5.9\times10^{11}$ & $-7.8\times10^{10}$ & $1.2\times10^{16}$ & $-2.2\times10^{15}$\\
    \hline 
    $\varkappa_{n}$ & $1.0$ & $-1.4912\times10^{-3}$ & $3515.73710$ & $-97.7480482$ & $-961350.317$ & $741927.368$ & $-1.3\times10^{10}$ & $-3.4\times10^{9}$ & $1.8\times10^{14}$ & $-1.6\times10^{14}$\\
    \hline 
    $C_{n}$ & $1.0$ & $-2.51\times10^{-5}$ & $1.0$ & $-4.689\times10^{-4}$ & $-7.77766\times10^{-2}$ & $1.0123\times10^{-3}$ & $-0.2973678$ & $-1.3108\times10^{-3}$ & $1.1699687$ & $-1.76138\times10^{-2}$\\
    \hline 
  \end{tabular}
}

\caption{Table of the moments $\alpha_n$ (main term, equation (\ref{mainterm})), cumulants $\varkappa_{n}$, and dimensionless normalized cumulants $C_{n}$ of the probability density function $P(\tau)$.}\label{tablemnxn}
\end{table}

\bigskip

\noindent
With the help of the orthogonality relation (see \cite{abramowitz}, p.775, and \ref{appC}),
\begin{equation}
\label{ortho}
\int_{-\infty}^{\infty} \e^{-\nu^2 / 2}\; {\rm He}_m (\nu )  {\rm He}_n (\nu ) \, {\mathrm d}\nu = \sqrt{2\pi}\;  n! \,\delta_{mn}\;\;,
\end{equation}
one derives from (\ref{Delta2}) the following integral representation for the $a_P (n)$'s ($n \ge 3$):
\begin{equation}
\label{AP2}
a_P (n) = \frac{1}{\sqrt{2\pi}} \int_{-\infty}^{\infty} \Delta_P (\sigma_0\nu )\; {\rm He}_n (\nu ) \, {\mathrm d}\nu \;\;.
\end{equation}
If the discrepancy function $\Delta_P$ is known, one can compute from (\ref{AP2}) the non--Gaussianity parameters $a_P (n)$ and then, from (\ref{AP1}) and (\ref{A1Bell}), the normalized cumulants $\gamma_1 , \gamma_2 , C_n$. Vice versa, one can compute the $a_P (n)$'s
from (\ref{AP1}) once the cumulants have been computed from the moments $\alpha_n = m_n$ of the PDF $P(\tau )$, using equations
(\ref{alphan}) and (\ref{kappan}), or directly from a theory of the primordial CMB fluctuations as, e.g., given by the ansatz (\ref{nlcorr}).
With ${\rm He}_{2n+1} (0) = 0$ and ${\rm He}_{2n} (0) = [(-1)^n (2n)!] / 2^n n!$, we obtain from (\ref{Delta2}) and (\ref{AP1}) for the 
discrepancy function at $\tau =0$:
\begin{eqnarray}
\label{DeltaPin0}
\fl\qquad\quad
\Delta_P (0) &= \sum_{n=2}^{\infty} \frac{(-1)^n a_P (2n)}{2^n\, n!} = \frac{a_P (4)}{8} - \frac{a_P (6)}{48} + \frac{a_P (8)}{384} 
- \frac{a_P (10)}{3840} + \frac{a_P (12)}{46080} \mp \cdots \nonumber \\
&= \frac{\gamma_2}{8} - \frac{5}{24}\gamma^2_1 - \frac{C_6}{48} + \frac{35}{384}\gamma_2^2 + \frac{7}{48}\gamma_1 C_5 + \frac{C_8}{384}\mp \cdots \;\;.
\end{eqnarray}

In table \ref{tableaP} we present the values for the first $a_P (n)$'s (see equations (\ref{AP1}) and (\ref{AP2})), respectively for $\gamma_1$, $\gamma_2$, and $C_5$ for the $\Lambda$CDM model (model and computational details are given in \ref{appA}).

\begin{table}[!htb] 
\centering

\resizebox{\columnwidth}{1.5cm}{
  \begin{tabular}{|c|c|c|c|c|c|c|c|c|c|c|} 
      \hline
    \multicolumn{11}{|c|}{$\Lambda$CDM sample~ Full individual map range, no mask, 2$^{\circ}$fwhm, bin 13$\mu$K
    ~ {\em c.f.} eqs. (\ref{AP1}), (\ref{AP2}) and model in \ref{appA}}\\
    \hline
    $n$ & $3$ & $4$ & $5$ & $6$ & $7$ & $8$ & $9$ & $10$ & $11$ & $12$ \\
    \hline
    $a_{P}(n)$ & $-5.1872\times10^{-4}$ & $5.825\times10^{-5}$ & $-1.6184\times10^{-4}$ & $-2.41112\times10^{-3}$ & $2.9116\times10^{-4}$ & $5.13035\times10^{-3}$ & $1.153302\times10^{-2}$ & $-1.964794\times10^{-2}$ & $7.123993\times10^{-2}$ & $0.41622908$\\
    \hline
    $a_{P}(n)$ & $-5.1872\times10^{-4}$ & $5.830\times10^{-5}$ & $-1.6184\times10^{-4}$ & $-2.40919\times10^{-3}$ & $2.9116\times10^{-4}$ & $5.19401\times10^{-3}$ & $1.153302\times10^{-2}$ & $-1.785385\times10^{-2}$ & $7.123994\times10^{-2}$ & $0.45721793$\\
    \hline 
    \hline
    \hline 
    \multicolumn{11}{|c|}{$\Lambda$CDM sample ~ Equal temperature range ~ (ETR~$\pm$ 201$\mu K$), U73 mask, 2$^{\circ}$fwhm, bin 6$\mu$K}\\
    \hline
    $n$ & $0$ & $1$ & $2$ & $3$ & $4$ & $5$ & $6$ & $7$ & $8$ & $$\\
    \hline
    $a_{P}(n)$ & $$ & $$ & $$ & $-5.4435\times10^{-4}$ & $-7.777654\times10^{-2}$ & $1.01233\times10^{-3}$ & $-2.9736480\times10^{-1}$ & $1.7101\times10^{-4}$ & $1.38165948$ & $$\\
    \hline
    $a_{P}(n)$ & $9.9534\times10^{-4}$ & $-2.8\times10^{-7}$ & $1.002019\times10^{-2}$ & $-4.7175\times10^{-4}$ & $5.20385\times10^{-3}$ & $1.00469\times10^{-3}$ & $-1.989114\times10^{-2}$ & $9.19\times10^{-6}$ & $-2.1129824\times10^{-1}$ & $$\\
    \hline 
  \end{tabular}
}
  \caption{Table of coefficients $a_{P}(n)$ for the $\Lambda$CDM model, computed from table \ref{tablemnxn} using equation (\ref{AP1}) in $1^{st}$ line, then using equation (\ref{AP2}) in $2^{nd}$ line.}
\label{tableaP}
\end{table}

\noindent
One observes that $a_P (6) =10 \gamma_{1}^{2} + C_{6} < 0$ implying $C_{6} < 0$ in agreement with 
table~\ref{tablemnxn}. This is in contrast to hierarchical ordering (as assumed in perturbation theory \cite{matsu2}) 
where $a_{P}^{\rm HO}(J,6)=6 a_{0}^{\rm HO}(J,5)=10\gamma_{1}^{2}>0$ in second-- and third--order $(J=2,3)$ (see equations (\ref{a0HOg2}) and (\ref{a0HOg3})). Only at fourth and higher order (i.e. $J\ge4$) it holds that $a_{P}^{\rm HO}(J,6)=a_{P}(6)$ (see (\ref{a0HOg4})).
This will be discussed more in detail in subsection~\ref{HO}.

\begin{figure}[!htb]
\includegraphics[height=0.6\textheight,width=1.00\textwidth]{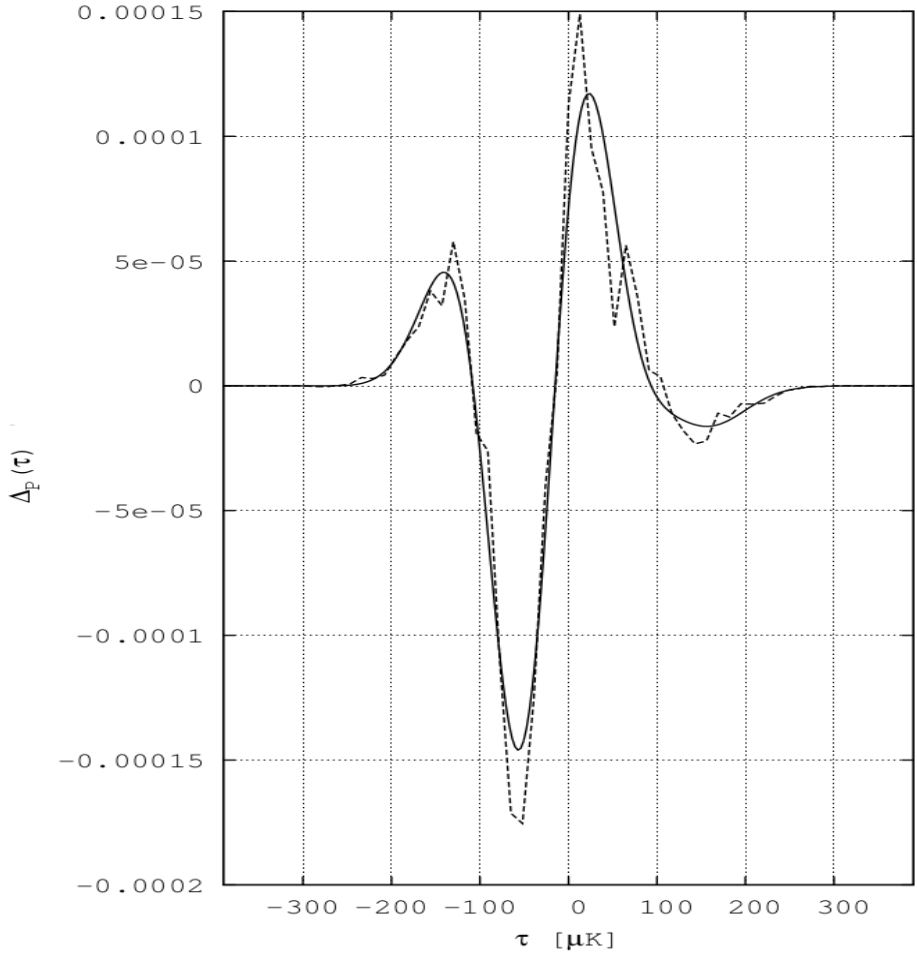}
\caption{$\Lambda$CDM map sample, $N_{\rm side}$=128, $\ell_{\rm range}$=[2,256] without mask, 2$^{\circ}$ fwhm, for a 13$\mu$K temperature bin width over the smallest temperature range covering all the sample maps ($\pm$396.5$\mu K$). We plot the discrepancy function (black dashed line) of the PDF for the $10^5$ maps ensemble ($\sigma'_{0} = \sqrt{\alpha'_{2}}=59.65166\muK$ being used here to calculate $\Delta_{P}$ -- see \ref{appA:discretization} for details). The black solid line shows the Hermite expansion according to equation (\ref{Delta2}) for $n=3$ to $n=8$.}
\label{PFR_Fig2}
\end{figure}

Figure \ref{PFR_Fig2} displays the discrepancy function $\Delta_{P}$ of the averaged PDF ($10^{5}$ map sample). $\Delta_{P}$ is calculated by equation (\ref{Delta1}) over the lattice defined by the mid--points of each segment in the histogram of $P(\tau)$ using $\sigma_{0}$ defined as the ``main term'' of $\alpha_{2}$ in equation (\ref{correction}). This figure also displays the expansion in Hermite polynomials (equation (\ref{Delta2}) limited to the order $8$). 

\begin{figure}[!htb]
\includegraphics[height=0.55\textheight,width=1.00\textwidth]{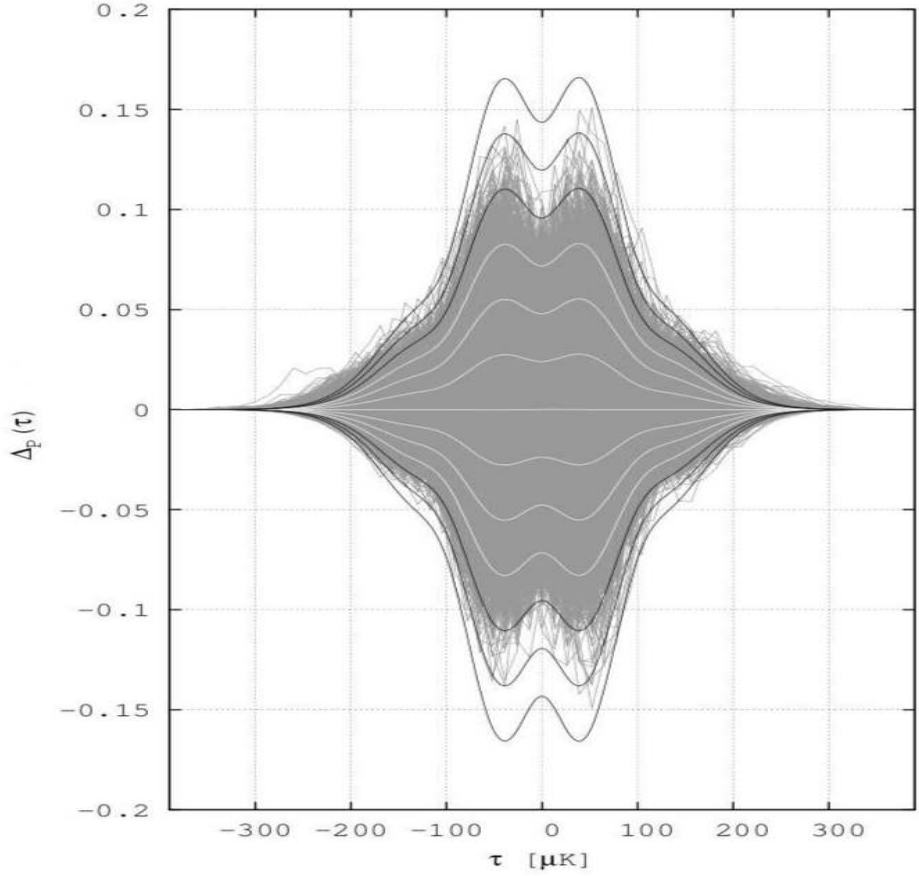}
\caption{$\Lambda$CDM map sample, $N_{\rm side}$=128, $\ell_{\rm range}$=[2,256] without mask, 2$^{\circ}$ fwhm, for a 13$\mu$K temperature bin width over the largest temperature range of the sample ($\pm$396.5$\mu K$). We plot the envelope of the $10^5$ individual discrepancy functions of the PDFs of figure \ref{PFR_Fig1}. The cosmic variance curves of the $10^5$ maps sample appear in white or in black from $1\sigma$ up to $6\sigma$. This proves the high level of non--Gaussianity of several maps. The quasi straight, white horizontal central line is the PDF discrepancy function of the ensemble average.}
\label{PFR_Fig3}
\end{figure}

Figure \ref{PFR_Fig3} shows the envelope of the $10^{5}$ discrepancy functions of the map sample.

\subsection{The first Minkowski Functional $\mv_{0}$}
\label{subsec:v0}

Let us discuss now the simplest morphological descriptor which is given by the first Minkowski Functional (MF). 
We consider the compact {\em excursion set} ${\mathcal Q}_{\nu} \in {\mathcal S}^2$ with boundary $\partial {\mathcal Q}_{\nu} (\nu : = \tau  / \sigma_0 )$:
\begin{equation}
\label{excursionset}
{\mathcal Q}_{\nu} : = \lbrace {\boldsymbol{\hat n}} \in {\mathcal S}^2 \ \Big\vert \ \delta T ({\boldsymbol{\hat n}}) \ge \sigma_0 \nu \rbrace \;\;.
\end{equation}
We define the MF $V_0 (\nu)$  as follows:
\begin{equation}
V_0 ({\mathcal Q}_{\nu}) : = \int_{{\mathcal Q}_{\nu}}\;{\mathrm d}a = {\rm area}({\mathcal Q}_{\nu})\;\;,
\end{equation}
with ${\mathrm d}a$ denoting the surface element on ${\mathcal S}^2$.
In the following we shall be using the {\em normalized MF} $\mv_0 (\nu)$ that is normalized with respect to ${\rm area}({\mathcal S}^2 )= 4\pi$, i.e.:
\begin{equation} 
\label{v0}
\fl
\mv_0 (\nu) := \frac{1}{4\pi} V_0 (\nu) = {\rm prob} (\delta T \ge \sigma_0 \nu ) = F_C (\sigma_0 \nu) = \int_{-\infty}^{\infty} \Theta (\tau' - \sigma_0 \nu) P(\tau') \; {\mathrm d}\tau' \;\;,
\end{equation}
where $F_C$ is the complementary cumulative distribution function (\ref{FC}).
From the definition (\ref{v0}) follows also the relation 
\begin{equation}
\label{Pv0}
P(\tau) = -\frac{1}{\sigma_0} \frac{{\mathrm d}\mv_0 (\nu)}{{\mathrm d}\nu }\Big\vert_{\nu = \tau / \sigma_0} \;\;.
\end{equation}
Thus, the MF $\mv_0 (\nu)$ can be seen as a ``smoothed'' (cumulative) version of the PDF $P(\tau)$. Using the Gaussian prediction
$P^{\rm G} (\tau)$ in equation (\ref{G}), we obtain the {\em Gaussian prediction for $\mv_0 (\nu)$} (setting $\mu = 0$):
\begin{equation}
\label{v0G_}
\mv_0^{\rm G} (\nu) := \frac{1}{2} {\rm erfc}\left(\frac{\nu}{\sqrt{2}}\right) \;\;,
\end{equation}
in terms of the complementary error function \cite{abramowitz}, from which one derives $\mv_0^{\rm G} (-\infty) = 1$,  $\mv_0^{\rm G} (\infty) = 0$,
$\mv_0^{\rm G} (0) = 1/2$, and the symmetry relation $\mv_0^{\rm G} (-\nu) = 1 - \mv_0^{\rm G} (\nu)$.
Furthermore, one has the asymptotic behaviour for $\nu \rightarrow \infty$,
\begin{equation}
\label{v0G}
\mv_0^{\rm G} (\nu) = \frac{1}{\sqrt{2\pi}} \frac{\e^{-\nu^2 /2}}{\nu} \left[ 1 - \frac{1}{\nu^2} + {\mathcal O} \left(\frac{1}{\nu^4}\right)\right]\;.
\end{equation}
In analogy to equation (\ref{Delta1}), we define the dimensionless {\em discrepancy function $\Delta_0 (\nu)$} as follows:
\begin{equation}
\Delta_0 (\nu) : = \sqrt{2\pi} \left( \mv_0 (\nu) - \mv_0^{\rm G} (\nu ) \right) \;\;,
\end{equation}
which in turn is expanded into Hermite polynomials,
\begin{equation}
\label{Deltahermite}
\Delta_0 (\nu) : = \e^{-\nu^2 /2} \, \sum_{n=2}^{\infty} \frac{a_0 (n)}{n! }\,{\rm He}_n (\nu) \;\;,
\end{equation}
with the coefficients
\begin{equation}
\label{A02}
a_0 (n) = \frac{1}{\sqrt{2\pi}} \int_{-\infty}^{\infty} \Delta_0 (\nu ) {\rm He}_n (\nu ) \, {\mathrm d}\nu \;\;.
\end{equation}
The above coefficients $a_0 (n),  \ n \ge 2$, measure a possible non--Gaussianity described by the MF $\mv_0 (\nu)$. With 
\begin{equation}
\frac{{\mathrm d}}{{\mathrm d}\nu }  \left[ \e^{-\nu^2 / 2}\ {\rm He}_n (\nu )  \right] = - \e^{-\nu^2 / 2}\ {\rm He}_{n+1} (\nu )\;\;,
\end{equation}
and the relation (see equations (\ref{Delta2}) and (\ref{Pv0})),
\begin{equation}
\frac{\mathrm d}{{\mathrm d}\nu}\, \Delta_0 (\nu) = - \Delta_P (\sigma_0 \nu)\;\;,
\end{equation}
one obtains the following relation between the coefficients $a_0 (n)$ and $a_P (n)$:
\begin{equation}
\label{APtoA0}
a_0 (n) = \frac{a_P (n+1)}{n+1}\;\;\; (n \ge 2 ) \;\;,
\end{equation}
and, thus, we arrive at the following exact coefficients (see equations (\ref{AP1}) and (\ref{A1Bell})):
\begin{eqnarray}
\label{AP3}
\fl
\qquad a_0 (2) = \frac{\gamma_1}{3} \quad;\quad
a_0 (3) = \frac{\gamma_2}{4} \quad;\quad
a_0 (4) = \frac{C_5}{5} \quad;\quad
a_0 (5) = \frac{5}{3} \gamma_1^2 + \frac{C_6}{6} \quad; \nonumber\\
\fl
\qquad a_0 (6) = 5 \gamma_1 \gamma_2 + \frac{C_7}{7} \quad;\quad
a_0 (7) = 7 \gamma_1 C_5 + \frac{35}{8}\gamma_2^2 + \frac{C_8}{8} \quad;\nonumber\\
\fl
\qquad a_0 (8) = \frac{280}{9} \gamma_1^3 + 14 \gamma_2 C_5 + \frac{28}{3} \gamma_1 C_6 + \frac{C_9}{9} \quad; \nonumber\\
\fl
\qquad a_0 (9) = 210 \gamma_1^2 \gamma_2 + 12 \gamma_1 C_7 + 21 \gamma_2 C_6 + \frac{63}{5} C_5^2 + \frac{C_{10}}{10} \quad;\nonumber\\
\fl
\qquad a_0 (10) = 525  \gamma_1 \gamma_2^2 + 420 \gamma_1^2 C_5 + 15 \gamma_1 C_8 + 30\gamma_2 C_7 + 42 C_5 C_6 + \frac{C_{11}}{11}  \quad;\nonumber\\
\fl
\qquad a_0 (11) = \frac{3850}{3} \gamma_1^4 + \frac{1925}{4} \gamma_2^3 + 2310\gamma_1 \gamma_2 C_5 + \nonumber\\ \qquad770 \gamma_1^2 C_6 + \frac{55}{3} \gamma_1 C_9 + \frac{165}{4} \gamma_2 C_8 + 66 C_5 C_7 + \frac{77}{2} C_6^2 + \frac{C_{12}}{12} \;\;.
\end{eqnarray}

\vspace{20pt}

\noindent
In table \ref{tablea0} we present the values for the first $a_0 (n)$'s of the $\Lambda$CDM model.

\begin{table}[!htb] 
\centering

\resizebox{\textwidth}{1.4cm}{
  \begin{tabular}{|c|c|c|c|c|c|c|c|c|c|} 
    \hline
    \multicolumn{10}{|c|}{$\Lambda$CDM ~ Full individual map range, no mask, 2$^{\circ}$fwhm, bin 13$\mu$K ~ {\em c.f.} equation (\ref{AP3}) and model in \ref{appA}}\\
    \hline
    $n$ & $0$ & $1$ & $2$ & $3$ & $4$ & $5$ & $6$ & $7$ & $8$\\
    \hline
    $a_{0}(n)$ & $$ & $$ & $-1.7295\times10^{-4}$ & $1.456\times10^{-5}$ & $-3.237\times10^{-5}$ & $-4.0185\times10^{-4}$ & $4.159\times10^{-5}$ & $6.4130\times10^{-4}$ & $1.28145\times10^{-3}$\\
    \hline
    \hline
    \hline 
    \multicolumn{10}{|c|}{$\Lambda$CDM ~ Equal temperature range ~ (ETR~$\pm$ 201$\mu K$), U73 mask, 2$^{\circ}$fwhm, bin 6$\mu$K }\\
    \hline
    $n$ & $0$ & $1$ & $2$ & $3$ & $4$ & $5$ & $6$ & $7$ & $8$\\
    \hline
    $a_{0}(n)$ & $$ & $$ & $-1.8145\times10^{-4}$ & $-1.944413\times10^{-2}$ & $2.0247\times10^{-4}$ & $-4.95608\times10^{-3}$ & $2.443\times10^{-5}$ & $1.7270744\times10^{-1}$ & $-1.54858\times10^{-3}$\\
    %
    \hline 
  \end{tabular}
}

  \caption{Table of coefficients $a_{0}(n)$, computed from table \ref{tablemnxn} using equation (\ref{AP3}).}
  \label{tablea0}
\end{table}

\begin{figure}[!htb]
\includegraphics[height=0.6\textheight,width=1.00\textwidth]{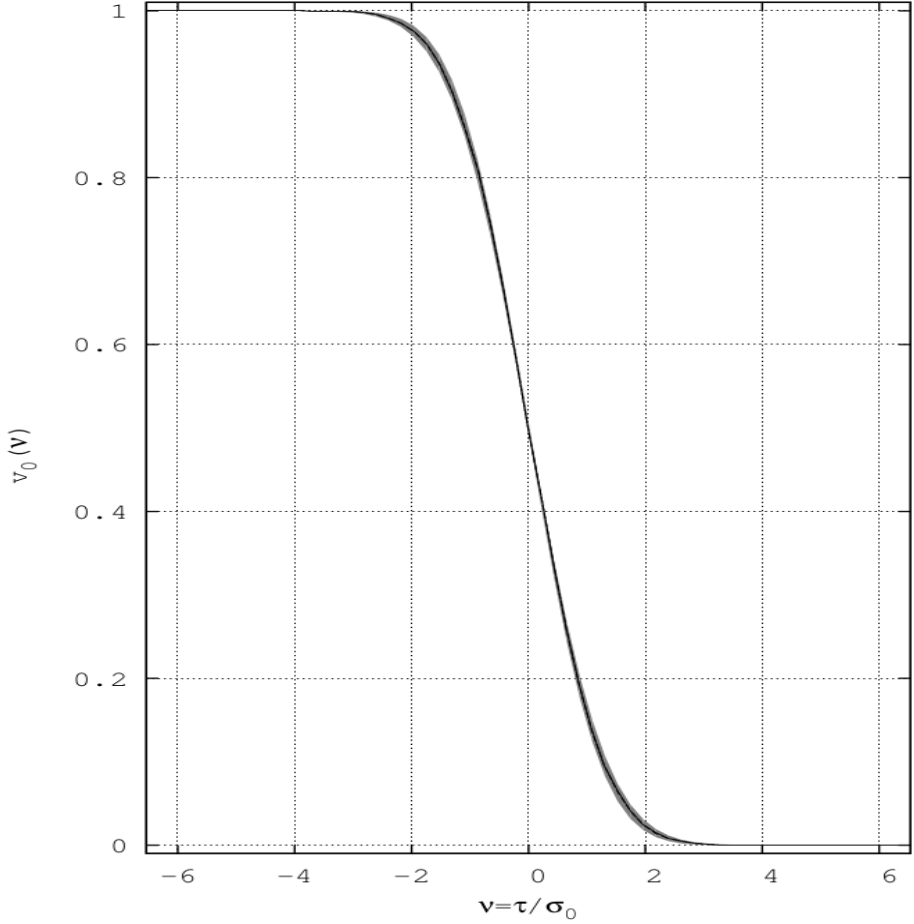}
\caption{$\Lambda$CDM map sample, $N_{\rm side}$=128, $\ell_{\rm range}$=[2,256] without mask, 2$^{\circ}$ fwhm, for a 13$\mu$K temperature bin width over the largest temperature range of the sample ($\pm$396.5$\mu K$). We plot the first Minkowski functional $\mv_{0}(\nu)$ as a black dashed line with its Gaussian premise in black solid line (almost coincident). The $1\sigma$ cosmic variance is displayed as a grey shaded area.}
\label{PFR_Fig4}
\end{figure}

\begin{figure}
\includegraphics[height=0.6\textheight,width=1.00\textwidth]{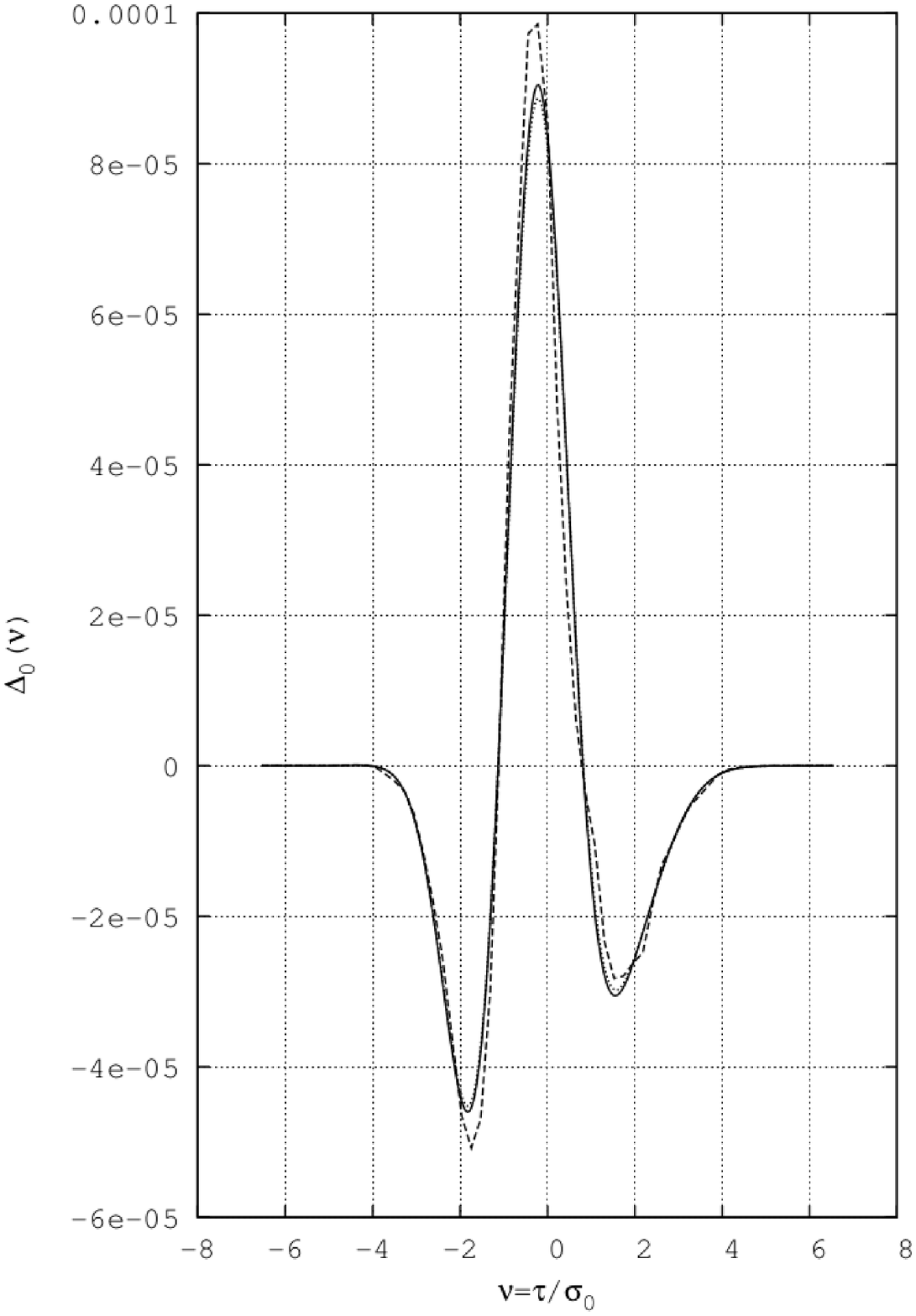}
\caption{We plot $\Delta_{0}(\nu)$, the discrepancy function of $\mv_{0}$ in dashed line ($\sigma_{0px}=59.53348\muK$, the variance from the moments of pixels being used here to calculate $\Delta_{0}$). The Hermite expansions (order $2$ to $5$) appear in black solid line using the coefficients $a_0 (n)$ computed from equation (\ref{A02}), and in dotted line from equation (\ref{AP3}). Coefficients are shown in table \ref{tablea0}.}
\label{PFR_Fig5}
\end{figure}

\noindent
Figure \ref{PFR_Fig4} shows the first Minkowski Functional $\mv_{0}(\nu)$
of the $\Lambda$CDM sample without mask, and figure \ref{PFR_Fig5} its discrepancy function $\Delta_{0}(\nu)$ together
with the Hermite expansion (\ref{Deltahermite}) of $\Delta_0 (\nu)$ (order $2$ to $5$) in $a_0 (n)$. 

\subsection{The accuracy of the Hermite expansion of the discrepancy functions}
\label{accuracy}

The Hermite expansions (\ref{Delta2}) and (\ref{Deltahermite}) for the discrepancy functions $\Delta_{P}(\tau)$ and $\Delta_{0}(\nu)$, respectively, provide very convenient parametrizations and quantitative measures of the CMB non--Gaussianities. They hold under very general conditions (as discussed in \ref{appC}) and do not require that the NGs have to be small. The main assumption is the existence of a unique probability density function $P(\tau)$ (a very natural assumption, indeed, from a physical point of view), which is equivalent to demanding that the ``Hamburger moment problem'' is determinate (see e.g. \cite{vorobyev,reed}). It is then guaranteed that the Hermite expansions are absolutely convergent. In subsections \ref{discrepancyandhermite} and \ref{subsec:v0} we considered polynomial approximations of degree $N$ in $\tau$, respectively in $\nu$ of $\Delta_{P}(\tau)$ and $\Delta_{0}(\nu)$ by truncating the Hermite expansions at $n=N$. A nice property of this truncation is that the accuracy of the approximation is well under control, since there is, for example for $\Delta_{0}(\nu)$, the exact {\em mean square error} of equations (\ref{error}) and (\ref{error2}),
\begin{equation}
\label{epsilon}
{\cal{E}}_{N} : = ||h_{0} - \sum_{n=2}^{N}{\frac{a_{0}(n)}{n!}{\rm He}_{n}}||^{2}~ = ~||h_{0}||^{2} - \sum_{n=2}^{N}{\frac{(a_{0}(n))^{2}}{n!}}\;,
\end{equation}
for $h_{0}(\nu):=e^{\nu^{2}/2} \Delta_{0}(\nu)$. The error ${\cal{E}}_{N}$ gets smaller and smaller if the degree $N$ increases and finally approaches zero in the limit $N \rightarrow \infty$ as a consequence of the completeness relation (Parseval's equation),
\begin{equation}  
\label{completeness}
\sum_{n=2}^{\infty}{\frac{(a_{0}(n))^{2}}{n!}} = ||h_{0}||^{2} := \frac{1}{\sqrt{2\pi}} \int_{-\infty}^{+\infty}{e^{-\nu^{2}/2}(h_{0}(\nu))^{2} {\mathrm d}\nu}\;,
\end{equation}
(here we use the weight function $w(\nu)=\frac{1}{\sqrt{2\pi}} e^{-\nu^{2}/2}$ in the definition of the inner product of the Hilbert space ${\rm I\!H}=L^{2}({\rm I\!R}, w(\nu){\mathrm d}\nu)$, see \ref{appC}.) As an example, let us consider the approximation of degree $N=5$ applied to $\Delta_{0}(\nu)$ for which the {\em mean square error} is explicitly given in terms of the skewness coefficient $\gamma_{1}$, the excess kurtosis $\gamma_{2}$ and the higher cumulants $C_{5}$ and $C_{6}$ by
\begin{equation}
\label{epsilon5}
{\cal{E}}_{5}=||h_{0}||^{2} - \left[\frac{\gamma_{1}^{2}}{18}+\frac{\gamma_{2}^{2}}{96}+\frac{C_{5}^{2}}{600}+\frac{(10\gamma_{1}^{2}+C_{6})^2}{4320}\right]\;,
\end{equation}
using the coefficients $a_{0}(2,3,4,5)$ of (\ref{AP3}). Figure \ref{PFR_Fig5} nicely illustrates that the polynomial approximation of degree 5 provides an excellent description of the discrepancy function $\Delta_{0}(\nu)$. Similar results exist for the discrepancy function $\Delta_{P}(\tau)$ as shown already in figure \ref{PFR_Fig2} at order $8$. 

\subsection{Hierarchical ordering and perturbation theory}
\label{HO}

Most of the commonly studied models of inflation predict weak primordial non--Gaussianities. Among these models there is 
a wide class where the normalized cumulants $C_n$ obey the additional property of {\em hierarchical ordering} (indexed below by {\rm HO}), i.e.,
$C_n \sim \sigma_0^{n-2}, n \ge 2$ (where we assume $C_1 =0, C_2 =1$). 
This suggests to expand the discrepancy function not into Hermite polynomials, but rather at fixed $\nu$ into a power series (perturbation theory) in $\sigma_0$. This approach has been pioneered by Matsubara \cite{matsu1,matsu2} who has derived second--order {\em perturbative formulae} in $\sigma_{0}$ for the discrepancy functions of the MFs.
In the following, we show that the perturbation theory applied to $\Delta_{0}(\nu)$ is a direct consequence of the general Hermite expansion (\ref{Deltahermite}) that can be easily carried out to any order $J$ in $\sigma_{0}$,
i.e., to the NG of order $\sigma_{0}^{J}$ (indexed below by ${\rm HO}(J), J \ge 1$). It turns out that the perturbation expansion ${\rm HO}(J)$ corresponds, at a given order $J$, to a double truncation of the Hermite expansion: i) a truncation of the Hermite expansion at $n=M_{0}(J):=3J-1$, ii) a truncation of the expansion coefficients $a_{0}(n)$ at order $J$ with respect to their expansion into a power series in $\sigma_{0}$. As an example, we give below the explicit formulae at second, third and fourth order, where the second--order formula is identical to Matsubara's result. Analogous perturbative formulae hold for $\Delta_{P}(\tau)$, which have not been given before. As another new result we also present the exact {\em mean square error} of the perturbative formulae and compare it with the corresponding error (\ref{epsilon}) of the original Hermite expansion. 

Let us assume that the hierarchical ordering holds and introduce the ``renormalized cumulants'' (or ``HO--cumulants''),
\begin{equation}
\label{hocumulants}
S_{n}:=\frac{C_{n}}{\sigma_{0}^{n-2}}, n \ge 3\;,
\end{equation} 
which are assumed to be of zeroth order in $\sigma_{0}$. Using the relation (\ref{APtoA0}) and the explicit expression (\ref{A1Bell}) for the coefficients $a_{P}(n)$ in terms of the complete Bell polynomials $B_{n}$, we obtain for $n \ge 2$,
\begin{eqnarray}
\label{Dn-z}
a_{0}(n)&=\frac{a_{P}(n+1)}{n+1}=\frac{1}{n+1} B_{n+1}(0,0,\gamma_{1},\gamma_{2},C_{5},C_{6},\cdots,C_{n+1}) \nonumber\\&
= \frac{1}{n+1} B_{n+1}(0,0,S_{3}\sigma_{0},S_{4}\sigma_{0}^{2},\cdots,S_{n+1}\sigma_{0}^{n-1}) \;, i.e. \nonumber\\& 
a_{0}(n)=D_{n-1}(\sigma_{0}), n \ge 2 \;,
\end{eqnarray}
where $D_{n}(z)$ is a polynomial of degree $n$ in $z$ with $D_{0}(z):=1$ and
\begin{equation}
\label{symmetryrel}
D_{n}(-z)=(-1)^{n} D_{n}(z)\;.
\end{equation}
Thus, $a_{0}(n)$ is a polynomial of degree $n-1$ in $\sigma_{0}$, 
\begin{equation}
\label{a0ntoggle}
a_{0}(n)=\sum_{j=j_{0}(n)}^{n-1}{D_{n-1,j}(S_{3},S_{4},\cdots,S_{n+1})\sigma_{0}^{j}}\;\;,\; n \ge 2 \;.
\end{equation}
Here, $j_{0}(n)$ denotes the smallest power of $\sigma_{0}$ which contributes to $a_{0}(n)$ and is given for $n \ge 2$ by
\begin{equation}
\label{j0}
j_{0}(n):=n(k,l)-(2l+1)=k+l-1 \ge l+1\;\;,
\end{equation}
where $n$ is parametrized as $n=n(k,l):=k+3l$ with $k=2,3,4$ and $l=0,1,2,\cdots$.
(As a consequence of the symmetry relation (\ref{symmetryrel}), the coefficients $D_{n-1,j}$ vanish for $j$ even or odd depending on whether $n$ is even or odd.) 
The first few polynomials are explicitly given by:
\begin{eqnarray}
\label{Dn}
D_{n}(z)&=\frac{S_{n+2}}{n+2}z^{n} ~~ (n=1,2,3)\;\;;\;\; D_{4}(z)=\frac{5}{3}S_{3}^{2}z^{2}+\frac{S_{6}}{6}z^{4}\;\;; \nonumber\\
D_{5}(z)&=5S_{3}S_{4}z^{3} + \frac{S_{7}}{7}z^{5}\;\;;\;\; D_{6}(z)=(7S_{3}S_{5}+\frac{35}{8}S_{4}^{2})z^{4}+\frac{S_{8}}{8}z^{6}\;\;.
\end{eqnarray}
(Note that the highest power in $\sigma_{0}$ is for all $D_{n}$ given by $\frac{S_{n+2}}{n+2}z^n$ for $n \ge 1$.) 
Explicit expressions for the polynomials $D_{n}(z)$ for $n \ge 7$ are easily obtained either from the recurrence relation (\ref{recurrenceB}) or from the combinatorial expression (\ref{combinatorialB}).

In order to derive the perturbative expansion ${\rm HO}(J)$ for the discrepancy function $\Delta_{0}$ at any order $J \ge 1$, we insert in the Hermite expansion (\ref{Deltahermite}) the polynomial relation 
(\ref{Dn-z}) for the expansion coefficients $a_{0}(n)$. At lowest order, $J=1$, one immediatly obtains from (\ref{Dn-z}) and (\ref{a0ntoggle}--\ref{Dn}) the simple result:
\begin{equation}
\label{Delta0HO}
\Delta_{0}^{{\rm HO}(1)}(\nu)=e^{-\nu^{2}/2}\frac{a_{0}(2)}{2}{\rm He}_{2}(\nu)=\frac{\gamma_{1}}{6}(\nu^{2}-1)e^{-\nu^{2}/2}\;\;,
\end{equation} 
which is completely determined by the skewness coefficient $\gamma_{1}=S_{3}\sigma_{0}$. To obtain the ${\rm HO}(J)$--expansion for $J \ge 2$, we decompose the exact (general) Hermite expansion into 
three terms ($M_{0}(J)=3J-1$):
\begin{eqnarray}
\label{Delta0decomp}
\Delta_{0}(\nu)=&e^{-\nu^{2}/2}\left[\sum_{n=2}^{J+1}\frac{D_{n-1}(\sigma_{0})}{n!}{\rm He}_{n}(\nu)+\sum_{n=J+2}^{M_{0}(J)}\frac{D_{n-1}(\sigma_{0})}{n!}{\rm He}_{n}(\nu) \right.\nonumber\\
&\left. +\sum_{n=M_{0}(J)+1}^{\infty}\frac{D_{n-1}(\sigma_{0})}{n!}{\rm He}_{n}(\nu)\right].
\end{eqnarray}
Since $D_{n-1}(\sigma_{0})$ is a polynomial of degree $n-1$ in $\sigma_{0}$, the first sum in (\ref{Delta0decomp}) represents (having $\nu$ fixed) a polynomial of degree $J$ in $\sigma_{0}$ and thus contributes to the 
${\rm HO}(J)$--expansion. The last infinite series in (\ref{Delta0decomp}), which is absolutely convergent, does not contribute at all to the ${\rm HO}(J)$--expansion, since the smallest power of $\sigma_{0}$ appearing in this 
series is already larger than $J$; (with $M_{0}(J)+1=n(3,J-1)$, equation (\ref{j0}) gives $j_{0}(M_{0}(J)+1)=J+1>J$.) We thus obtain the important result that the perturbative expansion at order $J$ necessarily 
implies a {\em truncation of the Hermite expansion} at $n=M_{0}(J)=3J-1$. It remains to discuss the second finite sum in (\ref{Delta0decomp}) where the summation runs over $J+2 \le n \le M_{0}(J)$. Inspection of the 
expression (\ref{a0ntoggle}) for the polynomials $D_{n-1}(\sigma_{0})$ (see also the explicit expressions (\ref{Dn})) shows that they will contribute (for $n \ge J+2$ and $j_{0}(n) \le J$) to this sum not 
only with powers $j \le J$, but also with higher powers that are not admitted in the ${\rm HO}(J)$--expansion. Thus, the polynomials $D_{n-1}$ have to be replaced by truncated ones. In the case where $j_{0}(n)>J$, the 
polynomials $D_{n-1}$ do not contribute at all. This leads us to define, for $n \ge J+2$, the {\em truncated polynomials} $D_{n-1}^{{\rm HO}(J)}(z)$, which now depend also on $J$:
\begin{equation}
\label{DHOJofz}
D_{n-1}^{{\rm HO}(J)}(z):= \left\{ \begin{array}{rcl} & \sum_{j=j_{0}(n)}^{J} D_{n-1,j} (S_{3},S_{4},\cdots,S_{n+1})z^{j} &,\; j_{0}(n) \le J\\ & 0 &,\; j_{0}(n) > J \;. \end{array}\right.
\end{equation}
We now explicitly give the {\em truncated polynomials} at order $J=2,3$ and $4$:

\noindent
$J=2, M_{0}(2)=5$:
\begin{eqnarray}
\label{DHO2ofz}
& D_{3}^{{\rm HO}(2)}(z)=0\nonumber\\
& D_{4}^{{\rm HO}(2)}(z)=\frac{5}{3} ~ S_{3}^{2} z^{2}\;\;;
\end{eqnarray}

\noindent
$J=3, M_{0}(3)=8$:
\begin{eqnarray}
\label{DHO3ofz}
& D_{4}^{{\rm HO}(3)}(z)=D_{4}^{{\rm HO}(2)}(z)\nonumber\\
& D_{5}^{{\rm HO}(3)}(z)=5 ~ S_{3} S_{4} z^{3}\nonumber\\ 
& D_{6}^{{\rm HO}(3)}(z)=0\nonumber\\
& D_{7}^{{\rm HO}(3)}(z)=\frac{280}{9} ~ S_{3}^{2} z^{3}\;\;;
\end{eqnarray}

\noindent
$J=4, M_{0}(4)=11$:
\begin{eqnarray}
\label{DHO4ofz}
& D_{5}^{{\rm HO}(4)}(z)=D_{5}^{{\rm HO}(3)}(z)\nonumber\\
& D_{6}^{{\rm HO}(4)}(z)=(7 ~ S_{3} S_{5} + \frac{35}{8} ~~ S_{4}^{2}) z^{4}\nonumber\\
& D_{7}^{{\rm HO}(4)}(z)=D_{7}^{{\rm HO}(3)}(z)\nonumber\\
& D_{8}^{{\rm HO}(4)}(z)=210 ~ S_{3}^{2} S_{4} z^{4}\nonumber\\
& D_{9}^{{\rm HO}(4)}(z)=0\nonumber\\
& D_{10}^{{\rm HO}(4)}(z)=\frac{3850}{3} ~ S_{3}^{4} z^{4}\;\;.
\end{eqnarray}
We then obtain from (\ref{Delta0decomp}) the general {\em perturbative formula} for the discrepancy function $\Delta_{0}(\nu)$ valid {\em at any order $J \ge 1$}:
\begin{equation}
\label{Delta0HOgeneral}
\Delta_{0}^{{\rm HO}(J)}(\nu):=e^{-\nu^{2}/2} \sum_{n=2}^{M_{0}(J)} \frac{a_{0}^{{\rm HO}}(J,n)}{n!} {\rm He}_{n}(\nu)\;,
\end{equation}
in terms of the $\rm HO$--Hermite expansion coefficients $a_{0}^{\rm HO}(J,n)$, which now also depend on $J$,
\begin{equation}
\label{a0HOgeneral}
a_{0}^{\rm HO}(J,n):= \left\{ \begin{array}{rcl} & a_{0}(n)=D_{n-1}(\sigma_{0}) & \mbox{, for} ~ 2 \le n \le J+1\\ & D_{n-1}^{{\rm HO}(J)}(\sigma_{0}) & \mbox{, for} ~ J+2 \le n \le M_{0}(J) \;. \end{array}\right.
\end{equation}

As an example, we give the expansion coefficients of the hierarchical ordering at second, third and fourth order:

\noindent
$J=2, M_0 (2) = 5$:
\begin{eqnarray}
\label{a0HOg2}
&a_0^{\rm HO} (2,2) = a_0 (2) = \frac{\gamma_1}{3} \nonumber\\
&a_0^{\rm HO} (2,3) = a_0 (3) = \frac{\gamma_2}{4} \nonumber\\
&a_0^{\rm HO} (2,4) = 0 \nonumber\\
&a_0^{\rm HO} (2,5) = \frac{5}{3} \gamma_1^2  \;\;.
\end{eqnarray}

\noindent
$J=3, M_0 (3) = 8$:
\begin{eqnarray}
\label{a0HOg3}
&a_0^{\rm HO} (3,2) = a_0 (2) = \frac{\gamma_1}{3} \nonumber\\
&a_0^{\rm HO} (3,3) = a_0 (3) = \frac{\gamma_2}{4} \nonumber\\
&a_0^{\rm HO} (3,4) = a_0 (4) = \frac{C_5}{5} \nonumber\\
&a_0^{\rm HO} (3,5) = \frac{5}{3} \gamma_1^2  \nonumber\\
&a_0^{\rm HO} (3,6) = 5 \gamma_1 \gamma_2  \nonumber\\
&a_0^{\rm HO} (3,7) = 0 \nonumber\\
&a_0^{\rm HO} (3,8) = \frac{280}{9} \gamma_1^3 \;\;.
\end{eqnarray}

\noindent
$J=4, M_0 (4) = 11$:
\begin{eqnarray}
\label{a0HOg4}
&a_0^{\rm HO} (4,n) = a_0 (n) \;\;\;(n = 2,3,4,5) \nonumber\\
&a_0^{\rm HO} (4,6) = 5\gamma_1 \gamma_2 \nonumber\\
&a_0^{\rm HO} (4,7) = 7\gamma_1 C_5 + \frac{35}{8}\gamma_2^2 \nonumber\\
&a_0^{\rm HO} (4,8) = \frac{280}{9}\gamma_1^3 \nonumber\\
&a_0^{\rm HO} (4,9) = 210 \gamma_1^2 \gamma_2  \nonumber\\
&a_0^{\rm HO} (4,10) = 0  \nonumber\\
&a_0^{\rm HO} (4,11) = \frac{3850}{3}\gamma_1^4 \;\;.
\end{eqnarray}
(Here, we have replaced the $S_{n}'s$ by the cumulants $C_{n}$ according to equation (\ref{hocumulants}).)

Being a direct consequence of the general Hermite expansion (\ref{Deltahermite}), the perturbative formula (\ref{Delta0HOgeneral}), valid at arbitrary order $J$ in
$\sigma_{0}$, has still the form of a Hermite expansion (truncated at $n=M_{0}(J)$). If we insert, however, for the Hermite coefficients $a_{0}^{\rm HO}(J,n)$ their definition 
(\ref{a0HOgeneral}) in terms of polynomials in $\sigma_{0}$, we obtain, by combining all terms of same power, the {\em alternative version of the perturbative formula}:
\begin{equation}
\label{Delta0HOalt}
\Delta_{0}^{{\rm HO}(J)}(\nu)=e^{-\nu^{2}/2} \sum_{j=1}^{J}{\mathrm w}_{0}^{{\rm HO}(J)}(j,\nu)\sigma_{0}^{j}\;\;,
\end{equation} 
which has now (at fixed $\nu$) the form of a power series in $\sigma_{0}$ with coefficient functions ${\mathrm w}_{0}^{{\rm HO}(J)}(j,\nu)$. (Since the series in (\ref{Delta0HOgeneral}) is finite, 
the rearrangement as a power series in $\sigma_{0}$ is of course always possible.) The coefficient functions ${\mathrm w}_{0}^{{\rm HO}(J)}(j,\nu)$ are given as a linear combination of a finite number 
of Hermite polynomials. Up to the second order, they have been calculated by Matsubara \cite{matsu1,matsu2}. It is straightforward to obtain them at arbitrary order $J$ using 
the equations (\ref{a0ntoggle}), (\ref{DHOJofz}) and (\ref{a0HOgeneral}). Here we give the coefficient functions up to fourth order:
\begin{eqnarray}
\label{v0HOg1}
J=1: &{\mathrm w}_{0}^{{\rm HO}(1)}(1,\nu)=\frac{S_{3}}{6}{\rm He}_{2}(\nu)\;,
\end{eqnarray}
\begin{eqnarray}
\label{v0HOg2}
J=2: &{\mathrm w}_{0}^{{\rm HO}(2)}(1,\nu)={\mathrm w}_{0}^{{\rm HO}(1)}(1,\nu)\nonumber\\
     &{\mathrm w}_{0}^{{\rm HO}(2)}(2,\nu)=\frac{S_{4}}{24}{\rm He}_{3}(\nu)+\frac{S_{3}^{2}}{72}{\rm He}_{5}(\nu)\;,
\end{eqnarray}
\begin{eqnarray}
\label{v0HOg3}
J=3: &{\mathrm w}_{0}^{{\rm HO}(3)}(1,\nu)={\mathrm w}_{0}^{{\rm HO}(1)}(1,\nu)\nonumber\\
     &{\mathrm w}_{0}^{{\rm HO}(3)}(2,\nu)={\mathrm w}_{0}^{{\rm HO}(2)}(2,\nu)\nonumber\\
     &{\mathrm w}_{0}^{{\rm HO}(3)}(3,\nu)=\frac{S_{5}}{120}{\rm He}_{4}(\nu)+\frac{S_{3}S_{4}}{144}{\rm He}_{6}(\nu)+\frac{S_{3}^{2}}{1296}{\rm He}_{8}(\nu)\;,
\end{eqnarray}
\begin{eqnarray}
\label{v0HOg4}
J=4: &{\mathrm w}_{0}^{{\rm HO}(4)}(1,\nu)= &{\mathrm w}_{0}^{{\rm HO}(1)}(1,\nu)\nonumber\\
     &{\mathrm w}_{0}^{{\rm HO}(4)}(2,\nu)= &{\mathrm w}_{0}^{{\rm HO}(2)}(2,\nu)\nonumber\\
     &{\mathrm w}_{0}^{{\rm HO}(4)}(3,\nu)= &{\mathrm w}_{0}^{{\rm HO}(3)}(3,\nu)\nonumber\\
     &{\mathrm w}_{0}^{{\rm HO}(4)}(4,\nu)= &\frac{S_{6}}{720}{\rm He}_{5}(\nu)+\frac{1}{720}(S_{3}S_{5}+\frac{5}{8}S_{4}^{2}){\rm He}_{7}(\nu)\nonumber\\
& &+\frac{S_{3}^{2}S_{4}}{1728}{\rm He}_{9}(\nu)+\frac{S_{3}^{4}}{31104}{\rm He}_{11}(\nu)\;.
\end{eqnarray}
The second--order formulae (\ref{v0HOg2}) agree with Matsubara's result \cite{matsu1,matsu2} (who writes $S:=S_{3}$ and $K:=S_{4}$). 

It is worthwhile to mention that the perturbative formula 
(\ref{Delta0HOalt}) closely resembles the so--called Edgeworth expansion (\ref{deltanx}) which gives a refinement of the classical central limit theorem (for details and references, see \ref{appC}).
However, (\ref{deltanx}) is an asymptotic expansion in the sense of Poincar\'e where the role of $\sigma_{0}$ is played by the small dimensionless parameter $1/\sqrt{n}$ which can be made arbitrarily 
small and does not have a fixed finite value as in the case of the standard deviation $\sigma_{0}$ of the CMB anisotropy. In fact, the central limit theorem is precisely the statement that the limit 
$n \rightarrow \infty$ is asymptotically exactly Gaussian and thus the NGs in the Edgeworth expansions have no fundamental meaning, they just determine the rate of convergence to the Gaussian limit. 
In contrast, the NGs of the CMB---if they are non--zero and of primordial origin---contain genuine information on the underlying model of inflation.

Having shown that the two perturbative formulae (\ref{Delta0HOgeneral}) 
and (\ref{Delta0HOalt}) are identical, we discuss in the following only the Hermite expansion (\ref{Delta0HOgeneral}). 

Let us compare the perturbative NG--coefficients, (\ref{a0HOg2}) respectively (\ref{a0HOg3}), with the complete NG--coefficients $a_0 (n)$ given in equation (\ref{AP3}). At {\em second order} ($J=2$) we see that the first two coefficients are identical to the complete coefficients $a_0 (n)$, i.e., $a_0^{\rm HO}(2,n) = a_0 (n)$ for $n=2,3$. The coefficient $a_0^{\rm HO}(2,4)$ vanishes because its complete value $C_5 / 5$ is of third order. Finally, the last coefficient $a_0^{\rm HO} (2,5)$ differs from $a_0 (5)$ by the term $C_6 / 6$, which is of fourth order. 
At {\em third order} ($J=3$) one observes that now the first three coefficients are identical to the complete expressions, i.e., $a_0^{\rm HO} (3,n) = a_0 (n)$ for $n=2,3,4$; $a_0^{\rm HO} (3,5) = a_0^{\rm HO}(2,5)$ still does not contain the term $C_6 / 6$; the coefficient 
$a_0^{\rm HO} (3,7)$ vanishes because the complete value for $a_0 (7)$ is of order $\sigma_0^4$; $a_0^{\rm HO}(3,6)$ does not contain the term $C_7 / 7$ (of order $5$), and $a_0^{\rm HO}(3,8)$ does not contain the terms of order $5$, respectively $7$. Note, in particular, that
a vanishing coefficient $a_0^{\rm HO}(J,n)$ implies that the associated contribution from the Hermite polynomial ${\rm He}_n (\nu)$ is absent in $\Delta_0^{\rm HO(J)}$ compared to $\Delta_0$. This is the case at second order with ${\rm He}_4 (\nu)$, and at third order with 
${\rm He}_7 (\nu)$. Since the ${\rm He}_n (\nu)$'s have exactly $n$ distinct zeros, the omission of one or several of them can have an important influence on the shape of the discrepancy function.

Finally, we come to the important question about the accuracy of the perturbative expansion. To this purpose we consider in analogy to equation (\ref{epsilon}) the {\em mean square error of the ${\rm HO}(J)$--expansion}:
\begin{eqnarray}
\label{epsilonHO}
&{\cal{E}}^{{\rm HO}(J)} &:= ||h_{0} - \sum_{n=2}^{M_{0}(J)}{\frac{a_{0}^{\rm HO}(J,n)}{n!}{\rm He}_{n}}||^{2}\nonumber\\
& &= ||h_{0}||^{2} - \sum_{n=2}^{M_{0}(J)}{\frac{(a_{0}(n))^{2}}{n!}} + \sum_{n=2}^{M_{0}(J)}\frac{(a_{0}(n)-a_{0}^{\rm HO}(J,n))^{2}}{n!}\;,
\end{eqnarray}
where we have used the identity (\ref{error2}). Here, the first two terms are identical to the error ${\cal{E}}_{M_{0}(J)}$, i.e. the error of the complete Hermite expansion truncated at order $N:=M_{0}(J)$ 
(see equation (\ref{epsilon})), which leads to the interesting formula:
\begin{eqnarray}
\label{epsilonHO'}
{\cal{E}}^{{\rm HO}(J)} = {\cal{E}}_{M_{0}(J)} + \sum_{n=J+2}^{M_{0}(J)}\frac{(a_{0}(n)-a_{0}^{\rm HO}(J,n))^{2}}{n!}\;. 
\end{eqnarray}
Here, we have used in the last sum $a_{0}(n)-a_{0}^{\rm HO}(J,n)\equiv 0$ for $2 \le n \le J+1$ (see (\ref{a0HOgeneral})). Since the last sum in (\ref{epsilonHO'}) is strictly positive, one infers that we have at any order in perturbation theory:
\begin{equation}
\label{HOgtHe}  
{\cal{E}}^{{\rm HO}(J)} > {\cal{E}}_{M_{0}(J)}\;,
\end{equation}
i.e., the {\em mean square error} of the ${\rm HO}(J)$--expansion is always larger than the {\em mean square error} of the complete Hermite expansion truncated at $n=M_{0}(J)$. For instance, at second--order of perturbation theory, 
$J=2$, we have ${\cal{E}}^{{\rm HO}(2)} > {\cal{E}}_{5}$, where ${\cal{E}}_{5}$ is explicitly given in (\ref{epsilon5}). 
Precisely, we obtain from (\ref{epsilonHO'}) through (\ref{AP3}) and (\ref{a0HOg2}):
\begin{equation}
\label{epsilonHO2}  
{\cal{E}}^{{\rm HO}(2)} = {\cal{E}}_{5}+\frac{C_{5}^{2}}{600}+\frac{C_{6}^{2}}{4320}\;.
\end{equation}
One observes that both errors approach zero in the limit $J \rightarrow \infty$ which reflects the fact that the 
perturbative expansion becomes identical to the general untruncated Hermite expansion in this limit. 

\begin{figure}[!htb]
\includegraphics[height=0.60\textheight,width=1.00\textwidth]{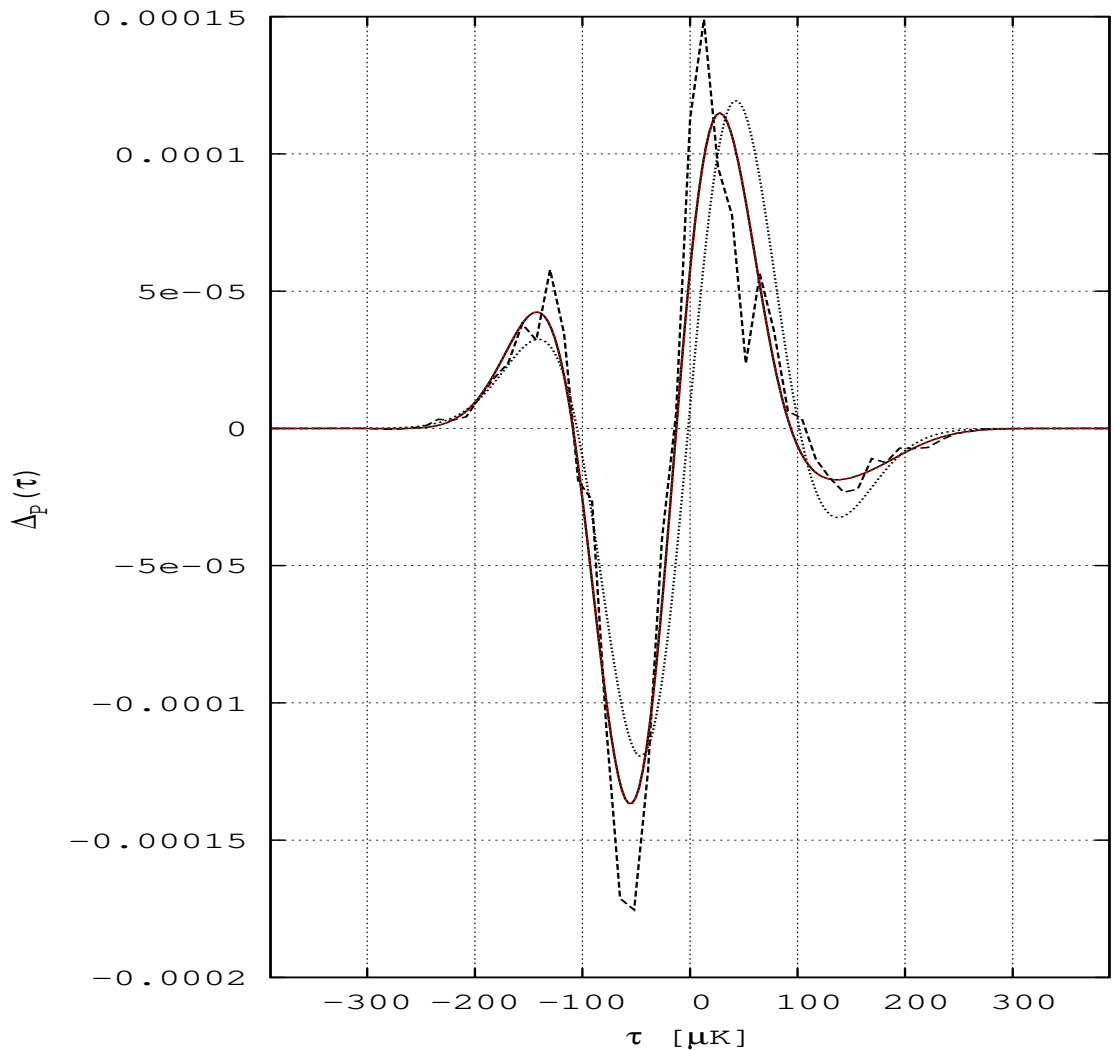}
\caption{$\Lambda$CDM map sample, $N_{\rm side}$=128, $\ell_{\rm range}$=[2,256] without mask, 2$^{\circ}$ fwhm, for a 13$\mu$K temperature bin width over the largest temperature range of the sample ($\pm$396.5$\mu K$). We plot the discrepancy function (black dashed line) of the PDF for the $10^5$ maps ensemble. As a black dotted line we show the perturbative expansion 
according to equation (\ref{DeltaP_PTmn}) limited to $n=6$ (hierarchical ordering and second--order perturbation theory); the red line shows the result for the ${\rm HO}(J=4)$ expansion of equation~(\ref{DeltaP_PTmn}) up to $n=12$ (the coefficients $a_P^{\rm HO}$ obtained from equations (\ref{APtoA0}) and (\ref{a0HOg4}),
which practically coincides with the black solid line showing the Hermite expansion at order $6$ according to equation~(\ref{DeltaPpart}).}
\label{PFR_Fig6}
\end{figure}

\subsubsection*{Test of the hierarchical ordering in perturbation theory}

In figure \ref{PFR_Fig6} we compare the discrepancy function $\Delta_{P}$ with its Hermite expansion in terms of the coefficients $a_{P}(n)$ (computed from (\ref{AP1}) and table \ref{tablemnxn}) for $n=3$ to $6$, according to the following equation:
\begin{equation}
\label{DeltaPpart}
\qquad\Delta_P (\tau ) = \e^{-\tau^2 / 2\sigma_0^2}\; \sum_{n=3}^{6} \frac{a_P (n)}{n!} \;{\rm He}_n ({\tau} / \sigma_0 ) \;\;.
\end{equation}
In addition, in figure \ref{PFR_Fig6}, we show the expansion of $\Delta_{P}$ with the assumption of hierarchical ordering (HO) in fourth--order perturbation theory, according to
\begin{equation}
\label{DeltaP_PTmn}
\Delta_P^{\rm HO(4)}(\tau) = \e^{-\tau^2 /2\sigma_0^2}\, \sum_{n=3}^{12} \frac{a_P^{\rm HO}(4,n)}{n!} \; {\rm He}_n ({\tau} / \sigma_0 ) \;,
\end{equation}
where the NG--coefficients $a_P^{\rm HO}(4,n)$ are obtained from equations (\ref{APtoA0}) and (\ref{a0HOg4}) using $\gamma_1$, $\gamma_2$ and the cumulants $C_{n}$ in table \ref{tablemnxn}. 

We verified that, at second and third--order, the HO--expansion displays an almost purely odd function (central symmetry), while $\Delta_{P}$ (black dashed line) does not pass by the central point ($\tau=0$, $\Delta_{P}(0)$, see equation (\ref{DeltaPin0})), and we note that $\Delta_{Px}(\tau)$ derived from moments of pixels, not shown here, confirms this behaviour.

At fourth--order, however, the HO--expansion fits well the non--Gaussianity of the sample and reveals the same shift from perfect central symmetry.
(For some further discussion on  the perturbative model with hierarchical ordering, we refer the reader to \ref{appC}).

\subsection{The Minkowski Functionals $\mv_1 (\nu)$ and $\mv_2 (\nu)$}
\label{sec:v1v2}

According to Hadwiger's theorem \cite{hadwiger}, there are three independent MFs on ${\mathcal S}^2$. The simplest case $V_0$, respectively $\mv_0$ has already been discussed in subsection~\ref{subsec:v0}. In this section, we discuss the remaining two which are defined again with respect to the excursion set ${\mathcal Q}_{\nu}$, see equation (\ref{excursionset}), and are given in their normalized form by
\begin{eqnarray}
\label{v_1}
& \mv_1 (\nu) : = \frac{1}{4\pi} V_1 (\nu) = \frac{1}{4\pi}\frac{1}{4}  \int_{\partial {\mathcal Q}_{\nu}} {\mathrm d}s = \frac{1}{16\pi} \, {\rm length}\;(\partial {\mathcal Q}_{\nu})\;\;,\\
\label{v_2}
& \mv_2 (\nu) = \frac{1}{4\pi} V_2 (\nu) = \frac{1}{4\pi}\frac{1}{2\pi} \int_{\partial {\mathcal Q}_{\nu}} \kappa (s) \ {\mathrm d}s \;\;,
\end{eqnarray}
where ${\mathrm d}s$ denotes the line element along $\partial {\mathcal Q}_{\nu}$, and $\kappa (s)$ the geodesic curvature of 
$\partial {\mathcal Q}_{\nu}$. The {\em Gaussian predictions} for the MFs, $\mv_k (\nu), k = 1,2$, have been computed by Tomita \cite{tomita1,tomita2,adler}
(we set $\mu = 0$):
\begin{equation}
\label{vg}
\mv_1^{\rm G} (\nu) : = \frac{1}{8\sqrt{2}} \frac{\sigma_1}{\sigma_0} \; \e^{-\nu^2 /2}\quad;\quad \mv_2^{\rm G} (\nu) = \frac{1}{2 (2\pi)^{3/2}} \frac{\sigma^2_1}{\sigma^2_0} \; \nu \, \e^{-\nu^2 /2} \;\;.
\end{equation}
Here, $\sigma_0$ denotes again the standard deviation of the CMB temperature anisotropy $\delta T ({\boldsymbol{\hat n}})$ on ${\mathcal S}^2$, where
${\boldsymbol{\hat n}} = {\boldsymbol{\hat n}}(\vartheta , \varphi)$ denotes a unit vector on ${\mathcal S}^2$ dependent on the coordinates $x^1 = \vartheta \in 
[0,\pi ]$ and $x^2 = \varphi \in [0,2\pi]$. Then, the line element on ${\mathcal S}^2$ is given by ${\mathrm d}s^2 = \gamma_{ij} {\mathrm d}x^i
{\mathrm d}x^j$ with $\gamma_{11} = 1$, $\gamma_{22} = \sin^2 \vartheta$, $\gamma_{ij} = 0$ otherwise, and $\gamma_{ik}\gamma^{kj} =
\delta_i^{\ j}$. Furthermore, $\sigma_1^2$ is the {\em variance of the gradient field} $\boldsymbol{\nabla}\delta T = 
(\nabla^1 \delta T, \nabla^2 \delta T )$, i.e.,
\begin{equation}
\label{sigma1_pxs}
\langle \nabla_i \delta T({\boldsymbol{\hat n}}) \nabla_j \delta T ({\boldsymbol{\hat n}}) \rangle = \frac{\sigma_1^2}{2}\,\gamma_{ij}\;\;.
\end{equation}
From the MFs $\mv_1$ and $\mv_2$ one can form, by a linear combination, two further interesting measures, the {\em Euler characteristic}
$\chi (\nu)$, respectively the {\em genus} $g(\nu) := 1 - \frac{1}{2}\chi (\nu)$. On the excursion set ${\mathcal Q}_{\nu}$ with smooth boundary $\partial {\mathcal Q}_{\nu}$ the {\em Gauss--Bonnet theorem} holds ($K = 1/R^2 \equiv 1$ is the Gaussian curvature 
on the unit sphere with radius $R=1$):
\begin{equation}
\int_{{\mathcal Q}_{\nu}} K\, {\mathrm d}a + \int_{\partial {\mathcal Q}_{\nu}} \kappa (s) {\mathrm d}s = V_0 (\nu) + 2\pi V_2 (\nu) = 2\pi \; \chi (\nu) \;\;,
\end{equation} 
which gives 
\begin{equation}
\chi (\nu) = 2 \mv_0 (\nu) + 4\pi \mv_2 (\nu) \quad{\rm and}\quad g(\nu) = 1 - \mv_0 (\nu) - 2\pi \mv_2 (\nu) \;\;.
\end{equation}
As a measure of possible non--Gaussianities based on the MFs $\mv_k (\nu), k=1,2$, we define, in analogy to the discrepancy functions
$\Delta_P (\nu)$ (equation (\ref{Delta2})) and $\Delta_0 (\nu)$ (equation (\ref{Deltahermite})), the following {\em discrepancy functions} (we here include also the case $k=0$):
\begin{equation}
\Delta_k (\nu) := \frac{\mv_k (\nu) - \mv^{\rm G}_k (\nu)}{N_k} \;\;\; (k = 0,1,2) \;\;,
\end{equation}
with $N_0 = 1/ \sqrt{2\pi}$, $N_1 = {\rm max} \lbrace \mv_1^{\rm G} \rbrace = [1/(8\sqrt{2})] \sigma_1 / \sigma_0$, and $N_2 = [
1/(2 (2\pi)^{3/2})] \sigma_1^2 / \sigma_0^2$.
Under the same assumptions made before (for $\Delta_{P}$ respectively $\Delta_{0}$, see \ref{appC}), we can expand the $\Delta_k$'s into a convergent {\em Hermite expansion},
\begin{equation}
\label{Deltaknu}
\Delta_k (\nu) = \e^{-\nu^2 /2} \sum_{n = n_k}^{\infty} \frac{a_k (n)}{n!} \, {\rm He}_n (\nu) \;\;;\;\;k=0,1,2 \;\;,    
\end{equation}
with the dimensionless NG--coefficients $a_k (n)$ and $n_0 = 2$, $n_1 = n_2 =0$. With the help of (\ref{ortho}) we obtain the integral representation:
\begin{equation}
\label{aknDelta}
a_k (n) = \frac{1}{\sqrt{2\pi}} \int_{-\infty}^{\infty} \Delta_k (\nu) \, {\rm He}_n (\nu) \,{\mathrm d}\nu \;\;.
\end{equation}
Again, it is assumed that the series (\ref{Deltaknu}) can be truncated at a low value $n=N$, where $N$ may depend on $k$. In the following, 
we shall again compare the general expansion (\ref{Deltaknu}) with the one derived under the assumption of hierarchical ordering ($k=0,1,2$),
\begin{equation}
\label{deltak57}
\Delta_{k}^{{\rm HO}(J)}(\nu)=\e^{-\nu^2 /2} \sum_{n = n_k}^{M_{k}(J)} \frac{a_{k}^{\rm HO}(J,n)}{n!} {\rm He}_{n}(\nu)  \;\;,
\end{equation}
for which the second order NG--coefficients $a_{k}^{\rm HO}(2,n)$ have also been calculated by Matsubara \cite{matsu2,hikage2} for $k=1$ and $k=2$.
The highest degree of the Hermite polynomials contributing for $J=2$ is given by $M_{1}(2)=6$ and $M_{2}(2)=7$.

\noindent
The NG--coefficients are then given for $\Delta_{1}^{{\rm HO}(J)}(\nu)$ as follows:

\noindent
$J=2, M_{1}(2)=6$:
\begin{eqnarray}
\label{A1HO2}
&a_1^{\rm HO} (2,0) = -\frac{K_{3}}{16} \nonumber\\
&a_1^{\rm HO} (2,1) = -\frac{S_{1}}{4} \nonumber\\
&a_1^{\rm HO} (2,2) = -\frac{1}{6}(K_{1}+\frac{3}{8}S_{1}^{2})\nonumber\\
&a_1^{\rm HO} (2,3) = \gamma_{1} \nonumber\\ 
&a_1^{\rm HO} (2,4) = \gamma_{2} - \gamma_{1} S_{1} \nonumber\\
&a_1^{\rm HO} (2,5) = 0 \nonumber\\
&a_1^{\rm HO} (2,6) = 10 \gamma_{1}^{2} \;\;.
\end{eqnarray}
and for $\Delta_{2}^{{\rm HO}(J)}(\nu)$:

\noindent
$J=2, M_{2}(2)=7$:
\begin{eqnarray}
\label{A2HO2}
&a_2^{\rm HO} (2,0) = -S_{2} \nonumber\\
&a_2^{\rm HO} (2,1) = -\frac{1}{2}(K_{2}+S_{1}S_{2}) \nonumber\\
&a_2^{\rm HO} (2,2) = -S_{1} \nonumber\\
&a_2^{\rm HO} (2,3) = -K_{1}-\gamma_{1}S_{2} \nonumber\\ 
&a_2^{\rm HO} (2,4) = 4\gamma_{1} \nonumber\\
&a_2^{\rm HO} (2,5) = 5\gamma_{2}-10\gamma_{1}S_{1} \nonumber\\
&a_2^{\rm HO} (2,6) = 0 \nonumber\\
&a_2^{\rm HO} (2,7) = 70\gamma_{1}^{2} \;\;.
\end{eqnarray}
Here, we introduced the three dimensionless skewness parameters $\gamma_{1}$, $S_{1}$, $S_{2}$ and the four 
dimensionless kurtosis parameters $\gamma_{2}$, $K_{1}$, $K_{2}$ and $K_{3}$ using the notation 
$\tau=\tau(\boldsymbol{\hat{n}}):=\delta T(\boldsymbol{\hat{n}})$ ($\gamma_{1}$ and $\gamma_{2}$ as in equations (\ref{skewcoeff}) and (\ref{gamma2}), respectively):
\begin{eqnarray}
\label{matsu_skew_kurt}
S_{1}:=\frac{\left<\tau^{2}{\boldsymbol{\nabla}}^{2}\tau\right>_{C}}{\sigma_{0} \sigma_{1}^{2}} \;\;;\;\;
S_{2}:=\frac{\left<|\boldsymbol{\nabla}\tau|^{2}\boldsymbol{\nabla}^{2}\tau\right>_{C}\sigma_{0}}{\sigma_{1}^{4}}\;\;;\nonumber\\
\fl
\quad K_{1}:=\frac{\left<\tau^{3}\boldsymbol{\nabla}^{2}\tau\right>_{C}}{\sigma_{0}^{2}\sigma_{1}^{2}} \;\;;\;\;
K_{2}:=\frac{2\left<\tau|\boldsymbol{\nabla}\tau|^{2}\boldsymbol{\nabla}^{2}\tau\right>_{C} + \left<|\boldsymbol{\nabla}\tau|^{4}\right>_{C}}{\sigma_{1}^{4}} \;\;;\;\;
K_{3}:=\frac{\left<|\boldsymbol{\nabla}\tau|^{4}\right>_{C}}{\sigma_{1}^{4}} \;.
\end{eqnarray}
(Note that the products of the field $\tau(\boldsymbol{\hat{n}})$ respectively of its derivatives are taken at the same point $\boldsymbol{\hat{n}}$ on ${\mathcal S}^{2}$.)

\medskip\bigskip
\noindent
Table \ref{tablea1a2} shows the coefficients $a_{1} (n)$ and $a_{2} (n)$ calculated from (\ref{aknDelta}).

\begin{table}[!htb] 
\centering

\resizebox{\columnwidth}{0.9cm}{
  \begin{tabular}{|c|c|c|c|c|c|c|c|c|c|}
    \hline
    \multicolumn{10}{|c|}{$\Lambda$CDM sample~ Full individual map range, no mask, 2$^{\circ}$fwhm, bin 13$\mu$K ~ {\em c.f.} \ref{appA}}\\
    \hline
    $n$ & $0$ & $1$ & $2$ & $3$ & $4$ & $5$ & $6$ & $7$ & $8$\\
    \hline
    $a_{1}(n)$ & $-2.21\times10^{-6}$ & $-2.808\times10^{-5}$ & $3.97274\times10^{-3}$ & $-5.0873\times10^{-4}$ & $-5.823\times10^{-5}$ & $-6.643\times10^{-5}$ & $-1.97922\times10^{-3}$ & $5.059\times10^{-5}$ & $6.37408\times10^{-3}$\\
    \hline 
    \hline
    \hline 
    $a_{2}(n)$ & $1.927\times10^{-5}$ & $-2.148\times10^{-5}$ & $-1.7313\times10^{-4}$ & $1.183631\times10^{-2}$ & $-1.53488\times10^{-3}$ & $-4.8941\times10^{-4}$ & $-1.20822\times10^{-3}$ & $-9.58844\times10^{-3}$ & $-7.33397\times10^{-3}$\\
    \hline 
    %
    %
    %
    %
  \end{tabular}
}
\caption{Table of coefficients $a_{1}(n)$ and $a_{2}(n)$.}\label{tablea1a2}
\end{table}

\begin{figure}[!htb]
\includegraphics[height=0.6\textheight,width=1.00\textwidth]{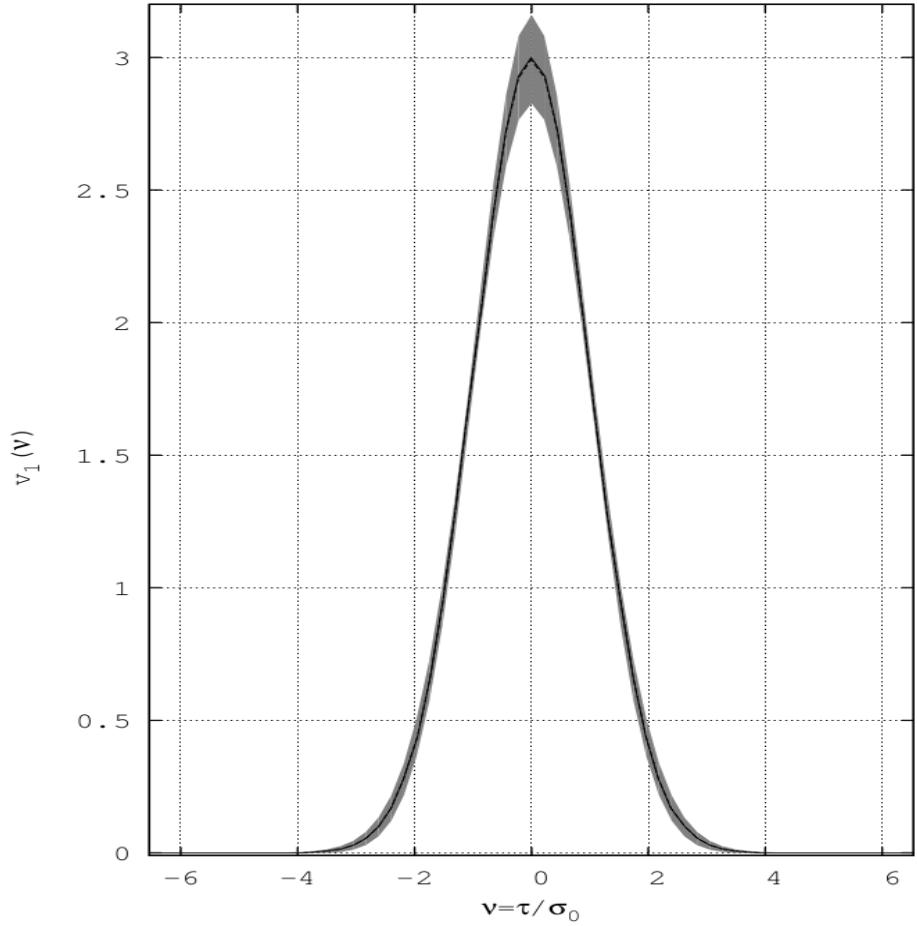}
\caption{$\Lambda$CDM map sample, $N_{\rm side}$=128, $\ell_{\rm range}$=[2,256] without mask, 2$^{\circ}$ fwhm, for a 13$\mu$K temperature bin width over the largest temperature range of the sample ($\pm$396.5$\mu K$). We plot the second Minkowski Functional $\mv_{1}(\nu)$ as a black dashed line with its Gaussian premise as a black solid line (almost coincident). The $1\sigma$ cosmic variance is the grey shaded area. (We use $\sigma^2_{0{\rm px}}$ and $\sigma^2_{1{\rm px}}$ -- see, respectively, 2$^{nd}$ equation in (\ref{sigma0_mnpx_sq}) and equation (\ref{sigma1_pxs_sq}) as well as table \ref{tablemusigmas}.)}
\label{PFR_Fig7}
\end{figure}

\begin{figure}
\includegraphics[height=0.6\textheight,width=1.00\textwidth]{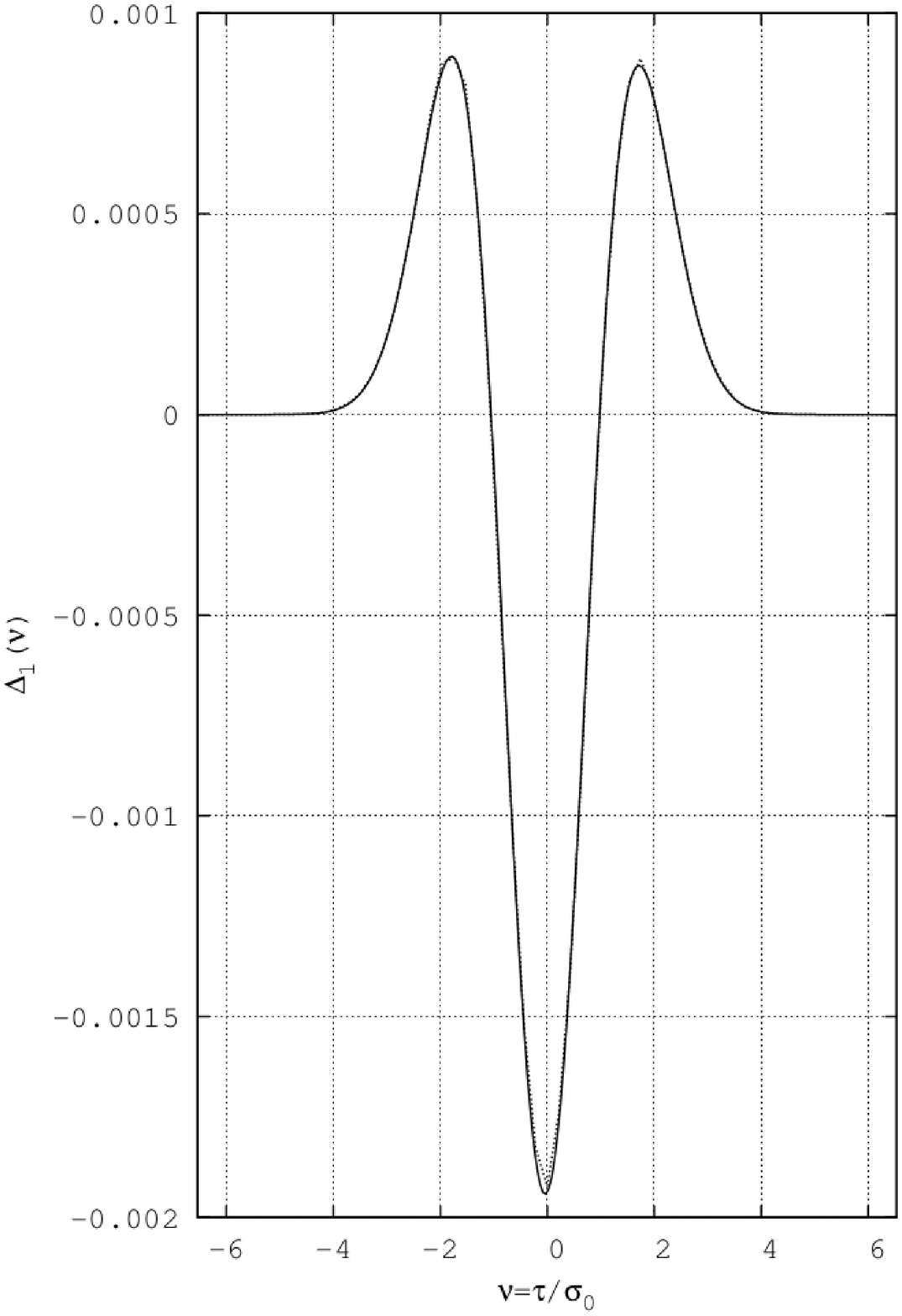}
\caption{We plot $\Delta_{1}(\nu)$, the discrepancy function of $\mv_{1}$ as a dotted line, and as a solid line the Hermite expansion to order $8$. (For the $\sigma-$values used, see caption to figure \ref{PFR_Fig7}.)}
\label{PFR_Fig8}
\end{figure}

\begin{figure}[!htb]
\includegraphics[height=0.6\textheight,width=1.00\textwidth]{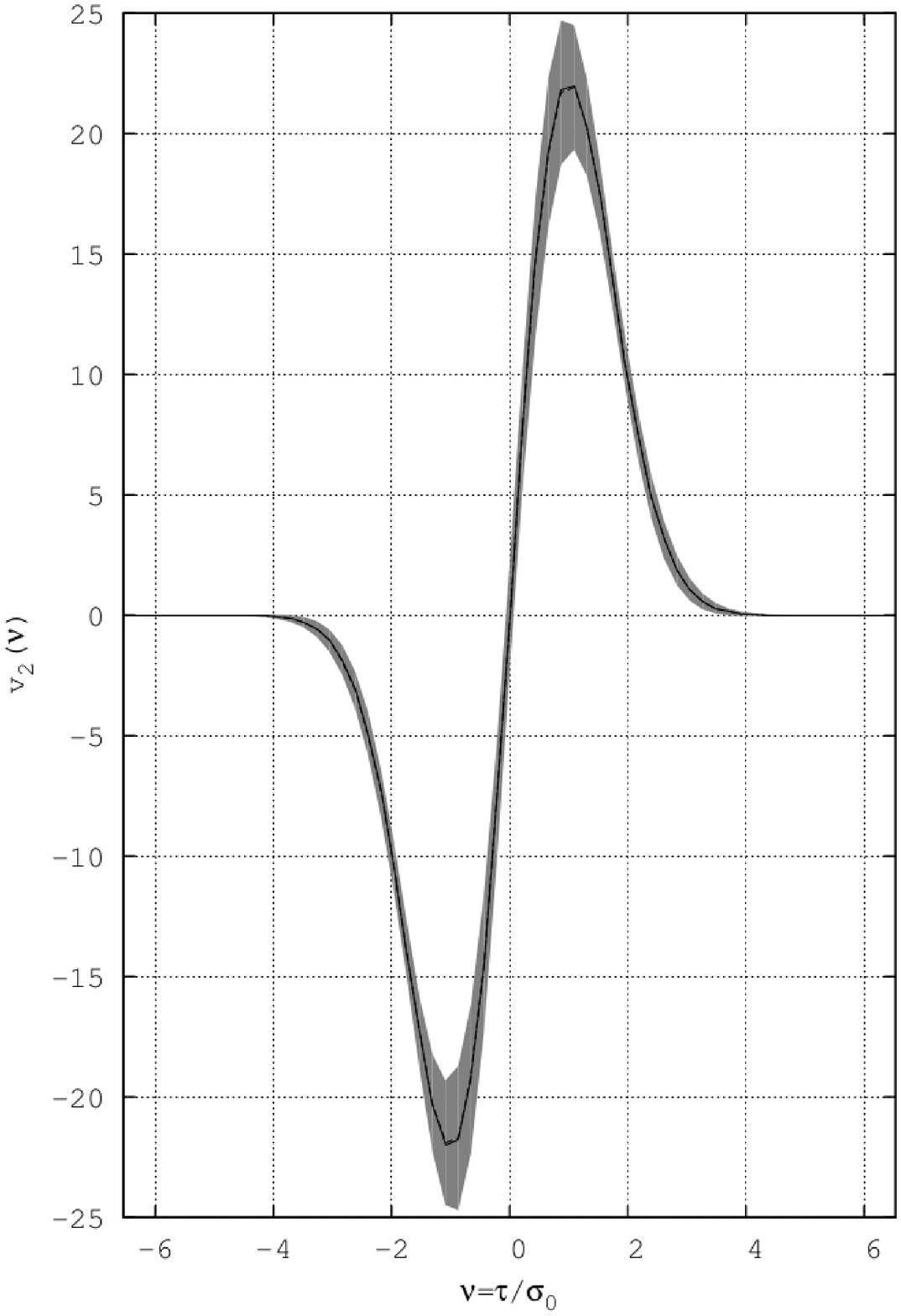}
\caption{$\Lambda$CDM map sample, $N_{\rm side}$=128, $\ell_{\rm range}$=[2,256] without mask, 2$^{\circ}$ fwhm, for a 13$\mu$K temperature bin width over the largest temperature range of the sample ($\pm$396.5$\mu K$). We plot the third Minkowski Functional $\mv_{2}(\nu)$ as a black dashed line with its Gaussian premise as a black solid line (almost coincident).
The $1\sigma$ cosmic variance is the grey shaded area. (For the $\sigma-$values used, see caption to figure \ref{PFR_Fig7}.)}
\label{PFR_Fig9}
\end{figure}

\begin{figure}[!htb]
\includegraphics[height=0.6\textheight,width=1.00\textwidth]{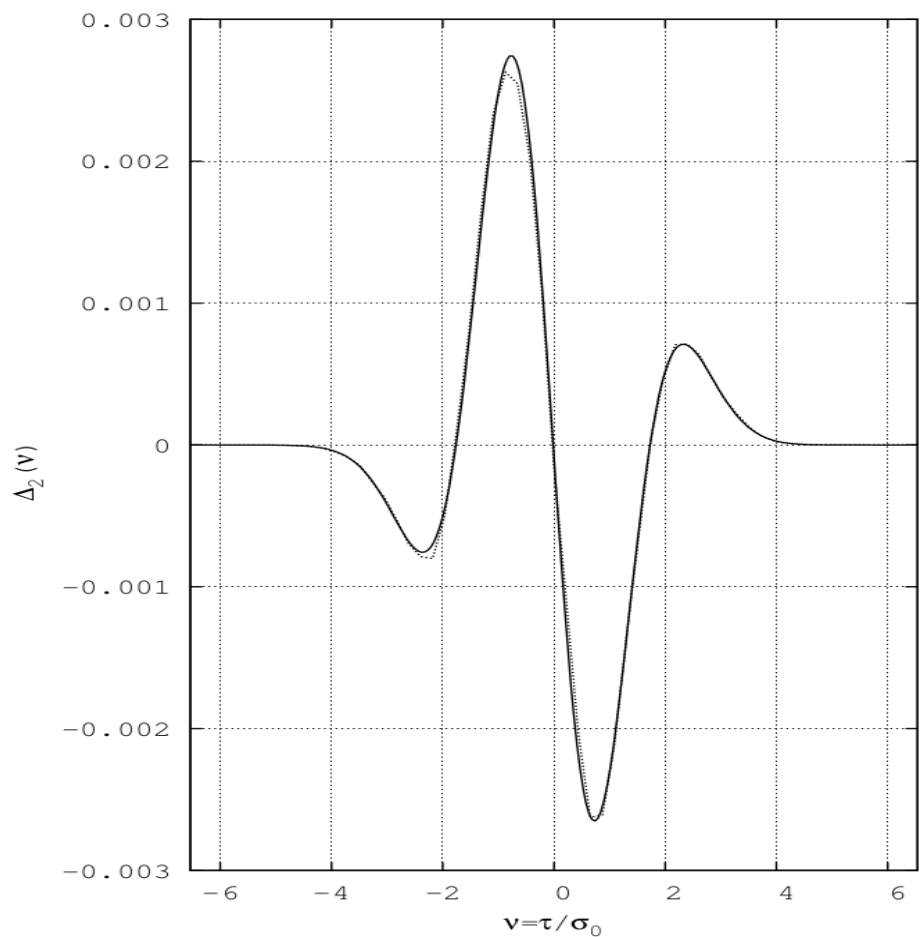}
\caption{We plot $\Delta_{2}(\nu)$, the discrepancy function of $\mv_{2}$ as a dotted line, and as a solid line the Hermite expansion to order $8$. (For the $\sigma-$values used, see caption to figure \ref{PFR_Fig7}.)}
\label{PFR_Fig10}
\end{figure}

\noindent
Figure \ref{PFR_Fig7} shows the second Minkowski Functional, and figure \ref{PFR_Fig8} the discrepancy function $\Delta_{1}$ together with the Hermite expansion to order $8$. The third Minkowski Functional is shown in figure \ref{PFR_Fig9}, and the discrepancy function $\Delta_{2}$ together with the Hermite expansion to order $8$ in figure \ref{PFR_Fig10}.

\clearpage

\subsection{Discrepancy functions and ${\rm Df}-$differences}
\label{discrepancy}

Gaussian random fields have a specific signature (the Gaussian prediction) depending only on the choice of the 
descriptor. Non--Gaussian processes 
may generate strong departures from the Gaussian prediction as in the formation of large--scale structure; however, attempts to find general and specific analytic signatures of a statistical property sufficiently far away from Gaussianity are most of the 
time unsuccessful in the context of CMB analyses. It is clear that the 
values $\sigma_{0 {\rm C\ell}}$ and $\sigma_{1 {\rm C\ell}}$ are model--dependent and that their use in the formulae for the Gaussian 
prediction, equations (\ref{G}), (\ref{v0G}) and (\ref{vg}), biases the reference of Gaussianity 
in general, so that the $\sigma-$values from the moments of pixels (denoted by subscripts $\rm px$, $\rm x$) or from the moments of the PDF should rather be used in the {\em discrepancy functions} $\Delta_{k}(...)$
we defined above. As a reminder, we list here the whole set of {\em discrepancy functions}:
\begin{eqnarray}
\label{deltapdf}
{\Delta_P (\tau)}:=\frac{P(\tau)-P^{\mathrm G}(\tau)}{N_{\rm P}},\ &{\rm with} \ \ N_{\rm P}=\frac{1}{\sigma_{0}\sqrt{2 \pi}} \;;\\
\label{delta0}
{\Delta_{0}(\nu)}:=\frac{\mv_{0}(\nu)-\mv^{\mathrm G}_{0}(\nu)}{N_{0}}, \ &{\rm with} \ \ N_{0}=\frac{1}{\sqrt{2 \pi}}, \ &{\rm}  \ \ \nu=\frac{\tau}{\sigma_{\rm 0 {\rm px}}} \;;\\
\label{delta1}
{\Delta_{1}(\nu)}:=\frac{\mv_{1}(\nu)-\mv^{\mathrm G}_{1}(\nu)}{N_{1}}, \ &{\rm with} \ \ N_{1}=\frac{1}{8 \sqrt{2}} \frac{\sigma_{1 {\rm px}}}{\sigma_{0 {\rm px}}}, \ &{\rm}  \ \ \nu=\frac{\tau}{\sigma_{\rm 0 {\rm px}}} \;;\\
\label{delta2}
{\Delta_{2}(\nu)}:=\frac{\mv_{2}(\nu)-\mv^{\mathrm G}_{2}(\nu)}{N_{2}}, \ & {\rm with} \ \ N_{2}=\frac{1}{2 (2 \pi)^{3/2}} \frac{\sigma_{1 {\rm px}}^{2}}{\sigma_{\rm 0 {\rm px}}^{2}}, \ &{\rm}  \ \ \nu=\frac{\tau}{\sigma_{\rm 0 {\rm px}}}\;.
\end{eqnarray}
These discrepancy functions are self--consistent and model--independent given that the terms used, $\sigma_{\rm S}$ 
and $\sigma_{\rm S px}$ are model--independent; these four formulae ($\Delta_{k}(...)$) are straightforwardly applicable to a 
single data or sample map. In the case of determination for a sample $\rm S$ of maps, we define the {\em sample discrepancy functions} this way:
\begin{eqnarray}
\label{densfctS}
&{P_{\mathrm S}^{\mathrm G}(\tau)}:=P^{\mathrm G} \left( \left<\mu \right>_{\mathrm S}, \sqrt{\left<\sigma_{\rm 0}^{2}\right>_{\mathrm S}},~\tau \right) ,\\
\label{vGiS}
&{\mv_{i {\mathrm S}}^{\mathrm G}(\nu)}:=\mv_{i}^{\mathrm G} \left( \left<\mu_{\rm px}\right>_{\mathrm S}, \sqrt{\left<\sigma_{1 {\rm px}}^{2}\right>_{\mathrm S}}\ , \sqrt{\left<\sigma_{0 {\rm px}}^{2}\right>_{\mathrm S}}\ ,~\nu \right) \;.
\end{eqnarray}
As mentioned at the end of section~\ref{context}, the $Planck$ collaboration applies 
a different method of calculation for the non--Gaussianity. While the analytic 
formulae are not given in the various papers \cite{komatsu2,eriksen,ducout,modest,planck2,planck3}, we can rebuild them here from 
the explicit formulations in the text. In the case of a single data map $d$, 
the {\em differences} of the normalized MFs (denoted ${\rm Df}$) are given by
\begin{equation}
\label{diffnzmfs_d}
{\mathrm Df}^d_{i}(\nu):=\frac{\mv^d_{i}(\nu)}{N^d_{i}} - \left< \frac{\mv_{i}(\nu)}{N_{i}} \right>_{\mathrm S} \;,
\end{equation}
where the reference map sample $\mathrm S$ is Gaussian by construction. Compared with the discrepancy functions,
the ${\rm Df}$'s measure the distance to a Gaussian premise which may be different in amplitude and shape from our analytical one. 
The ${\rm Df}$'s may, under certain conditions, provide a direct statistical measure of the departure from a given cosmological model statistics, which is an interesting feature. 

Let us make some remarks on this methodology addressing some important issues. 
The variance terms $\sigma_S$, entering the left and right denominators of equation (\ref{diffnzmfs_d}),
may differ in the sample map generation, if no special precaution is taken.
We have compared the results of this ${\rm Df}-$method (using Gaussian $\Lambda$CDM map samples) with our 
discrepancy functions method, using the analytical Gaussian premises, and found no significant difference of non--Gaussianity. But,
obviously, the dependence on the statistical properties of the map sample and the number of maps may 
yield different results. It is not clear how the non--Gaussianity of the reference map sample is 
calculated as the direct application of the formula (\ref{diffnzmfs_d}) gives zero and in this case, the analytic Gaussian premise is probably used. If instead
our discrepancy functions are applied to the sample, we do not employ 
different methods for comparing the non--Gaussianity of a CMB data map with a CMB map sample. 

\vspace{-5pt}

\subsection{Discrepancy functions of the $\Lambda$CDM sample maps versus $Planck$ maps}
\label{planck15}

The next figures, figure~\ref{PFR_Fig11}, figure~\ref{PFR_Fig12}, figure~\ref{PFR_Fig13} and figure~\ref{PFR_Fig14} show the statistical and morphological behaviour of each of the $10^{5}$ sample maps 
in the $\Lambda$CDM model compared with the four $Planck$ maps $NILC$, $SEVEM$, $SMICA$ and $Commander$---$Ruler$,
all with the $U73$ mask and a bin width of $6\muK$. The average quantities $\mu$, $\sigma_{0}$ and 
$\sigma_{1}$, as well as the discrepancy functions are calculated over the largest common and centered 
(on $\nu=0$) temperature range $[-201\muK,+201\muK]$, as given by the $Planck$ 2015 $SMICA$ map. In these figures,
clearly, the non--Gaussianity of the average sample is weak.  
For each descriptor, many $\Lambda$CDM sample maps show a strong non--Gaussianity; several individual sample maps
lying well beyond the 5$\sigma$ cosmic variance limit. 

The benchmark study of $Planck$ maps and the $\Lambda$CDM sample with U73 mask (figures~\ref{PFR_Fig11} to \ref{PFR_Fig14}) compares all the different maps over the same temperature range ($\pm 201 \mu K$) for a $6 \mu K$ binning, and shows very populated (almost saturated) 
$4 \sigma$--sample envelopes for each of the four discrepancy functions. In the case of $\Delta_{P}$ we count $240$ maps (among $100000$) peaking beyond the $6 \sigma$--envelope. Peaking beyond $6 \sigma$ we also count $354$ maps for $\Delta_{0}$.

\vspace{-5pt}

\begin{figure}[!htb]
\includegraphics[height=0.75\textheight,width=1.00\textwidth]{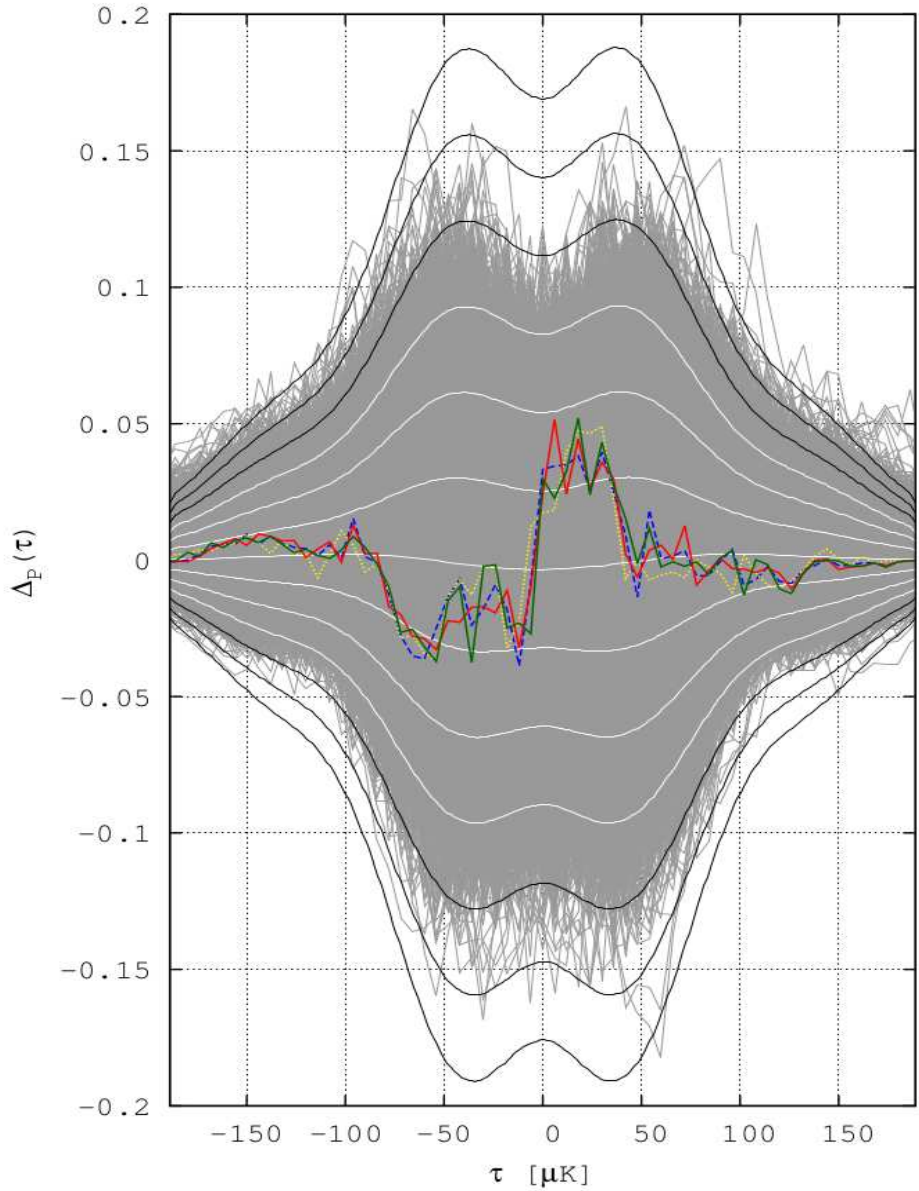}
\caption{Envelope of each $\Delta_{P}$ function in grey, $1\sigma$ to $6\sigma$ cosmic variances in white or black solid lines. The central and almost horizontal white line is the discrepancy function of the average sample. $Planck$ maps are in black dashed (resp. blue dashed) line for $NILC$, white dotted (resp. yellow dotted) line for $SEVEM$, in black solid (resp. red solid) line for $SMICA$, and dark grey solid (resp. green solid) line for $Commander$--$Ruler$, as calculated for the $\Lambda$CDM map sample, $N_{\rm side}$=128, $\ell_{\rm range}$=[2,256] with $U73$ mask, 2$^{\circ}$ fwhm, for a 6$\mu$K temperature bin width over the equal temperature range ($\pm$201$\mu K$). (Lines in colour in the online version.)}
\label{PFR_Fig11}
\end{figure}

\begin{figure}[!htb]
\includegraphics[height=0.75\textheight,width=1.00\textwidth]{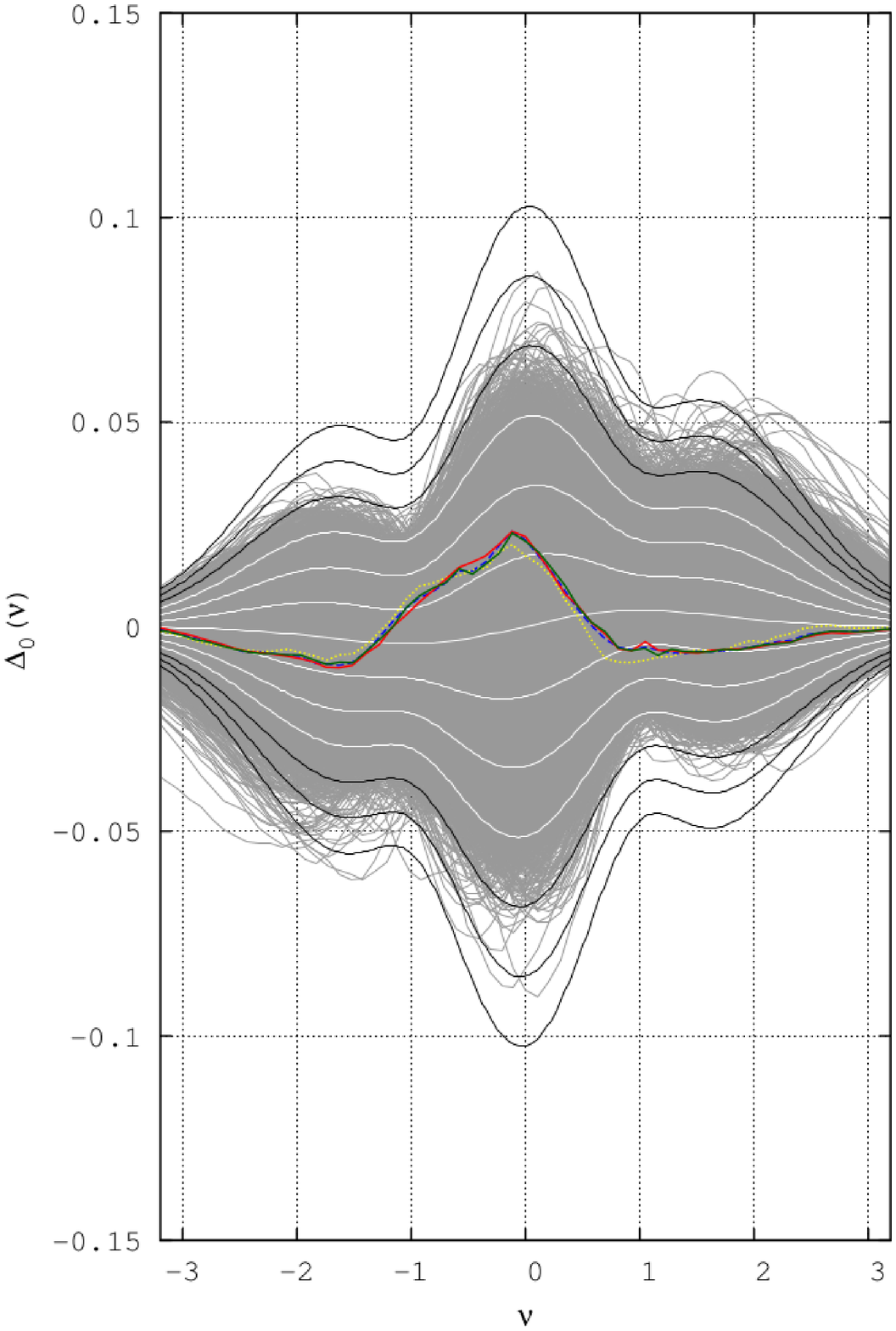}
\caption{Envelope of each $\Delta_{0}$ function. Same figure caption as in figure~\ref{PFR_Fig11}. }
\label{PFR_Fig12}
\end{figure}

\begin{figure}[!htb]
\includegraphics[height=0.75\textheight,width=1.00\textwidth]{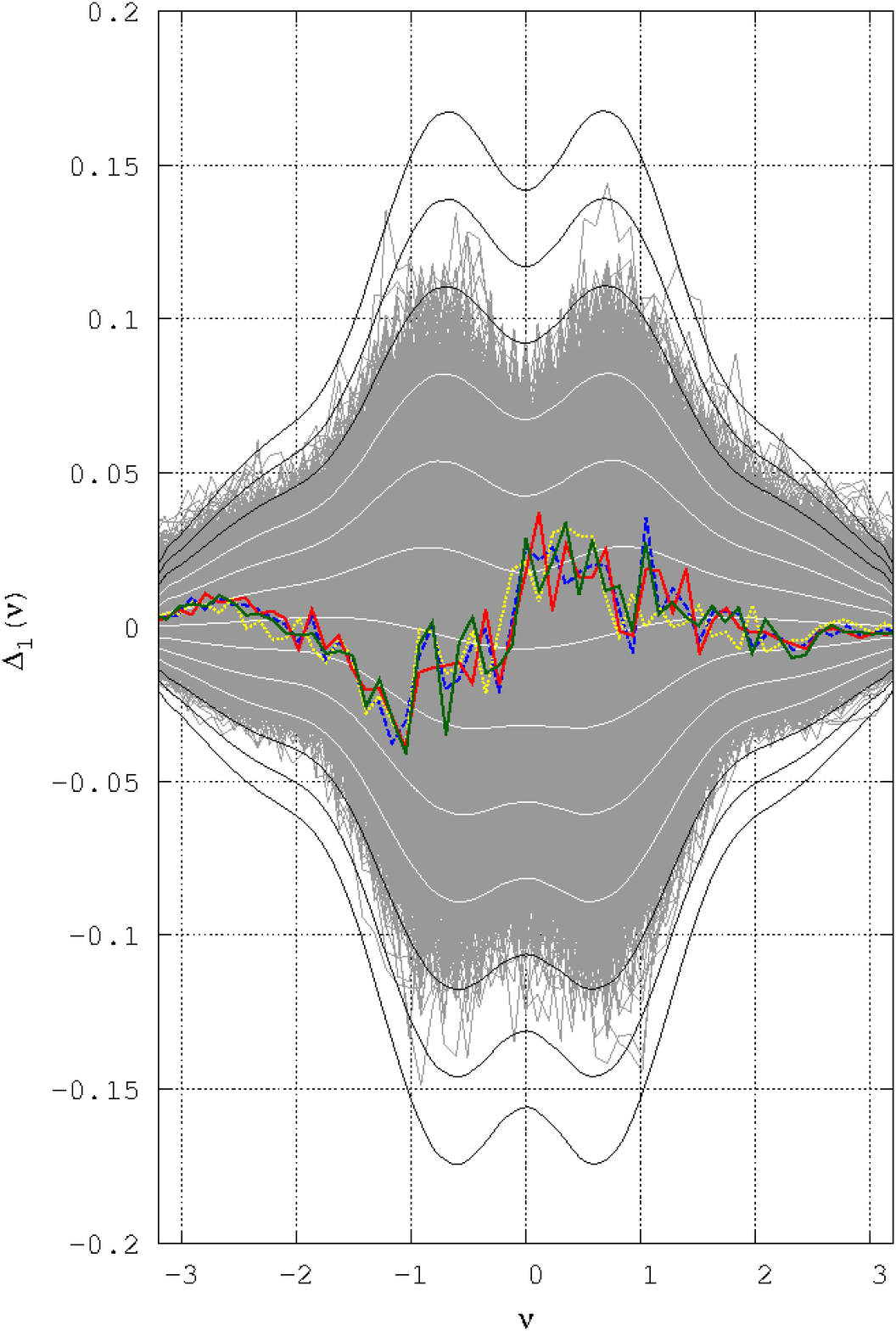}
\caption{Envelope of each $\Delta_{1}$ function. Same figure caption as in figure~\ref{PFR_Fig11}.}
\label{PFR_Fig13}
\end{figure}

\begin{figure}[!htb]
\includegraphics[height=0.75\textheight,width=1.00\textwidth]{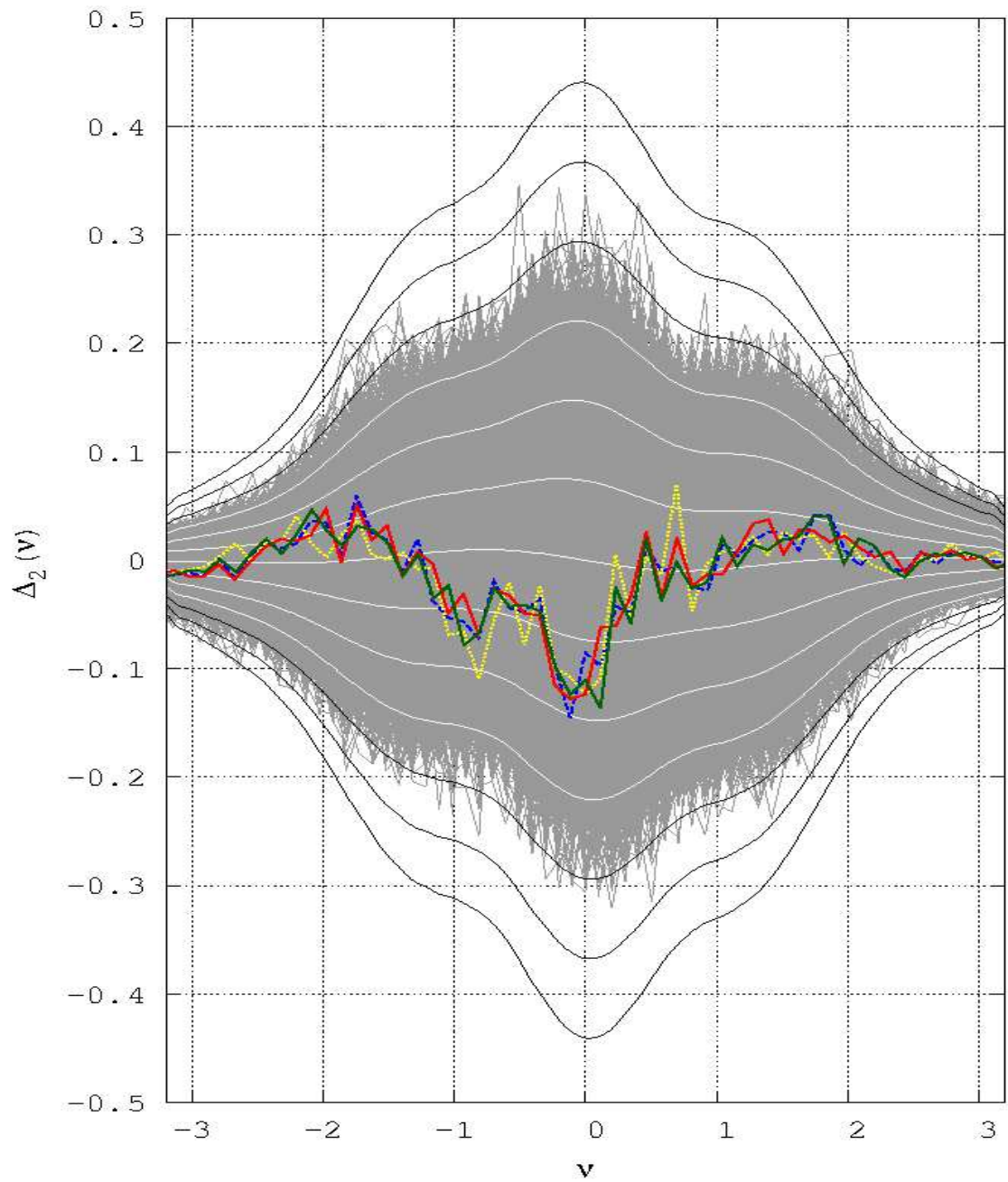}
\caption{Envelope of each $\Delta_{2}$ function. Same figure caption as in figure~\ref{PFR_Fig11}.}
\label{PFR_Fig14}
\end{figure}

\clearpage

\subsection{Discussion: origin of non--Gaussianities}
\label{discussion NGs}

Contrary to other works to unveil CMB non--Gaussianity, no specific model of non--Gaussianity has been put into the $100000$ maps of the $\Lambda$CDM sample probed all along this study (see \ref{appA}). However, already without mask, as in the initial analysis of the sample explored in this paper, we find that the four discrepancy functions show small but clear departures from Gaussianity, and this for a map ensemble that is supposed to be highly if not completely Gaussian, since it is computed from the $\Lambda$CDM model using purely Gaussian initial conditions. What is the origin of these NGs? For a given ensemble of realizations one expects several extraneous mechanisms that generate supplementary NG such as: an increasing smoothing scale beyond the $1^{\circ}$ horizon angular scale on the surface of last scattering increases the NG amplitude (increasing the dependence between neighbouring pixels acts in the sense of enlarging the causal horizon radius); we also checked that smoothing leaves the shape of the discrepancy functions unaffected, while rising the amplitude of NG.  
Our present study shows that these numerical NGs are of non-negligible amplitude even for a $100 000$ map sample without mask. Our previous analyses using different parameters $N_{\rm side}$, $l_{\rm max}$, $\rm fwhm$, without or with mask leave the shape of the descriptors $\Delta_1$ and $\Delta_2$ invariant, but with different amplitudes (we have no such invariance for $\Delta_P$ and $\Delta_0$, they detect the mask). We reached the same conclusion using different cosmological parameters (e.g. $WMAP$ 7yr) for the generation of the power spectrum to get a $100000$ map sample. 
The comparison with the NG of the observed CMB (figures~\ref{PFR_Fig11} to \ref{PFR_Fig14}) shows that numerous individual $\Lambda$CDM maps can possibly be of the same amplitude and shape of NG than the $Planck$ data map. These assessments would say that without any addition of supplementary NGs in each sample map, some sources of non--Gaussianity existing in the power spectrum used to generate the maps may well be detected by the discrepancy functions. The census of components is in \ref{appA:CMBcomponents}. Among them, probably some cannot be detected because we limit the {\em l--{\rm range}} to $[2,256]$, but the Sachs--Wolfe effect, the Doppler effect and the integrated Sachs--Wolfe effect would contribute to the non--Gaussianity detected in the individual maps of the ensemble. A systematic study is needed to disentangle the NG effects of the various components.

\subsection{Discussion: the case of $\sigma_{0}$}
\label{discussion sigma0}

When $\mu$, $\sigma_{0}$, and $\sigma_{1}$ are evaluated over $[-\tau_m , \tau_m ]$, while the discrepancy functions are calculated over a smaller interval in $\tau$ (resp. in $\nu$), this leads to an error on their shape and to an overestimate of the magnitude of NG, as we tested for the $100000$ map sample. Such a problem may arise when one works with the normalized temperature $\nu=\tau/\sigma_{0}$. While $\sigma_{0}(\tau_{max1})$ is known for the temperature range $\tau_{max1}$ giving $\nu_{max1}=\tau_{max1}/\sigma_{0}(\tau_{max1})$, the need to impose a $\nu$--range ($\nu_{max2}$) different from $\nu_{max1}$ (e.g. for comparing with another map sample) is equivalent to solve for an implicit function $f$, $\sigma_{0}(\nu_{max2})=f(\sigma_{0}(\nu_{max2}))$, simply because two different map samples or two different maps will have in general different variances and variances of the gradient, even if they are explored over the exactly same temperature range. The estimation of the unknown $\sigma_{0}$ corresponding to the new $\nu$--range can be made using iterative methods, or by referring to a model that predicts the behaviour of $\sigma_{0}$ for given changes of the range. For the comparison (with U73 mask) of $Planck$ maps to the $\Lambda$CDM sample, $P(\tau)$ and $\Delta_{P}(\tau)$ show no difficulties as abscissas are in $\tau$ and impose the same range to data and to the simulation maps. 

For $\Delta_{0,1,2}$ we first calculated $\mu$'s and $\sigma$'s over the full $\tau$--range ($\pm$ 201 $\muK$) of the four different $Planck$ maps. For each of these we obtain the following $\sigma_{0}$ values: $NILC$ 51.532$\muK$, $SEVEM$ 51.750$\muK$, $SMICA$ 51.576$\muK$, $Commander$---$Ruler$ 51.794$\muK$, while we obtain $\sigma_{0}=$59.294$\muK$ for the $\Lambda$CDM simulations over exactly the same ($\pm$ 201 $\muK$) range! This illustrates well the ``anomalously'' low variance of $Planck$ data already observed with $WMAP$ compared to a $\Lambda$CDM model map ensemble (see \cite{monteserin,cruz,bennett11,bennett13,hinshaw13,gruppuso,schwarz} and \cite{planck3,addison16}). When not treated correctly, the impact of this anomaly is significant for the calculation of the Gaussian premises $P^{\rm G}$, equation (\ref{G}), and $\mv^{\rm G}_{k}$, equations (\ref{v0G_}) and (\ref{vg}), as they are functions of the standard deviation $\sigma_{0}$. Regarding $\mv_{1}$ and $\mv_{2}$ we made two interesting observations: firstly, the Gaussian premises $\mv^{\rm G}_{1}$ and $\mv^{\rm G}_{2}$ are functions of the ratio $\rho= \sigma_{1} / \sigma_{0}$ in the prefactor of the exponential, and we verified that this ratio is very similar for the four $Planck$ maps, $\rho=36.657 \pm 0.031$, even if the $\sigma_{0}$'s are not very similar, $\sigma_{0}=51.663 \pm 0.131 \muK$. For the simulation sample, $\rho_{\Lambda CDM}$ has a much smaller value: $34.162$. Secondly, we observed that the ratio $\rho$ is very stable for the sample with U73 mask and for $Planck$ maps with U73 mask when passing from the largest common temperature range ($\pm$ 201 $\muK$) to the smallest temperature range covering all the maps ($\pm$ 396.5 $\muK$). On the other hand we noticed (see at the beginning of \ref{appA}) that the ratio $\rho$ is no more stable but increasing when passing from no mask to U73 mask. 

In summary, in the present work, the issue of the fair comparison of different CMB maps, 
(i) is treated by using exactly the same temperature range for $P$, $P^{\rm G}$ and $\Delta_{P}$, (ii)
for $\mv_{0}$, $\mv^{\rm G}_{0}$ and $\Delta_{0}$ it required some iterations upon different temperature ranges to reach close to the $\nu_{\rm max}$ values, and (iii) is simplified for $\mv_{1}$, $\mv^{\rm G}_{1}$, $\mv_{2}$, $\mv^{\rm G}_{2}$, $\Delta_{1}$ and $\Delta_{2}$ by assuming $\rho=\sigma_{1} / \sigma_{0}$ to be constant for the given change of temperature range.

\subsection{Discussion: $f_{NL}$ results}
\label{discussion fNL}

Non--linear correction terms within the standard perturbation approach are commonly investigated by constraining the coefficients (bi--spectrum with $f_{\rm NL}$ and tri--spectrum with $g_{\rm NL}$). Applied to Bardeen's curvature these constraints allow to decide whether inflation models such as with single slow--roll scalar field are rejected or not. A scalar random field of the primordial period, the gravitational potential $\Phi({\boldsymbol{\hat n}})$ (${\boldsymbol{\hat n}}=\boldsymbol{\hat n} (\vartheta,\varphi)$), is commonly expanded in real space 
around a Gaussian random field $\phi({\boldsymbol{\hat n}})$ (of vanishing mean), using the expansion:
\begin{equation}
\label{nlcorr}
\fl
{\Phi({\boldsymbol{\hat n}})} = \phi({\boldsymbol{\hat n}}) + f_{\rm NL}^{\rm (local)} (\phi^{2}({\boldsymbol{\hat n}}) - \left<\phi^{2}({\boldsymbol{\hat n}})\right>) + g_{\rm NL}^{\rm (local)} (\phi^{3}({\boldsymbol{\hat n}}) - \left<\phi^{3}({\boldsymbol{\hat n}})\right>) + \cdots \;.
\end{equation}
In 2010, the limit derived from $WMAP$ 7yr was $f_{\rm NL}^{\rm (local)} \approx 30 \pm 20$ at $1\sigma$, a very small non--Gaussianity almost compatible with a zero value allowing for single--scalar fields \cite{komatsu1}. 
$Planck$ 2013 data, expected to give smaller error bars, have narrowed this to
$f_{\rm NL}^{\rm (local)}=2.7\pm 5.8$, $f_{\rm NL}^{\rm (equilateral)}=-42\pm 75$, and $f_{\rm NL}^{\rm (orthogonal)}=-25\pm 39$ \cite{planck4}.
Then, $Planck$ 2015 combined temperature and polarization data provided
$f_{\rm NL}^{\rm (local)}=0.8\pm 5.0$, $f_{\rm NL}^{\rm (equilateral)}=-4\pm 43$, and $f_{\rm NL}^{\rm (orthogonal)}=-26\pm 21$ ($68\%$ CL, statistical \cite{planck1}). 
All these very similar outcomes are consistent with a vanishing $f_{NL}$, which is itself consistent with a very weak primordial non--Gaussianity. However, the interpretation of this characterization of CMB Gaussianity depends on the cosmological model and, in particular, on the type of inflation mechanism that is assumed in this model. 

\section{Conclusion}
\label{conclusion}

Detecting non--Gaussianity in the CMB temperature maps is still a great challenge. Whatever the descriptor, the discrepancy functions of the CMB by $Planck$ 2015 data are not zero 
but stay within $1\sigma$ or slightly leak into the $2\sigma$ cosmic variance band of current (model--dependent) ensembles. 
In our search for CMB non--Gaussianities we find systematic signatures of weak non--Gaussianity which would be of importance if the ensemble cosmic variance would be re-evaluated at smaller amplitudes. Compared to the $Planck$ maps we find that the $\Lambda$CDM simulations commonly used offer a systematically larger value for the variances $\sigma_{0}^{2}$ and $\sigma_{1}^{2}$ as observed in \cite{monteserin,cruz,bennett11,bennett13,hinshaw13,gruppuso,schwarz} and \cite{planck3,addison16}.
This anomaly of the variances and its interpretation
has to be explored in more detail in the prescriptions of model simulations. Furthermore, after the first main mask is applied, further different possible
superimposed foreground and source masks have a big impact upon the magnitude of non--Gaussianities, showing that numerous tiny field sources contribute to
residual foreground contamination and may imply a noticeable change in the values of the variances \cite{chingangbam_park}. Several possible strategies should 
be explored to select ensembles on the basis of the actually observed statistical parameters within a constrained random field approach, in order to reach a 
deeper understanding of a statistical comparison with a single realization, or to accordingly constrain cosmological models.

\vspace{14pt}

{\em Acknowledgements:}
\small{
This work was conducted within the ``Lyon Institute of Origins'' under grant ANR--10--LABX--66. 
A part of this project was funded by the National Science Centre, Poland, under grant 2014/13/B/ST9/00845. 
The authors wish to thank David Wiltshire and Fran\c{c}ois Bouchet for the invitation to this Focus Article,
as well as Giuseppe Fanizza, Takahiko Matsubara, David Wiltshire, Wen Zhao, and an anonymous referee for valuable comments on the manuscript. We also wish to thank Pierre Mourier and Boud Roukema for interesting discussions,
and we are grateful to L\'eo Michel--Dansac for his help concerning various computing facilities, and to 
Sven Lustig for a fruitful collaboration at an early stage of this work.

This work is based on observations obtained with $Planck$ (\href{http://www.esa.int/Planck}{http://www.esa.int/Planck}), 
an ESA science mission with instruments and contributions directly funded by ESA Member States, NASA, and 
Canada. 
Some of the results in this paper have been derived using the HEALPix package \cite{gorski2005}, available at \href{http://healpix.sourceforge.net}{http://healpix.sourceforge.net}. The CMB 
power spectra are calculated from the cosmological parameters using $CAMB$ software written by Lewis and Challinor  
(\href{http://lambda.gsfc.nasa.gov/toolbox/tb_camb_form.cfm}{http://lambda.gsfc.nasa.gov/toolbox/tb camb form.cfm}) from the original 
Boltzmann codes by Bertschinger, Ma and Bode resumed by Seljak and Zaldarriaga. 
The $CAMB$ ReadMe (2016) is available here (\href{http://camb.info/readme.html}{http://camb.info/readme.html})
and $CAMB$ Notes by A Lewis (2014) here (\href{http://cosmologist.info/notes/CAMB.pdf}{http://cosmologist.info/notes/CAMB.pdf}).



We acknowledge the use of the gfortran compiler and the gnuplot graphics utility software under various 
Linux operating systems. 
}

\appendix

\section{Definitions and notations for the CMB analysis and CMB map generation}
\label{appA}

Statistical and morphological descriptors of a random field are {\em a priori} based on model--independent 
definitions for a given support manifold. But these descriptors are bias-- and error--dependent according to the impact they have on the amplitude and shape of the non--Gaussianity, problems we shall 
touch upon in this appendix. 
Numerical biases and errors originate from the pixelation and finite--temperature resolution 
of the CMB data maps and sample maps, but also from the way infinitesimal calculations are translated into discrete 
algorithms. These numerical issues affect the descriptors, the fundamental quantities 
of the field and also the Gaussian predictions.

In what follows we recall some fundamental quantities of the CMB random field as they are used in the numerical analysis.
The temperature anisotropy $\delta T$ we use in this paper is dimensionful, $\delta T := T - T_{0}$ in units of $\mu K$ with $T$ the absolute local temperature, and $T_0 = 2.7255 \pm 0.0006$K, is the mean CMB temperature, which can be considered as the monopole
component of the CMB maps. The dimensionless CMB temperature anisotropy $(T - T_{0}) / T_{0}$ is used in linear perturbation theory, in the Sachs--Wolfe formula (see for instance \cite{aurich5,aurich6}). The temperature anisotropy $\delta T$ derives from the measure of the spectral radiance $I(\nu({\rm GHz}))$ by the instruments $HFI$ and $LFI$ of the $Planck$ probe \cite{notari}, and depend on the modelling of several effects: relativistic Doppler--Fizeau, and Sunyaev--Zel'dovich. The magnitude of these effects is a function of the cosmological model, therefore, the CMB spectral radiance and consequently the CMB temperature rest--frame estimations are model--dependent.
The suffixes, $_{\rm px}$ or $_{\rm C\ell}$ apply to the quantities calculated: from the moments of the probability distribution function (no suffix), $P(\tau)$, directly from the pixels, or from the angular power spectrum, respectively. 

\subsection{Basic quantities of the CMB random field}
\label{basics}

We consider a Gaussian window function
with smoothing scale $\theta_{\rm fwhm} (^{\circ})$, 
\begin{equation}
\label{sigmag}
{\sigma_{\mathrm G}}:=\theta_{\rm fwhm}\frac{\pi}{180}\frac{1}{\sqrt{8 \ln 2}}\;.
\end{equation}
For $\sigma_{\mathrm G}^{2} \ll 1$ the Gaussian kernel is approximated by
\begin{equation}
\label{gausskernel}
{\rm {W}_{\ell}}=\exp{(-\frac{\sigma_{\mathrm G}^{2}}{2}\ell(\ell+1))} \;,
\end{equation}
and once the $a_{\ell,m}$ coefficients of the expansion of $\delta T$ on ${\cal S}^{2}$ in spherical harmonics are calculated from the map, 
given the pixelation correction ${\rm pxc}(\ell)$, the corrected multipole moments are given by
\begin{equation}
\label{C_l}
{\rm C}_{\ell} := \left< \frac{1}{2\ell+1} \sum_{m=-\ell}^{+\ell} |a_{\ell,m}|^{2} \right>\;,
\end{equation}
\begin{equation}
\label{clcorr}
{{\rm Cw}_{\ell}}={\rm C}_{\ell} \ {\rm pxc}(\ell)^{2} \exp{(-\ell(\ell+1)\sigma_{\mathrm G}^{2})} \;, 
\end{equation}
and the angular power spectrum by
\begin{equation}
\label{apsl}
(\delta {\rm T}_{\ell})^{2}:= (\ell(\ell+1) / 2\pi ){\rm Cw}_{\ell}.
\end{equation}

\smallskip
\smallskip

We consider the mean values,
\begin{equation}
{\mu}:=\alpha_{1} \ ; \
{\mu_{\rm px}}:=\left<\delta T\right>_{\rm px}\;,
\end{equation}
and the variances,
\begin{equation}
\label{sigma0_mnpx_sq}
{\sigma_{0}^{2}}:=\alpha_{2}-\mu^{2} \ ; \
{\sigma_{0 {\rm px}}^{2}}:=\left<(\delta T)^{2}\right>_{\rm px} - \left<\delta T\right>_{\rm px}^{2}\;.
\end{equation}
For a homogeneous and isotropic Gaussian random field, $\sigma_{\rm 0 C\ell}^{2}$ is independent of the 
absolute direction of light provenance; the variance ensemble average is
\begin{equation}
\label{sigma0_Cl_sq}
{\sigma_{0 {\rm C\ell}}^{2}}:=\sum_{\ell=2}^{\ell_{\rm max}}{{\rm Cw}_{\ell}\frac{(2\ell+1)}{4 \pi}} \;.
\end{equation}
The variance of the local gradient is (compare the definition in equation (\ref{sigma1_pxs}))
\begin{equation}
\label{sigma1_pxs_sq}
{\sigma_{1 {\rm px}}^{2}}:=\left<(\nabla_{1} \delta T ({\boldsymbol{\hat n}}))(\nabla^{1} \delta T ({\boldsymbol{\hat n}})) + (\nabla_{2} \delta T ({\boldsymbol{\hat n}}))(\nabla^{2} \delta T ({\boldsymbol{\hat n}}))\right>\;,
\end{equation}
and, only in a homogeneous and isotropic model, for a Gaussian random field, the ensemble average
of $\sigma_{1 {\rm C\ell}}^{2}$ is
\begin{equation}
\label{sigma1_Cl_sq}
{\sigma_{1 {\rm C\ell}}^{2}}:=\sum_{\ell=2}^{\ell_{\rm max}}{{\rm Cw}_{\ell}\frac{(2\ell+1)\ell(\ell+1)}{4 \pi}} \;.
\end{equation}
The measures of the variance and variance of the local gradient, equations (\ref{sigma0_mnpx_sq}) and (\ref{sigma1_pxs_sq}), contain an averaged statistical, respectively, morphological information of the random field. Since the model--dependent expressions, equations (\ref{sigma0_Cl_sq}) and (\ref{sigma1_Cl_sq}), are valid only for homogeneous--isotropic Gaussian fields,
any significant difference between these two methods indicates a preliminary detection of non--Gaussianity and the degree of conformity with the model for the CMB. For the $\Lambda$CDM sample we notice such a difference between the variances calculated from the pixels and the variances calculated from the angular power spectrum.
Model--dependent predictions for $\sigma_{0 C\ell}$ and $\sigma_{1 C\ell}$ and further discussions can be found in \cite{aurich4}.

\bigskip

Table \ref{tablemusigmas} shows the averaged values of $\mu$, $\sigma_{0}$ and $\sigma_{1}$ over the CMB map sample in the $\Lambda$CDM model.
Two cases are shown in this table and all the further tables for our $\Lambda$CDM $10^{5}$ map sample at $N_{\rm side} =128$, $\ell_{\rm max} =256$, $2^{\circ}$fwhm: 
without mask at bin width $13\mu$K and over the full temperature range of the sample; with $U73$ mask subtraction at bin width $6\mu$K and 
over the limited temperature range fixed by the $Planck$ $SMICA$ map. The effect of passing from no mask to the mask $U73$ is upon $\mu$, $\sigma_{0}$ and $\sigma_{1}$. 
And the variation of the ratio $\sigma_{1}$/$\sigma_{0}$ is noticeable (e.g. $34.16218 / 33.93190 = 1.006787$) and affects the magnitude of the Gaussian premise of the Minkowski Functionals $\mv_{1}$ and $\mv_{2}$.

\begin{table}[!htb] 
\centering

\resizebox{\columnwidth}{1.4cm}{
  \begin{tabular}{|c|c|c|c|}
    \hline
    \multicolumn{4}{|c|}{(Units of $\mu K$) ~ ~ $\Lambda$CDM sample~ Full individual map range, no mask, 2$^{\circ}$fwhm, bin 13$\mu$K}\\
    \hline
    $\mu_{px}$ & $\sigma_{0 px}$ & $\sigma_{1 px}$ & $\sigma_{1 px}/\sigma_{0 px}$\\
    \hline
    $-3.4\times10^{-7}$ & $59.53348$ & $2020.08398$ & $33.93190$\\
    \hline 
    \hline
    \hline 
    \multicolumn{4}{|c|}{(Units of $\mu K$) ~ ~ $\Lambda$CDM sample ~ Equal temperature range ~ (ETR~$\pm$ 201$\mu K$), U73 mask, 2$^{\circ}$fwhm, bin 6$\mu$K}\\
    \hline
    $\mu_{px}$ & $\sigma_{0 px}$ & $\sigma_{1 px}$ & $\sigma_{1 px}/\sigma_{0 px}$\\
    \hline
    $-1.4534\times10^{-3}$ & $59.13275$ & $2020.10372$ & $34.16218$\\
    \hline
  \end{tabular}
}
  \caption{Table of $\mu$ and $\sigma$ values. }\label{tablemusigmas}
\end{table}
\vspace{0.1 true cm}

\subsection{The cosmic microwave background in the $\Lambda$CDM model.}

In order to construct the model map sample, the CMB angular power spectrum is generated with $CAMB$ from the cosmological parameters of the $\Lambda$CDM model according to $Planck$ 2015 \cite{planck5}, p.31, table 4, last column, and Review of particle physics \cite{revpartphys}:

$H_{0}=67.74 \pm 0.46$ km$s^{-1}$Mpc$^{-1}$, the Hubble constant today

$h=H_{0}$/($100$kms$^{-1}$Mpc$^{-1}$) $=0.6774$, the normalized Hubble constant today

$T_{0}=2.7255 \pm 0.0006$ K, present--day CMB temperature

$\Omega_{b} h^{2}=0.02230 \pm 0.00014$, baryon density

$\Omega_{c} h^{2}=0.1188 \pm 0.0010$, Cold Dark Matter density

$\Omega_{\nu} h^{2}=0.00209$, neutrino density 

$\Omega_{k}=0$, constant--curvature density parameter

$Y_{P}=0.249_{-0.026}^{+0.025}$, helium fraction

$N_{\rm eff}=3.04 \pm 0.33$, number of massless neutrinos

$\sum m_{\nu}<$ 0.194 eV, neutrino mass eigenstates 

$\tau$=0.066 $\pm$ 0.012, reionization optical depth 

$z_{re}$=8.8 $_{-1.1}^{+1.2}$, redshift of the reionization

$n_{S}$=0.9667 $\pm$ 0.0040, scalar spectral index

$x_{e}$=1, ionization fraction

$z_{*}$=1089.90 $\pm$ 0.23, redshift of decoupling

$\Omega_{\Lambda}$=0.6911 $\pm$ 0.0062, cosmological constant density

$\Omega_{m}$=0.3089 $\pm$ 0.0062, total matter density

$Age$= 13.799 $\pm$ 0.021 Gyr, age of Universe

\subsection{The CMB radiation components}
\label{appA:CMBcomponents}

The full Boltzmann physics is implemented in the $CAMB$ software; thus, the maps of the CMB sample in the $\Lambda$CDM model include the following effects: ordinary Sachs--Wolfe effect / Doppler effect / Silk damping / Reionization / Polarization of photons / Neutrinos / Integrated Sachs--Wolfe effect ($ISW$) / Lensing.
The weak lensing effect upon the CMB is treated in \cite{durrer} and in \cite{CMBlensing} beyond the Born approximation, which is usually applied at first--order of perturbation theory. For a second--order treatment the deflection angles, still assumed small if $\ell \leqslant 2500$\footnote{Two times the typical angular deviation equals $4$ arc minutes (CMB lensing due to structures at $z < 20$ in a flat $FLRW$ model); this corresponds to $\ell_{max}=180^{\circ}/(4/60)=2700$.}, are no more Gaussianly distributed, but the post--Born corrections would be weakly detectable in low noise CMB temperature maps merging the contributions of hundreds of $\ell$--values between $\ell=1000$ and $\ell=2500$. Our present study uses an $\ell$--range [2,256], well below $\ell = 1000$.
One notices that an evaluation of the Sunyaev--Zeldovich effect ($SZ$) on the CMB radiation is not implemented in $CAMB$, the sources responsible of the $SZ$ effect being more or less excluded when using an appropriate foreground mask. It is clear that, in this list, some of the effects are not primordial but depend on model--dependent knowledge of the matter distribution and the physical assumptions in the concordance model, i.e., mainly the $ISW$ effect, the lensing and the neutrinos. 

\subsection{Map ensemble and statistical stability}

From the power spectrum we generate with $synfast$ a $10^5$ map ensemble. Given the map resolution ($N_{\rm side} =128$, $\ell_{\rm max} = 256$, and $2^{\circ}$fwhm), the number of $10^5$ maps allows a statistical stability in the sense of the second averaged MF $\mv_1$, which becomes smooth and stable around $10^5$ maps, a second different sample of $10^5$ maps gives very similar results and doubling the sample brings no visible improvement in the shape of $\mv_1$. A number of $10^5$ maps is satisfactory in the sense of the statistical stability for the MFs themselves, but less for the PDF. However, once we come to the discrepancy functions, the stability is definitely impaired when using a smaller number of maps, and $10^5$ maps must be considered as a minimum requirement. We shall not develop more on these studies in the present work.

\subsection{Discretization: definition of the $\tau$--lattice}
\label{appA:discretization}

One observes that in the case of $10^5$ realizations there is an interval in $\tau$, $[\tau_- , \tau_+ ]$ such that $P(\tau ) = 0$
{\em for all} realizations, if $\tau \notin [\tau_- , \tau_+ ]$.
In order to simplify the problem, we use a symmetric interval $[ -\tau_m , + \tau_m ]$, where $\tau_m = {\rm max}\lbrace \vert\tau_- \vert, \tau_+ \rbrace$.
The interval $[ -\tau_m , + \tau_m ]$ is divided into $2L + 1$ bins of equal bin width $\Delta \tau : = 2\tau_m / (2L+1)$, where the mid--points
of the bins are given by the ``$\tau-$lattice'', $\tau_{l} = - \tau_m + (2l - 1) (\Delta\tau / 2)$, $l = 1,2, \cdots, (2L +1)$, in such a way that the $(L+1)^{\rm th}$ bin is centered at $\tau_{L+1} = 0$.
(Example: $\tau_m = 396.5\; \muK$, $L=30$, $\Delta\tau = 13 \mu K$ $\Rightarrow (2L + 1) = 61$ bins.)
Then, the {\em discretized PDF} for a given realization, which is now a step--function, i.e. piecewise continuous, can be represented as a {\em histogram} that is 
defined by ($l = 1, 2, \cdots , (2L+1)$):
\begin{equation}
P(\tau_{l}): = \frac{\#{\rm pixels} \lbrace \delta T \in \lbrack \tau_{l} - \frac{\Delta\tau}{2} , \tau_{l} + \frac{\Delta\tau}{2} )\rbrace }{N_{\rm tot} \Delta\tau} \;\;,
\end{equation}
with $N_{\rm tot}$ the total number of pixels, and
for $\tau$ within the $l^{\rm th}$ bin, i.e., $\tau \in \left[\tau_{l} - \frac{\Delta\tau}{2} , \tau_{l} + \frac{\Delta\tau}{2}\right)$ (half open interval!).

The {\em ensemble average (mean)} $\langle P(\tau )\rangle$ is given by the arithmetic mean of all the PDFs of the $10^5$ realizations.
$\langle P(\tau )\rangle$ is still constant in a given bin, denoted by $\langle P(\tau_{l} )\rangle$ in the $l^{\rm th}$ bin and, thus, the derivative ${\mathrm d} \langle P(\tau )\rangle/ {\mathrm d}\tau$ is zero almost everywhere, but $\langle P(\tau )\rangle$ will have, in general,
$(2L + 2)$ jumps, namely at the points $\tau_{l} + \Delta\tau / 2$, $l = 1,2, \cdots 2L$ and at the points $\mp \tau_m$ (iff $\langle P(\tau )\rangle$ possesses this special property!) In general, there will be a {\em first jump} at a value $\tau_0 \in [ - \tau_m , \tau_1 + \Delta\tau /2 )$,
and a {\em last jump} at a value $\tau_{2L+2} \in (\tau_{2L+1} - \Delta\tau /2 , \tau_m ]$, and these contributions have to be treated separately. 

If we ignore the last subtlety, the expectation value (\ref{E}) of a given random field $f(\delta T)$ is exactly given by the finite sum
\begin{equation}
\langle f (\delta T )\rangle = \sum_{l = 1}^{2L+1}  \langle P(\tau_{l}) \rangle \int_{\tau_{l}- \Delta\tau/2}^{\tau_{l}+ \Delta\tau/2}  f (\tau) \, {\mathrm d}\tau \;\;.
\end{equation}
As an important example, this yields with $f(\tau) = \tau^n$ the {\em exact formula for the moments $\alpha_n$ of $\delta T$} (see equation (\ref{alphan}) for $n=0,1,2,\cdots $):
\begin{equation}
\label{exactformula}
\alpha_n = \frac{1}{n+1} \sum_{l = 1}^{2L+1} \langle P(\tau_{l}) \rangle \left[ (\tau_{l}+ \Delta\tau/2)^{n+1} -  (\tau_{l}- \Delta\tau/2)^{n+1} \right] \;\;.
\end{equation}
For $n=0$ and $n=1$ one obtains the expected values
\begin{equation}
\alpha_0 = \sum_{l = 1}^{2L+1}\langle P(\tau_{l}) \rangle\, \Delta\tau = 1 \quad ({\rm normalization}) \;\;,
\end{equation}
\begin{equation}
\alpha_1 = \mu = \sum_{l = 1}^{2L+1}\, \tau_{l} \langle P(\tau_{l}) \rangle \, \Delta\tau = 0 \;\;,
\end{equation}
whereas, for $n \ge 2$, one gets a decomposition into a ``main term'' and an ``exact correction'' in the form of a finite series in the bin width $\Delta\tau$:
\begin{equation}
\label{correction}
\fl
\alpha_n = \sum_{l = 1}^{2L+1}\, \tau^n_{l} \langle P(\tau_{l}) \rangle \, \Delta\tau + \frac{2}{n+1}  
\sum_{k=1}^{[n/2]} \left( \begin{array}{c} n+1 \\ 2k+1 \end{array} \right) \sum_{l =1}^{2L+1} \tau_{l}^{n-2k} 
\langle P(\tau_{l}) \rangle \, \left(\frac{\Delta\tau}{2}\right)^{2k+1}\;\;.
\end{equation}
(Example for $n=2$: $\sigma_0^2 : = \alpha_2 =  \sum_{l =1}^{2L+1}\tau_{l}^2 \langle P(\tau_{l}) \rangle \Delta\tau + (\Delta\tau)^2 / 12 $.)

In many papers on MFs the integral in equation (\ref{E}) and similar integrals are approximated by the ``main term'' and, thus, the results suffer from an {\em error}, which is exactly given (in the case of (\ref{E})) by the ``correction'' in (\ref{correction}), if it is not taken into account. For a discussion of the correction term in the case of the MFs $\mv_1$ and $\mv_2$, see \cite{lim_simon}. This error can be made small, iff $\Delta\tau$ or $\Delta\nu$ is chosen small enough in principle, which is not the case, i.e., for $\Delta\tau = 13\mu K$. Looking at the ``main term'' (mentioned above) without correction,  
\begin{equation}
\label{mainterm}
\alpha'_n = \sum_{l = 1}^{2L+1}\, \tau^n_{l} \langle P(\tau_{l}) \rangle \, \Delta\tau \;,
\end{equation}
we obtain that $\sigma_{0}'=\sqrt{\alpha'_{2}}=59.6516\muK$ to be compared with $\sigma_{0 \rm px}=59.5335\muK$ (second equation in (\ref{sigma0_mnpx_sq}) and see also table \ref{tablemusigmas}).

\section{The generating functions of the moments and cumulants, and a closed expression for the NG--parameters ${\bf a}_{\bf P} {\bf (n)}$}
\label{appB}

From the definition (\ref{Delta1}) of the discrepancy function $\Delta_P (\tau)$ and the Hermite expansion (\ref{Delta2}), one obtains with (\ref{M}) for the generating function $M(x)$ of the moments $\alpha_n$ (with $\alpha_0 = 1$, $\alpha_1 = \mu = 0$, $\alpha_2 = \sigma_0^2$):
\begin{equation}
\label{MX}
\fl
M(x) = \sum_{n=0}^{\infty} \frac{\alpha_n}{n!}  x^n = \langle \e^{x \delta T}\rangle = \int_{-\infty}^{\infty} e^{x\tau}  \left[ P^{\rm G} (\tau) + \frac{1}{\sqrt{2\pi}\sigma_0} \Delta_P (\tau ) \right]  {\mathrm d}\tau = : M^{\rm G} (x) + M^{\Delta} (x) , 
\end{equation}
with 
\begin{equation}
\label{MGx}
M^{\rm G} (x) : = \frac{1}{\sqrt{2\pi}\sigma_0} \int_{-\infty}^{\infty} \e^{x \tau - \tau^2 / 2\sigma_0^2} \; {\mathrm d}\tau \;=\;\e^{\sigma_0^2 x^2 / 2} \;\;,  
\end{equation}
and
\begin{eqnarray}
\label{MDelta}
M^{\Delta} (x)&:= \frac{1}{\sqrt{2\pi}\sigma_0}\sum_{n=3}^{\infty} \frac{a_P (n)}{n!}  \int_{-\infty}^{\infty} \e^{x \tau - \tau^2 / 2\sigma_0^2} \; {\rm He}_n (\tau / \sigma_0) \; {\mathrm d}\tau \nonumber\\&
= \frac{1}{\sqrt{2\pi}} \e^{\sigma_0^2 x^2 / 2} \sum_{n=3}^{\infty} \frac{a_P (n)}{n!}  \int_{-\infty}^{\infty} \e^{-\nu^2 /2} \e^{(\sigma_0 x) \nu - \frac{1}{2} (\sigma_0 x)^2} {\rm He}_n (\nu)\, {\mathrm d}\nu \;.\nonumber\\ 
\end{eqnarray}
Using the generating function of the Hermite polynomials \cite{magnus},
$$
\e^{z\nu-\frac{1}{2}z^2} = \sum_{m=0}^{\infty} {\rm He}_m (\nu) \frac{z^m}{m!}\;\;,
$$
equation (\ref{MDelta}) can be rewritten as
$$
M^{\Delta} (x) =  \frac{1}{\sqrt{2\pi}} \e^{(\sigma_0 x)^2 / 2} \sum_{n=3}^{\infty} \frac{a_P (n)}{n!} \sum_{m=0}^{\infty} 
\frac{(\sigma_0 x)^m}{m!} 
\int_{-\infty}^{\infty} \e^{-\nu^2 /2} \;{\rm He}_m (\nu) \, {\rm He}_n (\nu)\; {\mathrm d}\nu \;\;,
$$
which leads with the orthogonality relation (\ref{ortho}) to
\begin{equation}
M^{\Delta} (x) = M^{\rm G} (x) \sum_{n=3}^{\infty} \frac{a_P (n)}{n!} (\sigma_0 x)^n \;\;,
\end{equation}
and, thus, with (\ref{MX}), (\ref{MGx}), to the {\em factorization}
\begin{equation}
\label{factorize}
M(x) = M^{\rm G} (x) K (x) \;\;,
\end{equation}
with 
\begin{equation}
K (x) : = 1+ \sum_{n=3}^{\infty} \frac{a_P (n)}{n!} \, (\sigma_0 x)^n \;\;.
\end{equation}
The cumulant generating function, defined in (\ref{C}), is then {\em additive},
\begin{equation}
C(x) = \sum_{n=2}^{\infty} \varkappa_n \frac{x^n}{n!} = C^{\rm G} (x) + \Delta C (x) \;\;,
\end{equation}
as given in equations (\ref{CX}) and (\ref{CX_}) of the main text. It is convenient to write $\Delta C (x):= \ln K (x)$ in terms of 
the dimensionless variable $z: = \sigma_0 x$,
\begin{equation}
\label{A8}
\Delta C \left(\frac{z}{\sigma_0}\right) = \ln \left[ 1 + \sum_{n=3}^{\infty} \frac{a_P (n)}{n!} \; z^n \right] = \sum_{n=3}^{\infty} \frac{C_n}{n!}
\; z^n \;\;,
\end{equation}
where $C_n:= \varkappa_n / \sigma_0^n$ are the {\em normalized} (dimensionless) {\em cumulants}. To obtain closed expressions for the $\rm NG$--coefficients $a_P (n)$ in terms of the cumulants $C_n$, we recall the {\em generating function of the complete Bell polynomials} $B_n$
\cite{comtet} ($B_0 = 1$):
\begin{equation}
\label{A9}
\exp \left[ {\sum_{n=1}^{\infty} \frac{x_n}{n!}\;z^n} \right] = 1 + \sum_{n=1}^{\infty} \frac{B_n (x_1, x_2, \cdots, x_n )}{n!} \; z^n \;\;.
\end{equation}
By expanding the exponential and comparing in  (\ref{A9}) the terms of the same power in $z$, it is not difficult to obtain, e.g., the first
four Bell polynomials: $B_1 (x_1 ) = x_1$, $B_2 (x_1,x_2) = x_1^2 + x_2$, $B_3 (x_1, x_2, x_3) = x_1^3 + 3x_1 x_2 + x_3$, $B_4 (x_1,x_2,x_3,x_4) = x_1^4 + 6 x_1^2 x_2 + 4 x_1 x_3 + 3 x_2^2 +x_4$. A comparison between (\ref{A8}) and (\ref{A9}) 
yields the closed expression ($x_1 = x_2 = 0, x_3 = \gamma_1 , x_4 = \gamma_2 , x_n = C_n$ for $n\ge5$):
\begin{equation}
\label{APcumulants}
a_P (n) = B_n (0,0, \gamma_1, \gamma_2, C_5 ,\cdots , C_n)\;\;,\;\; n \ge 3 \;\;, 
\end{equation}
as given in the main text in equation (\ref{A1Bell}). The explicit expressions for $B_3$ and $B_4$ give 
$a_P (3) = B_3 (0,0,\gamma_1) = \gamma_1$, $a_P ( 4) = B_4 (0,0,\gamma_1, \gamma_2) = \gamma_2$, in agreement with
(\ref{AP1}). In order to obtain the higher Bell polynomials, one can either use the {\em recurrence relations},
\begin{equation}
\label{recurrenceB}
B_{n+1} (x_1,x_2,\cdots , x_{n+1}) = \sum_{m=0}^{n} \left( \begin{array}{c} n \\ m \end{array} \right)\, B_{n-m} (x_1,x_2, \cdots , x_{n-m}) \;x_{m+1}\;,
\end{equation}
or the {\em combinatorial expression}
\begin{equation}
\label{combinatorialB}
B_n (x_1, x_2, \cdots , x_n ) = \sum_{\pi(n)} \frac{n!}{a_{1}! a_{2}! \cdots a_{n}!} \left(\frac{x_{1}}{1!}\right)^{a_{1}} \left(\frac{x_{2}}{2!}\right)^{a_{2}} \cdots \left(\frac{x_{n}}{n!}\right)^{a_{n}}\;\;.
\end{equation}
Here, the sum is over all partitions $\pi(n)$ of $n$, i.e., over all positive integers $a_{m}$ such that $\sum_{m=1}^{n} m\; a_{m} = n$. The multinomial coefficients
\begin{equation}
(n; a_{1}, a_{2}, \cdots, a_{n})':= \frac{n!}{(1!)^{a_{1}} a_{1}! (2!)^{a_{2}} a_{2}! \cdots (n!)^{a_{n}} a_{n}!}\;\;,
\end{equation}
are given, for $n=1, 2, \cdots , 10$, in table 24.2 in \cite{abramowitz}. The Bell polynomials have the nice property that their coefficients
are integers and, therefore, the NG coefficients $a_{P}(n)$ of the PDF discrepancy function are linear combinations of the normalized cumulants with integer coefficients, as seen in equation (\ref{AP1}). Since the NG--coefficients $a_{0}(n)$ of the discrepancy function $\Delta_{0}$ are related to the $a_{P}(n)$'s by $a_0 (n) = \frac{a_P (n+1)}{n+1}, n \ge 2 $, it follows that the coefficients of the $a_{0}(n)$'s are, in general, rational numbers (see equation (\ref{AP3})).

Finally, we give an alternative closed formula for the expansion coefficients $a_P (n)$, which does not express them in terms of the cumulants $C_n$ as in equation (\ref{APcumulants}), but rather in terms of the {\em normalized moments} ${\hat\alpha}_n$ (see equation (\ref{alphan})),
\begin{equation}
{\hat\alpha}_n : = \frac{\alpha_n}{\sigma_0^n} = \Bigl\langle \left( \frac{\delta T}{\sigma_0}\right)^n \Bigr\rangle\;\;.
\end{equation}
Replacing $\Delta_P$ in equation (\ref{AP2}) by its definition (\ref{Delta1}), we obtain:
\begin{equation}
a_P (n) = \int_{-\infty}^{\infty} \lbrack P(\tau) - P^{\rm G}(\tau) \rbrack \, {\rm He}_n (\tau / \sigma_0)\;{\mathrm d}\tau \;\;,
\end{equation} 
which gives, with the definition (\ref{E}) and the orthogonality relation (\ref{ortho}), ($n=0,1,2,\dots$):
\begin{equation}
\label{APnew}
a_P (n) = \langle {\rm He}_n (\delta T / \sigma_0) \rangle - \delta_{n0} \;\;.
\end{equation}
From (\ref{APnew}) follows immediately $a_P (0) = 0$, $a_P (1) = {\hat\alpha}_1 = \mu / \sigma_0 = 0$, and
$a_P (2) = \alpha_2 / \sigma_0^2 - {\hat\alpha}_0 =0$, and for the required $a_P (n)$'s with $n\ge 3$ using the expansion \cite{abramowitz},
\begin{equation}
{\rm He}_n (x) = n! \, \sum_{k=0}^{[n/2]} \frac{(-1)^k}{2^k k! (n -2k)!}\; x^{n-2k} \;\;,
\end{equation}
the closed expression 
\begin{equation}
\label{APC}
a_P (n) = n! \, \sum_{k=0}^{[n/2]} \frac{(-1)^k}{2^k k! (n -2k)!}\; {\hat\alpha}_{n-2k} \quad (n \ge 3 )\;\;.
\end{equation}
By virtue of this formula, the integral in the original definition (\ref{AP2}) is replaced by a finite sum and, thus, one obtains in combination with the formula (\ref{alphan}) for the moments simple closed expressions for the Hermite expansion coefficients. 

\clearpage 
 
\section{General Hermite and Edgeworth expansions}
\label{appC}

Here, we summarize some well--known mathematical facts (see, e.g., \cite{couranthilbert,boasbuck,erdelyi,cramer1,cramer2,cramer3}), which are used in section \ref{sec:v0} for expanding the various discrepancy functions $\Delta_P (x)$, $\Delta_0 (x)$, etc., in terms of Hermite polynomials.

Let $f(x)$, $\mathbb{R} \rightarrow \mathbb{R}$, be square integrable with respect to a positive weight function $w(x)$, i.e., $f(x) \in \mathbb{H}: = L^2 (\mathbb{R}, w(x) {\mathrm d}x)$.  For $f,g \in \mathbb{H}$, we define the {\em inner product} 
\begin{equation}
(f,g):= \int_{-\infty}^{\infty} w(x) f(x) g(x) {\mathrm d}x \;\;,
\end{equation}
which satisfies the Schwarz inequality $\vert (f,g)\vert \le || f || \cdot || g ||$, where $|| f ||$ denotes the {\em norm of $f$}, $|| f || : = \sqrt{(f,f)}$. 
Let $\psi_0 (x), \psi_1 (x), \cdots  \in \mathbb{H}$ be an {\em orthonormal system} satisfying the {\em orthogonality relation} $(m,n \in \mathbb{N}_0)$,
\begin{equation}
(\psi_m , \psi_n) = \delta_{mn}\;\;,
\end{equation} 
and let $f \in \mathbb{H}$ be any function. Then, the numbers
\begin{equation}
b_n : = (f, \psi_n) \quad (n \in \mathbb{N}_0 )
\end{equation}
are called the {\em expansion coefficients} (``Fourier coefficients'') of $f$ with respect to the $\psi_n$'s. From the relation
\begin{equation}
0 \le \int_{-\infty}^{\infty} w(x) \left( f(x)- \sum_{n=0}^{N} b_n \psi_n (x)\right)^2  {\mathrm d}x = || f ||^2 - \sum_{n=0}^{N} b_n^2 \;\;,  
\end{equation}
one obtains $\sum_{n=0}^{N} b_n^2 \le || f ||^2$, and since the right--hand--side of the last inequalities is independent of $N$, we obtain
{\em Bessel's inequality}
\begin{equation}
\label{bessel}
\sum_{n=0}^{\infty} b_n^2 \le || f ||^2 \;\;.
\end{equation} 
This proves that the sum of the squares of the expansion coefficients $b_n$ always converges.

Consider now, for a given function $f(x)\in \mathbb{H}$ the following linear combination:
\begin{equation}
\label{FN}
F_N (x) : = \sum_{n=0}^{N} \gamma_n \psi_n (x) \;\;,
\end{equation}
with constant coefficients $\gamma_n$ and fixed $N$. Then, there arises the question under which conditions the approximation (\ref{FN})
can be considered as an {\em approximation ``in the mean''} such that the {\em mean square error}
\begin{equation}
\label{error}
{\cal E}_N : = || f - F_N ||^2 
\end{equation}
is as small as possible. From the identity
\begin{equation}
\label{error2}
{\cal E}_N  = || f ||^2 + \sum_{n=0}^{N} (\gamma_n - b_n )^2 - \sum_{n=0}^{N} b_n^2  
\end{equation}
it follows immediately that ${\cal E}_N$ takes on its least value for $\gamma_n = b_n \ (n=0,1,\cdots , N)$. If the error ${\cal E}_N$ converges for 
every {\em piecewise continuous} function $f \in \mathbb{H}$ to zero as $N$ goes to infinity, then the orthonormal system $\lbrace \psi_n \rbrace$ is said to be {\em complete}, i.e., it provides a {\em complete basis of $\;\mathbb{H}$}, and Bessel's inequality (\ref{bessel}) becomes an equality for every piecewise continuous function $f$,
\begin{equation}
\label{completeness}
\sum_{n=0}^{\infty} b_n^2 = || f ||^2 \;\;,
\end{equation} 
which is known as {\em completeness relation} (also called {\em Parseval's equation}).

It is important to note that the completeness of the system $\lbrace \psi_n \rbrace$, expressed by the equation
\begin{equation}
\label{lim}
\lim_{N \rightarrow \infty} \int_{-\infty}^{\infty} w(x) \left( f(x) - \sum_{n=0}^{N} b_n \psi_n (x)\right)^2 \, {\mathrm d}x =0\;\;,
\end{equation}
does not necessarily imply that $f(x)$ can be expanded in a series in the functions $\psi_n (x)$. The expansion
\begin{equation}
\label{C9}
f(x) = \sum_{n=0}^{\infty} b_n \psi_n (x) 
\end{equation}
is, however, valid if the series in (\ref{C9}) converges {\em uniformly} and, thus, the limit in (\ref{lim}) can be carried out under the integral.

In this paper, our main concern is not the {\em theoretical problem} to find an expansion (\ref{C9}) of $f(x)$, but rather the {\em practical problem} to obtain a representation (\ref{FN}) of $f(x)$ with $\gamma_n = b_n$ with a small number of terms, $N < 10$, say, which provides a fairly good approximation by minimizing the error (\ref{error}). To this end, it is convenient to choose an orthogonal basis $\lbrace \phi_n (x)\rbrace$ in $\mathbb{H}$, where each $\phi_n (x)$ is a polynomial of degree $n$. For the Gaussian weight function $w^{\rm G} (x) : = \exp{(-x^2 /2)} = \sqrt{2\pi} P^{\rm G} (x)$, it turns out that the polynomials are uniquely determined (up to a multiplicative constant in each polynomial) by the {\em Hermite polynomials} ${\rm He}_n (x)$ ({\em c.f.} \cite{abramowitz}), 
\begin{equation}
\phi_n (x) := {\rm He}_n (x) \;\;,
\end{equation}
which provide a complete orthogonal (not orthonormal) system in $\mathbb{H}= L^2 (\mathbb{R}, \exp{(-x^2 / 2)}\ {\mathrm d}x)$, satisfying the orthogonality relation (see (\ref{ortho}))
\begin{equation}
(\phi_m , \phi_n ) := h_n \;\delta_{mn}  \;\;\; (m,n = 0,1,2, \cdots )\;\;,
\end{equation}
with $h_n : = \sqrt{2\pi} n! \,$. In order that the piecewise continuous function $f(x)$ satisfies $|| f || < \infty$ in $\mathbb{H}$, it must obey the asymptotic condition
\begin{equation}
f(x) = O \left(\frac{\e^{x^2 / 4}}{|x|^{1/2 + \epsilon}}\right) \;\;, \;\; \epsilon > 0 \;\;,\;\; |x| \rightarrow \infty \;\;.
\end{equation}
The approximation (\ref{FN}) becomes then the {\em polynomial approximation}
\begin{equation}
\label{FNHe}
F_N (x) = \sum_{n=0}^{N} \frac{a (n)}{n!} \, {\rm He}_n (x) \;\;,
\end{equation} 
where the expansion coefficients $a(n) = \frac{1}{\sqrt{2\pi}} (f,\phi_n)$ are explicitly given by
\begin{equation}
a(n) = \frac{1}{\sqrt{2\pi}}\int_{-\infty}^{\infty} \e^{-x^2 / 2} \, f(x) \, {\rm He}_n (x) \, {\mathrm d} x \;\;\;(n = 0,1,2, \cdots ) \;\;.
\end{equation}
The {\em completeness relation} (\ref{completeness}) then reads:
\begin{equation}
\sum_{n=0}^{\infty} \frac{a^2 (n)}{n!} = \frac{1}{\sqrt{2\pi}} \int_{-\infty}^{\infty} e^{-x^2 /2}\, f^2 (x) \, {\mathrm d} x \;\;,
\end{equation}
which implies the asymptotic behaviour
\begin{equation}
\frac{|a(n)|}{n!} = O \left(\frac{1}{n^{1/2 + \delta}\sqrt{n!}}\right) \;\;,\;\; \delta> 0 \;\;, \;\; n \rightarrow \infty \;\;.
\end{equation}
It is important to bear in mind that, even in the case when the series (\ref{C9}) is uniformly convergent, it by no means follows that the 
$N^{th}$ partial sum (\ref{FNHe}) is the best selection of $N$ terms for representing the function $f(x)$. Even though (\ref{FNHe}) 
gives the best fit in the sense of {\em least squares} by minimizing the error (\ref{error}), it may be that some other measure of approximation is better suited for a given problem \cite{boasbuck}. All the more this may be the case if the series (\ref{C9}) is divergent, in which case
one may ask whether there exists an {\em asymptotic expansion} in the sense of Poincar\'e ({\em c.f.} \cite{erdelyi}).

In the particular case of the Hermite expansion (\ref{FNHe}), many authors have worked on asymptotic expansions since quite a long time,
mainly in the context of probability theory, statistics, number theory, and mathematical aspects of insurance risk ({\em c.f.} \cite{boasbuck,cramer1,cramer2,cramer3}, and references therein). The results relevant to us in this paper are connected with the attempts to give a refinement of the classical {\em central limit theorem} in probability theory, and are often, historically incorrect, referred to as {\em Charlier} or {\em Gram--Charlier A--series} and 
{\em Edgeworth expansion}, although they had been introduced by Tchebychev already before \cite{cramer3}. Since Matsubara's expansion of the MFs in 
\cite{matsu1,matsu2} (see also \cite{juszkiewicz} and \cite{blinnikov}), based on the assumption of hierarchical ordering (HO), is formally closely related to the Edgeworth expansion, we shall summarize the main properties of the latter.~It turns out that the Edgeworth expansion furnishes, in the case of the central limit theorem, a genuine asymptotic expansion with a well--defined remainder term.

Let $x_1,x_2, \cdots , x_n$ be independent and identically distributed random variables with a common continuous distribution function, such that every $x_k$ has zero mean, standard deviation $\sigma$, and a third absolute moment $\beta_3 = \langle|\delta T|^{3}\rangle$. Consider the standardized sum variable
\begin{equation}
y_n : = \frac{x_1 + x_2 + \cdots + x_n}{\sqrt{n}\sigma}\;\;,
\end{equation}
and let $P_n (x)$ be the PDF of $y_n$. The central limit theorem then asserts that, as $n \rightarrow \infty$, under appropriate conditions,
$P_n (x)$ tends to the Gaussian (normal) PDF, $P^{\rm G} (x) = \frac{1}{\sqrt{2\pi}}\exp{(-x^2 / 2})$. The Edgeworth expansion is a refinement 
of this by giving also the {\em rate of convergence} to the Gaussian limit. 
Define the {\em discrepancy function} 
\begin{equation}
\delta_n (x) : = \sqrt{2\pi} \left( P_n (x) - P^{\rm G} (x) \right) \;\;.
\end{equation}
Then, the {\em Edgeworth (E) expansion} \cite{boasbuck,cramer1,cramer2,cramer3} is the following {\em asymptotic expansion}:
\begin{equation}
\label{deltanx}
\delta^E_n (x) = \e^{-x^2 /2} \left[ \sum_{k=1}^K \frac{q_k (x)}{n^{k/2}} + {\mathcal O}\left(\frac{1}{n^{(K+1)/2}}\right)\right] \;\;\;(n \rightarrow \infty)\;\;,
\end{equation}
where $q_k (x)$ is a polynomial of degree $3k$, which only depends on the normalized cumulants $C_n \; (n \ge 3; C_3 = \gamma_1, C_4 = \gamma_2)$, and which can be expressed as a linear combination of the Hermite polynomials ${\rm He}_n (x)$. For $k=1,\cdots,4$ they
are explicitly given by:
\begin{eqnarray}
\label{hermiteq}
\fl\quad
q_1 (x) = \frac{\gamma_1}{3!} {\rm He}_3 (x) \quad;\nonumber \\
\fl\quad
q_2 (x) = \frac{\gamma_2}{4!} {\rm He}_4 (x) + \frac{10 \gamma_1^2}{6!} {\rm He}_6 (x) \quad;\nonumber \\
\fl\quad
q_3 (x) = \frac{C_5}{5!} {\rm He}_5 (x) + \frac{35 \gamma_1 \gamma_2}{7!} {\rm He}_7 (x) + \frac{280 \gamma_1^3}{9!} {\rm He}_9 (x)\quad;\nonumber \\
\fl\quad
q_4 (x) = \frac{C_6}{6!} {\rm He}_6 (x) + \frac{35 \gamma^2_2 + 56 \gamma_1 C_5}{8!} {\rm He}_8 (x) + \frac{2100 \gamma_1^2 \gamma_2}{10!} {\rm He}_{10} (x) + \frac{15400 \gamma_1^4}{12!} {\rm He}_{12} (x) \;\;.
\end{eqnarray}
The expansion (\ref{deltanx}), $\delta_n^E (x)$, taking the first $K$ terms into account, is an asymptotic expansion of $\delta_n (x)$ in powers of $1/\sqrt{n}$ with a remainder of the same order as the first term neglected (i.e., it is an ``asymptotic expansion to $K$ terms" as defined by Poincar\'e, {\em c.f.} \cite{erdelyi}). Thus, it gives a correction to the central limit theorem, with a well--defined error of order
$n^{-(K+1)/2}$, in the case where $n$ is finite but large $(1\ll n < \infty)$. Due to the Hermite expansions (\ref{hermiteq}) of the polynomials $q_k (x)$, the Edgeworth expansion (\ref{deltanx}) can be interpreted as a particular rearrangement of the following polynomial approximation:
\begin{equation}
\label{delta3K}
\delta_{n,3K} (x): = \e^{-x^2 /2} \left[ \sum_{m=3}^{3K} \frac{a^E (m)}{m!} {\rm He}_m (x) + {\mathcal O}\left(\frac{1}{n^{(K+1)/2}}\right) \right]
\;\;\;(n \rightarrow \infty ) \;,
\end{equation}
where the ordering is not with respect to powers of $1/\sqrt{n}$, but rather according to the order of the Hermite polynomials. (The 
$n-$dependence of the coefficients $a^E (m)$ is not explicitly noted.) By a comparison of (\ref{delta3K}) with (\ref{deltanx},\ref{hermiteq}) one immediately reads off the first $12$ expansion coefficients (here we choose $K=4$): 
\begin{eqnarray}
\label{AE}
\fl
\qquad a^E (3) = \frac{\gamma_1}{\sqrt{n}} \quad;\quad
a^E (4) = \frac{\gamma_2}{n} \quad;\quad
a^E (5) = \frac{C_5}{n^{3/2}} \quad;\quad
a^E (6) = \frac{10\gamma_1^2}{n} + \frac{C_6}{n^2}\quad;  \nonumber\\
\fl
\qquad a^E (7) = \frac{35 \gamma_1 \gamma_2}{n^{3/2}} \quad;\quad
a^E (8) = \frac{56 \gamma_1 C_5 + 35 \gamma_2^2}{n^{2}}  \quad;\quad
a^E (9) = \frac{280 \gamma_1^3}{n^{3/2}}  \quad;\nonumber\\
\fl
\qquad a^E (10) = \frac{2100\gamma_1^2 \gamma_2}{n^2} \quad;\quad
a^E (11) = 0 \quad;\quad
a^E (12) = \frac{15400 \gamma_1^4}{n^2}\;\;.
\end{eqnarray}
(Note that the coefficients $a^E (m)$, $m \ge 7$, change if one goes to higher order in $1/\sqrt{n}$, e.g., for $K=5$ they receive additional 
contributions of order $n^{-5/2}$ and $n^{-3}$.) If the coefficients (\ref{AE}) are compared with the coefficients $a_P (m)$ in equation (\ref{AP1}), derived for a {\em general} Hermite expansion for the non--Gaussianities as in (\ref{Delta2}), it is not difficult to derive the law that governs the size, with respect to powers of $1/\sqrt{n}$, of the coefficients $a^E (m)$ associated with the asymptotics (\ref{deltanx}) of the central limit theorem. It simply says that every cumulant $C_r$ in (\ref{AP1}) has to be replaced by $C_r / n^{(r-2)/2}$ (for fixed $n \gg 1, r \ge 3$).

This is precisely the law, rigorously proven for the central limit theorem, which in some models of inflation is built in as {\em hierarchical ordering} (HO). But then, the role of the small dimensionless expansion parameter $1/\sqrt{n}$ is played by the standard deviation
$\sigma_0$ which, for the CMB anisotropy, has a fixed value and cannot be made arbitrarily small as in the case of the central limit theorem.
It is, thus, important to perform numerical checks, as done in this paper. 
Furthermore, one should keep in mind that the Edgeworth expansion, applied to a given PDF, can suffer from the fact that the PDF 
is not correctly normalized and it can exhibit undesirable properties such as negative probabilities. In particular, the approximation 
deteriorates in the tails. 

\bigskip\bigskip

\section*{References}
\bibliographystyle{iopart-num}

\end{document}